\documentclass[3p,preprint,times]{elsarticle}
\usepackage{amssymb}
\usepackage[figuresright]{rotating}

\newcommand{\dd}{\mathrm{d}}
\newcommand{\ReN}{\operatorname{Re}}

\newcommand{\GeV}{\mathrm{GeV}}

\newcommand{\fm}{\mathrm{fm}}
\newcommand{\ii}{\mathrm{i}}

\usepackage[utf8]{inputenc}
\usepackage[T1]{fontenc}
\usepackage{tikz}
\usepackage{cancel}
\usepackage{graphicx,color} 
\usepackage{amssymb}
\usepackage{amsmath} 
\usepackage{longtable}
\usepackage{array}
\usepackage{overpic}
\usepackage{hyperref}

\begin{document}

\begin{frontmatter}




\title{Kramers's escape rate problem within a non-Markovian description}


\author[1]{B.\ Sch{\"u}ller}
\ead{schueller@th.physik.uni-frankfurt.de}

\author[1]{A.\ Meistrenko}
\ead{meistrenko@th.physik.uni-frankfurt.de}

\author[1]{H.\ van Hees}
\ead{hees@th.physik.uni-frankfurt.de}

\author[2]{Z.\ Xu}
\ead{xuzhe@mail.tsinghua.edu.cn}

\author[1]{and C.\ Greiner}
\ead{Carsten.Greiner@th.physik.uni-frankfurt.de}

\address[1]{Institut f{\"u}r Theoretische Physik, Goethe-Universit{\"a}t
  Frankfurt am Main, Max-von-Laue-Straße 1, 60438 Frankfurt am Main, Germany
}

\address[2]{Department of Physics, Tsinghua University and Collaborative
	Innovation Center of Quantum Matter, Beijing 100084, China}

\begin{abstract}
  We compare the thermal escape rates of a Brownian particle, initially
  trapped into one of the two wells of an asymmetric double-well
  potential, for thermal Markovian and non-Markovian noise. The Markovian
  treatment of this problem goes originally back to the studies of
  Kramers in 1940 and is therefore often referred to as ``Kramers's
  escape rate problem''. We solve the generalized Langevin equation for
  the trajectories of the particles numerically and analytically for
  both limiting cases, Markovian and non-Markovian thermal noise. We compute the
  escape rate and work out the fundamental differences arising from
  finite correlation times of the thermal noise.
\end{abstract}


\end{frontmatter}

% \linenumbers

\section{Introduction}
Since the seminal development of the theory of Brownian motion by
Einstein \cite{Einstein:1905} and Langevin's formulation in terms of a
stochastic process
\cite{Langevin:1908_brown,Lemons:1997_langevin_translation} this framework has found
applications in a very broad range of fields of physics, chemistry,
engineering, and finance mathematics \cite{Coffey:2004_langevin_eq}. Of
particular interest are also semi-classical descriptions of the dynamics
of open quantum systems
\cite{Caldeira:1981rx,Caldeira:1982uj,Weiss:1999_quantum_diss} and
non-equilibrium relativistic quantum field theory with applications in
(inflationary) cosmology and the early universe like thermalization,
decoherence and structure formation (see e.g. Ref.\
\cite{Calzetta-Hu:2008} and references within) and with applications in
the description of the hot and dense strongly interacting matter as
created in ultrarelativistic heavy-ion collisions like the Markovian and
non-Markovian dynamics of disoriented chiral condensates, heavy quarks,
the chiral phase transition, and baryon diffusion
\cite{GLEISER1993,Knoll:1995nz,Greiner:1996dx,Rischke:1998qy,Greiner:1998vd,Xu:1999aq,Farias:2009stochastic,Calzetta-Hu:2008,
  Dunkel-Haenggi:2008,FARIAS2008,Rapp:2009my,Andronic:2015wma,Nahrgang:2011mg,Herold:2013bi,Kapusta:2014dja}.

The general concept of a Brownian particle, initially trapped in a
metastable state and being able to escape from it via thermally
activated fluctuations can describe a large variety of phenomena from
different fields of science as for example the transport of electrons in
semiconductors, the diffusion of impurities bound in a harmonic lattice,
biophysical transport problems like the migration of ligands in
biomolecules and chemical reactions \cite{HanggiNM:1982,Hanggi:1983}.  After an empirical analysis of
various reaction-rate data in the late 19th century Svante Arrhenius
concluded that the rate of escape out of the metastable state obeys the
following law:
\begin{equation}
k=\nu\exp\left[-\frac{E_{\text{b}}}{k_{\text{B}}T}\right],
\label{eq:vantHoffArr}
\end{equation}
where $\nu$ is some prefactor, which will be specified later in the
course of this work, $E_{\text{b}}$ is the energy the Brownian particle
must attain to escape, $k_{\text{B}}$ is the Boltzmann constant, and $T$
denotes the temperature.  In the literature this general result for the
rate of escape from a metastable state is referred to as Van't
Hoff-Arrhenius law \cite{vantHoff:1884,Arrhenius:1889,Hanggi:1990}.

Subsequently, investigators tried to determine the actual form of the
prefactor $\nu$ in Eq.\ \eqref{eq:vantHoffArr} using different
approaches. One of them was Hendrik Antonie Kramers in 1940 in his work
on a diffusion model of chemical reactions \cite{Kramers:1940}.
%Besides, major contribution to this field can be attributed to H{\"a}nggi,
%Talkner and Borcovic

This work, based on B.S.'s Master's thesis \cite{MThesis_Schueller:2018}, is precisely focused
on this diffusion model, dealing with the thermally activated rate of
escape of a Brownian particle, initially trapped in a potential
well. Kramers's classical model, characterized by a Markovian thermal
noise, will be extended to the case of non-Markovian thermal noise
terms. Thereby, the main objectives will be computing Kramers's escape
rate for Markovian and non-Markovian noise numerically as a function of
the damping rate $\beta$ and working out the differences between these
two cases. Furthermore, an attempt will be made to explain the occurring
differences.

To this end, the generalized Langevin equation (GLE), Eq.\ \eqref{eq:genlang1}, is solved for an
asymmetric double-well potential, using a Markovian and three
non-Markovian thermal noise variants.

This work is organized as follows. In Sec.\ \ref{chap:genCN}, the
algorithm for the generation of non-Markovian noise, used for the
numerical simulations in this work, is presented.

Sec.\ \ref{chap:KEP} is devoted to Kramers's diffusion model.  Besides
the classical model, also extensions to it will be introduced, before
analytical results for the escape rate of the Markovian and one of the
non-Markovian thermal noise variants are reviewed.

Thereafter, Sec.\ \ref{seq:numStud} addresses the detailed numerical
simulations and the comparison of numerical with analytical results.
After presenting the actual numerical setup, Kramers's escape rate as a
function of the damping rate is presented for different correlation
functions and correlation times.

Finally, in Sec.\ \ref{chap:concl} the results of this work are
summarized.  These results and
methods are applicable in various physical surroundings, however,
motivated by high energy nuclear and particle physics natural units are used, $\hbar=c=k_{\text{B}}=1$ and
$\mathrm{fm}\,\mathrm{GeV}=0.197^{-1}$.

\section{Generating colored noise}
\label{chap:genCN}

This section is devoted to the method for the generation of stationary
Gaussian colored noise, the numerical simulations of this work are based
on. The method was developed in Ref.\ \cite{Xu:1999aq}
and recently employed in Ref.\ \cite{Schmidt:2014zpa}, where a detailed
instruction for the numerical implementation of this method is indicated
as well.

It should be noted here that the two terms, white and colored noise, 
which will be frequently used in the further course of this work, correspond 
to Markovian and non-Markovian noise, respectively. That terminology 
originates from considerations concerning the spectral density of the 
correlation function of the stationary Gaussian noise. While the spectral 
density is constant for a $\delta$-correlated Markovian noise, it is 
dependent on the frequency for non-Markovian noise \cite{Risken:1996}.

% It should be noted here that this section is partially based on
%Ref.\ \cite{BThesis:2015}.

Before the actual method is presented several preliminary considerations
are needed.  The starting point is a very general expression for a
centered stochastic process $\xi(t)$ which consists of $n$ random pulses
in a time interval $[0,T]$ \cite{Heer:1972}:
\begin{align}
\xi(t) =\sum_{i=1}^n a_i b(t-t_i), \quad t \in[0,T], 
% \left<\xi(t)\right> &=0 \Leftrightarrow \left<a_i\right>=0.
\label{eq:genProcess}
\end{align}
where $\left<\xi(t)\right> =0 \Leftrightarrow \left<a_i\right>=0$. While
$n$, $a_i$, and $t_i$ denote random variables, $b(t)$ designates an
arbitrary pulse shape. The number of pulses in the time interval $[0,T]$
is supposed to be Poisson-distributed with mean $\bar{n}=\mu T$, whereby
$\mu$ identifies with the mean rate of pulses in $[0,T]$.  Furthermore
$a_i$ is the random height of the $i$-th pulse and $t_i$ the random
instant of time for the occurrence of a pulse.

The next step is to find an expression for white noise. Since white
noise is $\delta$-correlated a reasonable choice for the pulse shape
$b(t)$ of white noise is \cite{Schmidt:2014zpa}
\begin{equation}
b(t)=\sqrt{\frac{D}{\mu \sigma^2}}\delta(t),
\label{eq:pulseWhiteNoise}
\end{equation}
where $D$ is an arbitrary positive real number, whose meaning will later be specified in a physical context and $\sigma^2$ denotes the variance of the pulse height $a_i$. With this pulse shape
for white noise the corresponding centered stochastic process
$\xi_{\text{w}}(t)$, where the subscript stands for white, reads
\begin{align}
\xi_{\text{w}}(t) &=\sqrt{D}\bar{\xi}_w(t), 
% \bar{\xi}_w(t) &= \sum_{i=1}^n \frac{a_i}{\sigma\sqrt{\mu}}\delta(t-t_i)= \sum_{i=1}^n \frac{\bar{a}_i}{\sqrt{\mu}}\delta(t-t_i), 
% \quad \bar{a}_i :=\frac{a_i}{\sigma}
\label{eq:whiteNoise}
\end{align}
where 
\begin{equation}
\begin{split}
\bar{\xi}_w(t) &= \sum_{i=1}^n \frac{a_i}{\sigma\sqrt{\mu}}\delta(t-t_i) \\
&= \sum_{i=1}^n \frac{\bar{a}_i}{\sqrt{\mu}}\delta(t-t_i), 
\quad \bar{a}_i :=\frac{a_i}{\sigma}.
\end{split}
\end{equation}
In the limit of a large rate of pulses $\mu$ ($\mu \to \infty$) and a
small variance $\sigma^2$ of the distribution function $p(a)$ of the
pulse height ($\sigma^2 \to 0$), the $\delta$-correlated white
stochastic process becomes Gaussian \cite{Xu:1999aq}.  It should be
noted, that by use of the central limit theorem the distribution
function for $a_i$ is optional and by definition of the white noise
\eqref{eq:whiteNoise} the prefactor $D$ of the pulse shape $b(t)$
\eqref{eq:pulseWhiteNoise} is identified with the strength of the
fluctuative force from the classical Langevin equation (LE) (see
Ref.\ \cite{Risken:1996}).

A centered Gaussian process is uniquely determined by its first two moments:
\begin{align}
\left<\xi(t)\right> &=0, \\
\left<\xi(t)\xi(t')\right> &=\mu\sigma^2\int_0^T b(t-s)b(t'-s)\dd s := C(t,t').
\end{align}
For the following considerations the correlation function $C(t,t')$ of
the Gaussian process needs to be stationary, meaning the correlation
function shall not be dependent on the times $t$ and $t'$ separately but
on the time difference $|t-t'|$, i.e. $C(t,t')=C(|t-t'|)$
\cite{Schmidt:2014zpa}. This can be attained by demanding a symmetric
correlation function \cite{Schmidt:2014zpa}. In what follows the purpose
is to determine the pulse shape $b(t)$ of a stationary Gaussian process
given a stationary correlation function. By use of the Wiener-Khinchin
theorem, stating that the spectral density $S_{\xi}(\omega)$ of a
stationary process is obtained by the Fourier transform of its
correlation function $C(t,t')$ \cite{Risken:1996}, one arrives at
\begin{equation}
S_{\xi}(\omega)=\mathcal{F}[C]=\mu\sigma^2 |\tilde{b}(\omega)|^2.
\label{eq:specDens}
\end{equation}
Without loss of generality $\tilde{b}(\omega)$ is set to be real and
positive ($\tilde{b}(\omega)\geq 0, \,\forall \omega\in\mathbb{R}$).  In
this way Eq.\ \eqref{eq:specDens} can be simply solved for
$\tilde{b}(\omega)$. Subsequent back-transform of
$\tilde{b}(\omega)$ leads to
\begin{align}
&\tilde{b}(\omega)=\frac{1}{\sigma\sqrt{\mu}}\sqrt{S_{\xi}(\omega)}, \\
&\Rightarrow b(t)=\mathcal{F}^{-1}\left[\tilde{b}(\omega)\right](t)=\frac{1}{\sigma\sqrt{\mu}}G(t), \quad G(t)=\mathcal{F}^{-1}\left[\sqrt{S_{\xi}(\omega)}\right](t).
\label{eq:pulseShape}
\end{align}
From this, the general expression for a stationary Gaussian process (see
Eq.\ \eqref{eq:genProcess}) is readily transformed into the
following form, using the definition \eqref{eq:whiteNoise} of a Gaussian
white noise and relation \eqref{eq:pulseShape} for $b(t)$
\begin{equation}
\begin{split}
\xi(t) &= \sum_{i=1}^n a_i b(t-t_i) \\
&= \sum_{i=1}^n \int_{-\infty}^{\infty} a_i b(t-t')\delta(t'-t_i)\dd t'=\int_{-\infty}^{\infty} b(t-t')\sum_{i=1}^n a_i \delta(t'-t_i)\dd t' \\
&= \int_{-\infty}^{\infty} b(t-t')\sigma\sqrt{\mu}\bar{\xi}_w(t')\dd t'=\int_{-\infty}^{\infty}G(t-t')\bar{\xi}_w(t')\dd t'.
\end{split}
\end{equation}
Hence, the method for generating stationary Gaussian colored noise,
described in this section, is primarily based on the determination of
the underlying pulse shape $b(t)$ of a stationary correlation function
$C(|t-t'|)$ and the subsequent convolution of this pulse shape $b(t)$
with a sequence of $\delta$-correlated Gaussian white noise
$\xi_{w}(t)$.

\begin{figure}[tb]
	\begin{center}
		\begin{overpic}[width=0.8\textwidth]{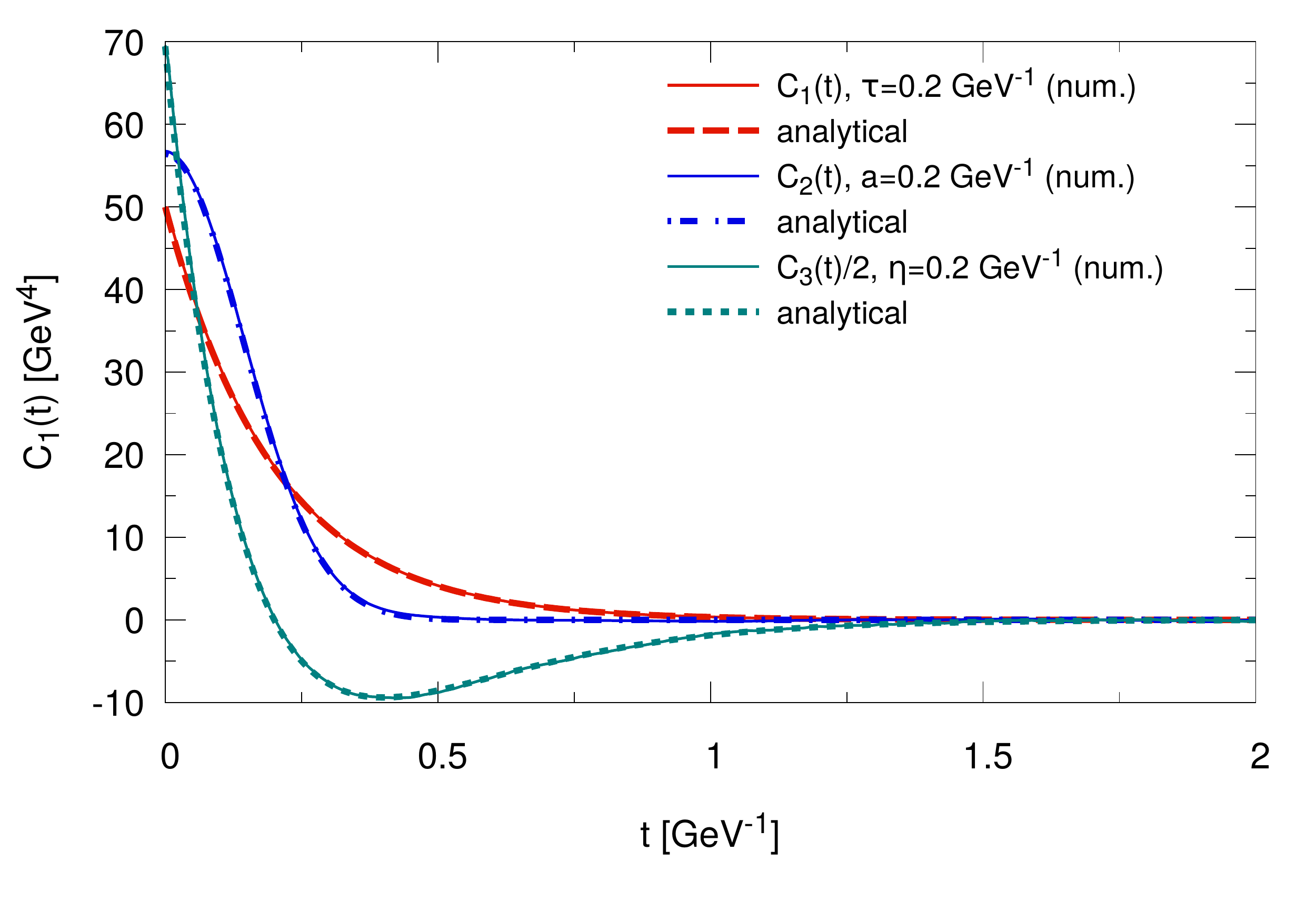}\end{overpic} 
		\caption{Comparison of the numerical correlations
                  $\left<\xi(t)\xi(0)\right>$ for $C_1$, $C_2$ and $C_3$
                  (red, blue and green line), averaged over $8\cdot10^5$
                  runs, with the appropriate analytical expressions
                  \eqref{eq:C1}-\eqref{eq:C3} (red long dashed line,
                  blue dashed/dotted line, green small dashed line), where
                  $m=1.11 \, \GeV$, $T=1 \, \GeV$, $D=20 \, \GeV^3$ for 
                  correlation functions $C_1(t)$ and $C_2(t)$ and $g=20$ for 
                  correlation function $C_3(t)$. The time is given in units of $\mathrm{GeV}^{-1}$ and can be converted 
                  to $\fm$ by means of relation $\mathrm{fm}\,\mathrm{GeV}=0.197^{-1}$.}
		\label{corrFunc}
	\end{center}
\end{figure}

In the course of this work various correlation functions are
investigated, which are listed below together with their corresponding
Fourier transforms,
\begin{align}
C_1(|t|):=\left<\xi(t)\xi(0)\right> &=\frac{D}{2\tau}\exp\left[-\frac{|t|}{\tau}\right], \label{eq:C1} \\
C_2(|t|):=\left<\xi(t)\xi(0)\right> &=\frac{D}{a\sqrt{\pi}}\exp\left[-\left(\frac{|t|}{a}\right)^2\right], \label{eq:C2} \\
C_3(|t|):=\left<\xi(t)\xi(0)\right> &=\frac{g}{4}k_{\text{B}}T\alpha^2\left(1-\frac{\alpha}{\sqrt{m}}|t|\right)
\exp\left[-\frac{\alpha}{\sqrt{m}}|t|\right] \label{eq:C3}
\end{align}
and
\begin{align}
\mathcal{F}[C_1]:&=\frac{D}{1+\tau^2\omega^2}, \label{eq:FourierC1} \\
\mathcal{F}[C_2]:&=D\exp^{-\frac{\alpha^2\omega^2}{4}}, \\
\mathcal{F}[C_3]:&=\frac{gk_{\text{B}}T\alpha^3\omega^2}{\sqrt{m}\left(\omega^2+\frac{\alpha^2}{m}\right)^2}, \label{eq:fourierC3}
\end{align}
where $\alpha$ is given by relation \eqref{eq:eta}, $g$ is a dimensionless coupling constant (see \ref{sec:C3}) and the following convention for the Fourier transform has been employed
\begin{align}
\Gamma(t) &=\frac{1}{2\pi}\int_{-\infty}^{\infty}e^{-\ii \omega t}\tilde{\Gamma}(\omega)\dd \omega,\label{eq:Fourier} \\
\tilde\Gamma(\omega) &=\int_{-\infty}^{\infty}e^{\ii\omega t}\Gamma(t)\dd t \label{eq:inverseFourier}.
\end{align}
For these correlation functions evidence of the validity of the
indicated method is given in Fig.\ \ref{corrFunc}.  Herein the
correlation of the colored noise $\left<\xi(t)\xi(0)\right>$, obtained
by numerically averaging an ensemble of particle trajectories %particles
, is compared to the appropriate analytical expression of the
correlation function.  The first two correlation functions $C_1$ and
$C_2$ do have an immediate intuitive interpretation, the first being an
exponential decay and the second being a Gaussian distribution. The
interpretation of the third correlation function $C_3$ is not as
trivial. Obviously, $C_3$ becomes slightly negative in the past time,
and the Fourier transform of $C_3$ vanishes for $\omega \rightarrow
0$. Such a dissipative kernel is rather typical in a quantum field
theoretical setting in a self-interacting theory like a scalar
$\Phi^4$-theory (see e.g. Ref.\ \cite{Xu:1999aq}). Some peculiarities of
this particular correlation function $C_3$ are given in Appendix C,
where for a free Brownian motion no full thermalization is observed.

\section{Kramers's escape rate problem} 
\label{chap:KEP}

\subsection{Classical Model}
\label{sec:classMod}
\begin{figure}
	\begin{center}
		\begin{tikzpicture}[xscale=3.5,yscale=0.52,samples=400]
		%geplottete funktion
		\draw[red, ultra thick, domain=-1.5:1.56] plot (\x, {2.5-3.75*\x-10*\x*\x+1.25*\x*\x*\x+5*\x*\x*\x*\x});
		\node [right] at (-1.5,6) {V(x)};
		
		%dashed lines + arrow
		\draw [dashed, ultra thick] (-1.4, 2.8495) -- (0,2.8495);
		\draw [dashed, ultra thick] (-1.4, 0) -- (0,0);
		\draw [<->,ultra thick] (-0.1875,0) -- (-0.1875,2.8495);
		\node [right] at (-0.1875,1.5) {$E_{\text{b}}$};
		
		%Legende
		\draw [->,ultra thick] (-2,-6.5) -- (2,-6.5) node [below right] {$X$};
		
		%a-well
		\draw [-,ultra thick] (-1,-6.5) -- (-1,-6.1);
		\node [below] at (-1,-6.5) {A};
		\node [above] at (-1,-6.1) {$x_a$};
		\node [above] at (-1,0.1) {$\omega_a$};
		
		%barrier
		\draw [-,ultra thick] (-0.1875,-6.5) -- (-0.1875,-6.1);
		\node [below] at (-0.1875,-6.5) {B};
		\node [above] at (-0.1875,-6.1) {$x_b$};
		\node [above] at (-0.1875,2.8495) {$\omega_b$};
		
		%c-well
		\draw [-,ultra thick] (1,-6.5) -- (1,-6.1);
		\node [below,thick] at (1,-6.5) {C};
		\node [above,ultra thick] at (1,-6.1) {$x_c$};
		\node [above] at (1,-5) {$\omega_c$};
		
		%Kreissegment für fluss
		\draw [<-,black, ultra thick] (0.3,3.5) arc [radius=0.5, start angle=20, end angle= 160];
		\node [above] at (-0.1875,3.8) {$k_{a\rightarrow c}$};
		
		%help lines
		%\draw[help lines] (-2,-5) grid (2,7);
		\end{tikzpicture}
	\end{center}
	\caption{Asymmetric double-well potential $V(x)$ of Kramers's
          classical escape rate problem, consisting of the A-well with
          frequency $\omega_a$ located around $x_a$ and
          the C-well with frequency $\omega_c$ located around
          $x_c$. Both wells are separated by a barrier around $x_b$ with
          barrier height $E_{\text{b}}$ and frequency $\omega_b$.
          Original figure from Ref.\ \cite{Hanggi:1990}.}
	\label{KramersPot}
\end{figure}
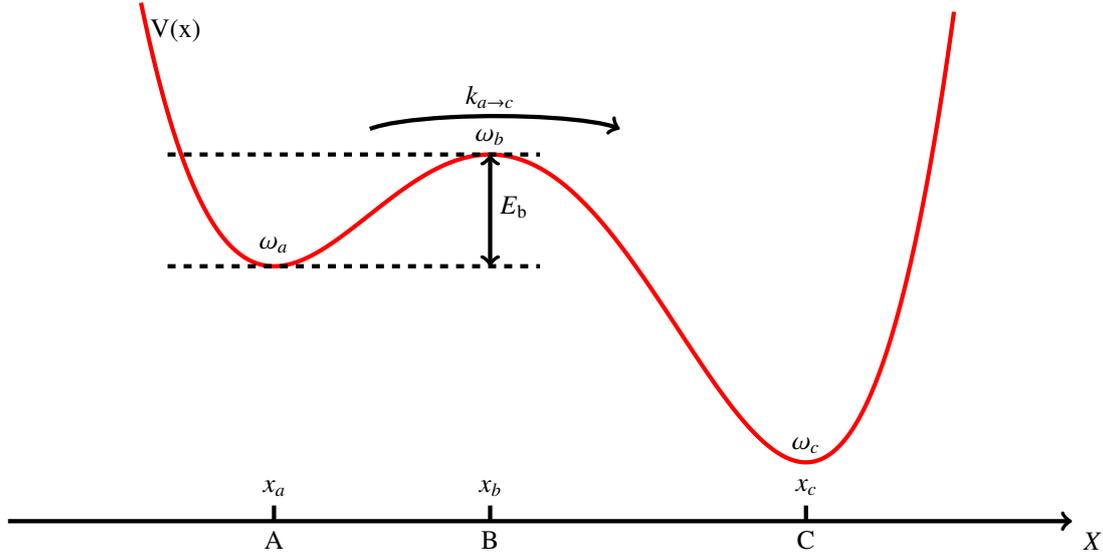

In 1940 Kramers established a model for chemical reactions in his paper on ``Brownian motion in a field of force and the diffusion model of chemical reactions'' (see Ref.\ \cite{Kramers:1940}). Herein, Kramers describes a chemical reaction by two metastable states divided by an intermediate state. The transition from one to the other state shall be thermally activated. 
This situation is then approximated by a classical Brownian particle of mass $m$ inside a
one-dimensional asymmetric double-well potential
\cite{Hanggi:1990,Kramers:1940} (see Fig.\ \ref{KramersPot}).
The two metastable states, corresponding in this model to the two wells of the asymmetric double-well potential, constitute the reactant and
product state located at $x_a$ and $x_c$, respectively. The intermediate
state represented by the maximum of the barrier between these two wells
at $x=x_b$ is designated as transition state \cite{Hanggi:1990}. The
position coordinate $x$ of the particle, describing the course of a
chemical reaction is fittingly referred to as reaction coordinate
\cite{Hanggi:1990}. 
Furthermore, the Brownian particle moving in the potential $V(x)$
is thought to be surrounded by a thermal environment in form of a heat
bath at temperature $T$. This heat bath, constituting a stochastic force
$\xi(t)$ and a friction force $F_r=-\gamma v$, has to be understood as a
consequence of the residual degrees of freedom of the system
\cite{Hanggi:1990}. The appropriate LE, describing the above
characterized dynamics of the Brownian particle is given by the
classical LE, complemented by the external potential field
$V(x)$,
\begin{equation}
\begin{split}
\dot{x} &= v , \\
\dot{v} &=-\frac{1}{m}\frac{dV(x)}{dx}-\beta v +\frac{\xi}{m}, \quad \left\langle \xi(t) \right\rangle=0, \quad \left\langle \xi(t)\xi(0) \right\rangle=D\delta(t)
\end{split}
\label{eq:langevinEq1}
\end{equation}
where $\xi(t)$ denotes a centered $\delta$-correlated and
Gaussian-distributed noise and $\beta=\gamma/m$.
The strength $D$ of the stochastic force $\xi(t)$ and $\beta$ in 
Eq.\ \eqref{eq:langevinEq1} are linked by the fluctuation-dissipation relation,
\begin{equation}
	\beta=\frac{\gamma}{m}=\frac{D}{2k_BTm}  \label{eq:gamma},
\end{equation}    
which states that both, frictional and stochastic force, originate from 
the same source. 

Dealing with an ergodic system, Kramers
considers an ensemble of particles, meaning an entirety of many similar
particles, all evolving independently from each other
\cite{Kramers:1940}.  Each of these particles is supposed to be
initially trapped in the potential well near the reactant state $x_a$.
Induced by many subsequent, thermally activated collisions with the
solvent molecules, constituting the thermal environment, the Brownian
particle will potentially, yet rarely be able to surmount the potential barrier at some point.

Kramers's escape rate problem is then to determine the probability for
this Brownian particles to overcome the barrier, whereby the barrier
height $E_{\text{b}}$ is supposed to be large compared to the energy
$E_{\mathrm{th}}=k_{\text{B}}T$ supplied by the thermal bath
\cite{Kramers:1940}:
\begin{equation}
k_{\text{B}}T \ll E_{\text{b}}.
\label{eq:Cond}
\end{equation}
In this way the Brownian particle will thermalize before escaping from
the initial well. Condition \eqref{eq:Cond}, furthermore, leads to a
clear-cut separation of time scales for $\tau_a:=2\pi\omega_a^{-1}$ and the
escape time
$\tau_e\approx\tau_a\exp\left[\frac{E_{\text{b}}}{k_{\text{B}}T}\right]\gg
\tau_a$, which always needs to hold when dealing with rate problems
\cite{Hanggi:1990}. Since under this condition the escape from the
initial well is very slow, Kramers assumes the diffusion process to be
quasi-stationary \cite{Kramers:1940}, which will be important for later
calculations (see Sec.\ \ref{sec:derivation}).

Thus, the quasi-stationary current from the initial well over the barrier
is given by the probability rate for the Brownian particles to leave the well,
$k_{A\rightarrow C}$, multiplied with the number of particles, $n_a$, being
located in this well \cite{Kramers:1940}:
\begin{equation}
j_b=k_{A\to C} n_a.
\label{eq:fluxKramers}
\end{equation}

The coupling strength of the considered Brownian particles, the thermal
bath and potential other degrees of freedom are completely determined by the
friction coefficient $\beta$ \cite{Hanggi:1990}. Depending on its
actual value Kramers differentiates between two regimes, the weak- and
strong-friction regime \cite{Kramers:1940}. While the weak-friction
regime is governed by an almost frictionless oscillation of the
respective Brownian particle in the bottom of the well, the
high-friction regime is determined by the spatial diffusive dynamics of
the Brownian particle around the barrier top
\cite{Hanggi:1990,Kramers:1940}.

\begin{figure}[tb]
	\begin{center}
		\begin{overpic}[width=1.0\textwidth]{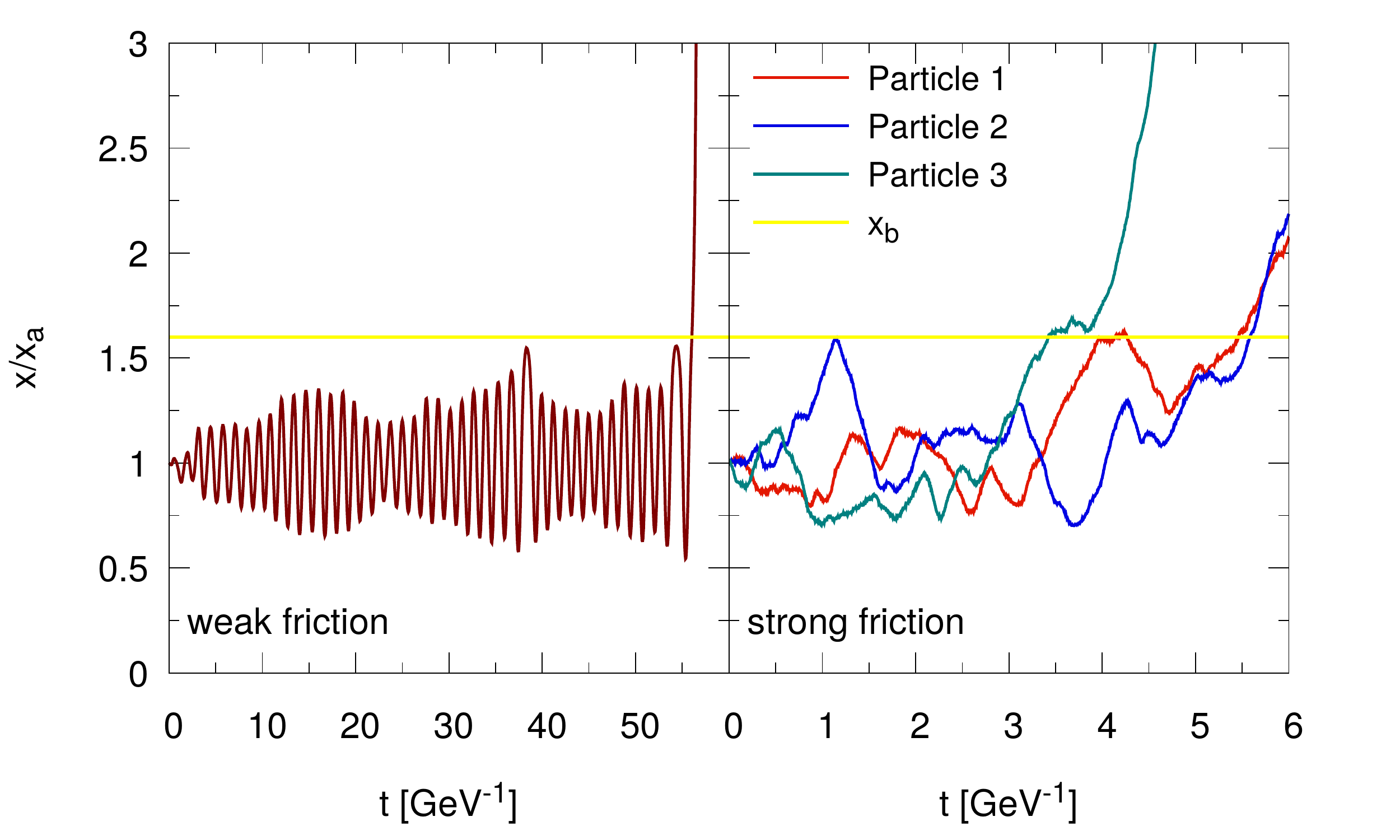}\end{overpic} 
		\caption{Position as a function of time for several Brownian particles and Markovian noise, Eq.\ \eqref{eq:C0}, moving in the potential V(x), Eq.\ \eqref{eq:potential}, where $E_{\text{b}}=2.5 \,\GeV$, $\omega_b=5 \,\GeV$,
			$m=1.11 \,\GeV$, $T=1 \,\GeV$, $\beta=0.03375 \,\GeV$ in the (very) weak- (left figure) or $\beta=9 \,\GeV$ in the strong-friction regime (right figure) and $x_a$ denotes the initial position of the Brownian particle, respectively. The above-mentioned potential V(x) is depicted
			in Fig. \ref{potSim}, where the barrier top is located at $x_b=1.6 x_a$.}
		\label{trajectory}
	\end{center}
\end{figure}

To visualize the processes connected to the different limiting regimes,
Fig. \ \ref{trajectory} shows typical trajectories of several Brownian
particles, one for weak and three for strong friction, being subjected
to an asymmetric double-well potential (see Figs. \ref{KramersPot} and
\ref{potSim}). Note that not only the shape of the curves but also the
time scale of escape, i.e. the time that elapses until a Brownian particle 
crosses the barrier located at $x_b$, is significantly different for
both limiting regimes.

In the weak-friction regime a particle oscillating in the A-well loses
almost no energy due to friction loss during the time of an oscillation
\cite{Kramers:1940}. The energy loss $\Delta E$ in this limiting regime
can be expressed in terms of the action $I$ \cite{Hanggi:1990},
\begin{equation}
\Delta E=\beta I(E),
\label{eq:energyLoss}
\end{equation}
where $I(E)$ defines the action at energy $E$ given by
\begin{equation}
I(E)=\oint p\dd x.
\end{equation}
Using relation \eqref{eq:energyLoss}, the weak-friction regime occurs
whenever the energy loss during an oscillation is much smaller than the
thermal energy provided by the heat bath \cite{Hanggi:1990}, i.e.
\begin{equation}
\beta I(E) \ll k_{\text{B}}T.
\label{eq:condLowFric}
\end{equation}
A particle eventually reaching the barrier top by successive
accumulation of small amounts of energy will relax towards the
C-well.  Hence, in this limiting regime the rate of escape is controlled
by energy diffusion \cite{Hanggi:1990}, described by the following
diffusion equation \cite{Hanggi:1990,Kramers:1940}:
\begin{equation}
\frac{\partial P(E,t)}{\partial t}=\beta\frac{\partial}{\partial E}I(E)\left[1+k_{\text{B}}T\frac{\partial}{\partial E} \right]
\frac{\omega(E)}{2\pi}P(E,t).
\label{eq:eDiffusion}
\end{equation}
This diffusion equation can be derived by performing a canonical
transformation from position and momentum coordinates to action and
angle coordinates, $(x,p)\to(I,\phi)$, and subsequent averaging over the
angle $\phi$ to obtain the diffusion equation for the probability
density of the action, starting from the Klein-Kramers equation
\cite{Hanggi:1990,Kramers:1940},
\begin{equation}
\frac{\partial P(I,t)}{\partial t}=\beta\frac{\partial}{\partial I}I
\left[1+\frac{2\pi k_{\text{B}}T}{\omega(I)}\frac{\partial}{\partial I} \right]P(I,t).
\label{eq:iDiffusion}
\end{equation}
Thereby energy and action are related through the angular frequency
$\omega(I)$ by \cite{Hanggi:1990}
\begin{equation}
\frac{\partial E}{\partial I}=\frac{\omega(I)}{2\pi}.
\label{eq:E-I}
\end{equation}
Using relation \eqref{eq:E-I} differential
Eq.\ \eqref{eq:iDiffusion} is readily transferred into the appropriate differential
equation for the energy, Eq.\ \eqref{eq:eDiffusion}. The corresponding
steady-state escape rate $k_{A\to C}$ is then given by
\cite{Hanggi:1990}
\begin{align}
k_{A\to C}
&=\beta\frac{ I(E_{\text{b}})}{k_{\text{B}}T}\frac{\omega_a}{2\pi}\exp\left[-\frac{E_{\text{b}}}{k_{\text{B}}T}\right], 
%   \quad \beta \to 0, \quad \frac{k_{\text{B}}T}{E_{\text{b}}}\ll 1, \quad \beta I_b\ll k_{\text{B}}T \\
%	    &=\beta\frac{E_{\text{b}}}{k_{\text{B}}T}\exp\left(-\frac{E_{\text{b}}}{k_{\text{B}}T}\right), 
\quad \beta \to 0, \, \ \frac{k_{\text{B}}T}{E_{\text{b}}}\ll 1, \, \ \beta I(E_{\text{b}})\ll k_{\text{B}}T. 
\label{eq:lowFric}
\end{align}
Gradually increasing the damping rate $\beta$ finally leads to a point,
where condition \eqref{eq:condLowFric} is no longer valid.  This limit,
which is characterized by the fact that the energy loss $\Delta E$
during the time of an oscillation is greater than the thermal energy,
i.e.
\begin{equation}
\beta I(E) > k_{\text{B}}T,
\label{eq:condLowFric1}
\end{equation}
is referred to as intermediate-to-strong-friction regime
\cite{Hanggi:1990}. Here the rate-determining mechanism is the dynamics
around the top of the barrier and the escape becomes controlled by
spatial diffusion, described by the Klein-Kramers equation \cite{Risken:1996},
\begin{equation}
\frac{\partial P(x,v,t)}{\partial t}=\left[-v\frac{\partial}{\partial x}-\frac{\partial}{\partial v}\left(-\frac{\gamma}{m}v - 
\frac{V'(x)}{m}\right)+\frac{D}{2m^2}\frac{\partial^2}{\partial v^2}\right]P(x,v,t),
\label{eq:FokkerPlanckEquation}
\end{equation}
which is a special Fokker-Planck equation (FPE).
Hereby it should be emphasized that a particle crossing the top of the
barrier $x_b$ will not necessarily be trapped into the neighboring
well. Instead, it can recross the barrier again and will, therefore,
reduce the escape rate.

The steady-state escape rate in the intermediate-to-strong-friction
regime, which will be explicitly derived in Subsec.\
\ref{sec:derivation}, is given by \cite{Hanggi:1990}:
\begin{equation}
k_{A \to C}=\frac{\lambda_\mathrm{M}}{\omega_b}\frac{\omega_a}{2\pi}\exp\left[-\frac{E_{\text{b}}}{k_{\text{B}}T}\right], \, \ \beta I(E_{\text{b}})> k_{\text{B}}T,
% \frac{\omega_a}{2\pi}\frac{\sqrt{\left(\frac{\beta}{2}\right)^2+\omega_b^2}-\frac{\beta}{2}}{\omega_b}\exp\left(-\frac{E_{\text{b}}}{k_{\text{B}}T}\right)
\label{eq:intermediate-to-strongFric}
\end{equation}
where 
\begin{equation}
\lambda_\mathrm{M}=\sqrt{\frac{\beta^2}{4}+\omega_b^2}-\frac{\beta}{2},
\label{eq:lamM}
\end{equation}
and the subscript M denotes the classical Markovian case. The expression
\eqref{eq:lamM} for the quantity $\lambda_\mathrm{M}$ will be motivated
later (see Subsec.\ \ref{sec:derivation}).  For large damping rates
$\beta$, that is $\beta\gg\omega_b$, Eq.\
\eqref{eq:intermediate-to-strongFric} can be expanded with respect to
$x:=\frac{\omega_b}{\beta}$ around $x\approx 0$, yielding
\begin{equation}
k_{A \to C}=\frac{\omega_b}{\beta}\frac{\omega_a}{2\pi}\exp\left[-\frac{E_{\text{b}}}{k_{\text{B}}T}\right], \quad \beta \to \infty.
\label{eq:strongFric}
\end{equation}
Altogether, it is to be stated that concerning $\beta$ there are two
limiting regimes, the weak- and the strong-friction regime, whereby the
escape rate $k_{A \to C}$ is proportional to $\beta$ in the weak- and
inversely proportional to $\beta$ in the strong-friction regime.

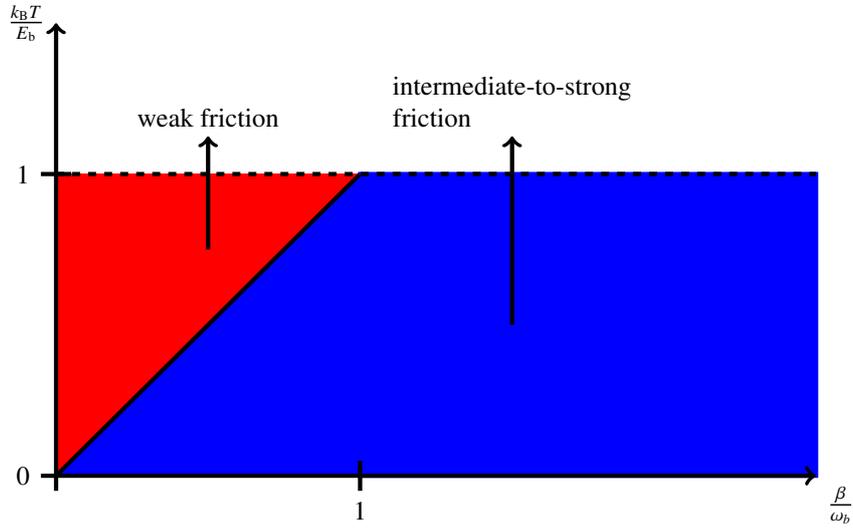
\begin{figure}
	\begin{center}
		\begin{tikzpicture}[xscale=4,yscale=4,samples=400]

		%roter Bereich
		\draw [red, fill=red] (-2,0) -- (-1,1) -- (-2,1) -- (-2,0);
		
		%blue Bereich
		\draw [blue, fill=blue] (-2,0) -- (-1,0) -- (-1,1) -- (-2,0);
		\draw [fill=blue,ultra thick,blue] (-1,0) rectangle (0.5,1);
		
		%geplottete funktion
		\draw [<->,ultra thick] (-2,1.5) node [left] {$\frac{k_{\text{B}}T}{E_{\text{b}}}$} -- (-2,0) -- (0.5,0) node [below right] {$\frac{\beta}{\omega_b}$};
		%		\draw [-,ultra thick] (-2,0) -- (-2,-2.5);
		\draw[ultra thick, domain=-2:-1] plot (\x, {\x+2});
		%		\draw[red, ultra thick, domain=-0.7:0.22] plot (\x, {2.5-13.888888887*(\x+0.4)*(\x+0.4)});
		
		%Beschriftung x-Achse 
		\draw [ultra thick] (-2,-0.05) -- (-2,0);
		\draw [ultra thick] (-1,-0.05) node[below]{1} -- (-1,0.05);
		
		%Beschriftung y-Achse
		\draw [ultra thick] (-2.05,0) node [left]{0}-- (-2,0);
		\draw [ultra thick] (-2.05,1) node [left]{1}-- (-1.95,1);

		%dashed lines + arrow
		\draw [-,dashed, ultra thick] (-2,1) -- (0.5,1);
		\draw [->, ultra thick] (-1.5,0.75) -- (-1.5,1.125);
		\node [above] at (-1.5,1.125) {weak friction};
		\draw [->, ultra thick] (-0.5,0.5) -- (-0.5,1.125);
		\node [align=left,above] at (-0.5,1.125) {intermediate-to-strong\\friction};

		%help lines
		%				\draw[help lines] (-2,0) grid (0.5,1.5);
		\end{tikzpicture}
	\end{center}
	\caption{Classical-rate phase diagram for the two dimensionless
          parameters $\frac{k_{\text{B}}T}{E_{\text{b}}}$ and
          $\frac{\beta}{\omega_b}$. The red area denotes the region of
          weak and the blue area represents the region of
          intermediate-to-strong friction. Original figure from
          \cite{Hanggi:1990}.}
	\label{phaseDiag}
\end{figure}

The range of validity of formulas \eqref{eq:lowFric},
\eqref{eq:intermediate-to-strongFric} and \eqref{eq:strongFric} can be
combined into one single diagram, the classical-rate phase diagram,
depicting the different regimes as a function of the dimensionless
parameters $\frac{k_{\text{B}}T}{E_{\text{b}}}$ and
$\frac{\beta}{\omega_b}$ \cite{Hanggi:1990} (see Fig.\ \ref{phaseDiag}).
The separating region, also often referred to as turnover region, of
weak- and intermediate-to-strong-friction regime can be pointed out by
considering condition \eqref{eq:condLowFric1}.  While the
intermediate-to-strong formula \eqref{eq:intermediate-to-strongFric} is
certainly valid for \eqref{eq:condLowFric1}, for the limiting case of
$k_{\text{B}}T=\beta I(E_{\text{b}})\approx
\beta\frac{E_{\text{b}}}{\omega_a}$ or equivalently for
$\frac{k_{\text{B}}T}{E_{\text{b}}}=\frac{\beta}{\omega_a}$, neither
\eqref{eq:lowFric} nor \eqref{eq:intermediate-to-strongFric} and
\eqref{eq:strongFric} are applicable.

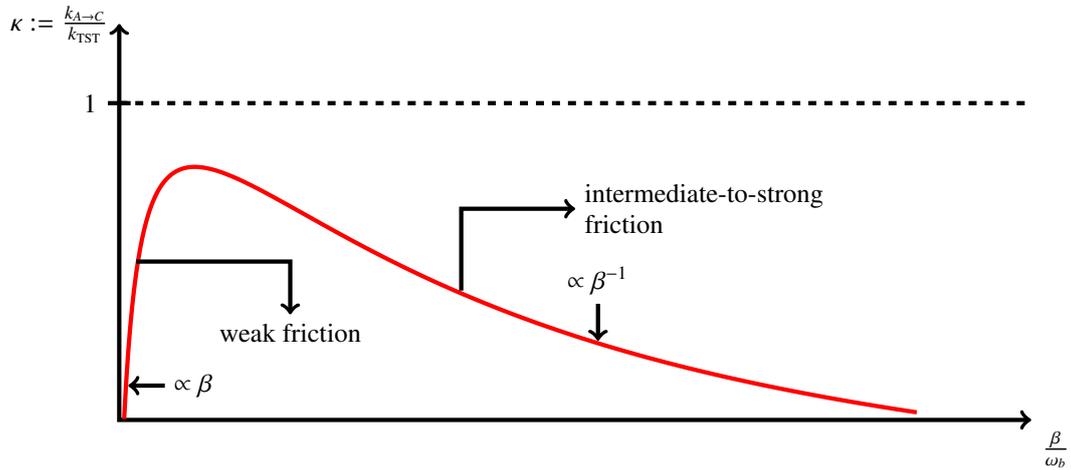
\begin{figure}
	\begin{center}
		\begin{tikzpicture}[xscale=3,yscale=7,samples=400]
		%geplottete funktion
		\draw [<->,ultra thick] (0,0.75) node [left] {$\kappa:=\frac{k_{A \to C}}{k_{\text{TST}}}$} -- (0,0) -- (4,0) node [below right] {$\frac{\beta}{\omega_b}$};
		%		\draw [-,ultra thick] (-2,0) -- (-2,-2.5);
		\draw[red, ultra thick, domain=0.0215:3.5] plot (\x, {pow(pow(5*pi*\x,-1.0)+pow((sqrt(pow(\x/2.0,2.0)+1)-\x/2.0),-1.0),-1.0)-0.25});

		%Beschriftung y-Achse
		\draw [ultra thick] (-0.05,0.6) node [left]{1}-- (0.05,0.6);
		
		%dashed lines + arrow
		\draw [-,ultra thick, dashed] (0,0.6) -- (4,0.6);
		\draw [->,ultra thick] (0.075,0.3) -- (0.75,0.3) -- (0.75,0.2);
		\node [align=left,below,black] at (0.75,0.2) {weak friction};
		\draw [<-,ultra thick] (0.04,0.065) -- (0.2,0.065);
		\node [align=left,right,black] at (0.2,0.065) {$\propto \beta$};
		\draw [->,ultra thick] (1.5,0.245) -- (1.5,0.4) -- (2,0.4);
		\node [align=left,right,black] at (2,0.4) {intermediate-to-strong\\ friction};
		\draw [<-,ultra thick] (2.1,0.15) -- (2.1,0.22);
		\node [above,black] at (2.1,0.22) {$\propto \beta^{-1}$};
		
		%help lines
		%				\draw[help lines] (0,0) grid (4,0.75);
		\end{tikzpicture}
	\end{center}
	\caption{Schematic representation of the bell-shaped curvature
          of the steady-state escape rate $k_{A \to C}$, normalized to the transition-state
          rate $k_{\text{TST}}$ (see Eq.\ \eqref{eq:KTST}), as a function of
          the dimensionless parameter $\frac{\beta}{\omega_b}$. Original figure
          from Ref.\ \cite{Hanggi:1990}.}
	\label{bellShape}
\end{figure}

Furthermore, given these two formulas it is not difficult to see that
both tend to zero in the limits of $\beta$ going to zero or $\beta$
going to infinity, respectively. From this, Kramers concluded that the
steady-state escape rate must possess a maximum between these two
limiting regimes \cite{Hanggi:1990,Kramers:1940}. The appearance of the
escape rate as a function of $\beta$ would therefore exhibit a
bell-shaped form, as depicted in Fig.\ \ref{bellShape}.
Ever since Kramers published his paper, researchers in this area tried
to find a way to join together the two limiting regimes within one single
formula, which yields the above-described bell-shaped form
%\cite{Grote:1980,Carmeli:1984,Hanggi1984,Matkowsky1984,Zawadzki:1985,Straub:1986}.
\cite{Carmeli:1984,Matkowsky1984,Zawadzki:1985,Straub:1986}.
A very simple and intuitive approach to give a bridging formula, only
using the already known formulas, Eqs.\ \eqref{eq:lowFric} and \eqref{eq:intermediate-to-strongFric}, reads \cite{Hanggi:1990}:
\begin{equation}
k_{A \to C}=\left(k^{-1}(\textrm{low damping})+k^{-1}(\textrm{moderate-to-strong damping})\right)^{-1}, \quad \forall\beta\in\mathbb{R}^{+}_0.
\label{eq:turnover}
\end{equation}
Before turning to the extensions of the classical model, special
attention has to be given to a term common to Eqs.\
\eqref{eq:lowFric}, \eqref{eq:intermediate-to-strongFric} and
\eqref{eq:strongFric} for the escape rate in the different limiting
regimes. This expression, given by
\begin{equation}
k_{\text{TST}}=\frac{\omega_a}{2\pi}\exp\left[-\frac{E_{\text{b}}}{k_{\text{B}}T}\right],
\label{eq:KTST}
\end{equation}
where the subscript, TST, stands for transition-state theory, denotes
the escape rate for the TST. The TST-rate is very similar to Kramers's escape
rate.  The substantial difference between these two rates is,
however, that TST considers a realization overcoming the potential
barrier will never return to the initial well \cite{Hanggi:1990}. Hence,
the TST-rate has to be always an upper bound to Kramers's escape rate
\cite{Hanggi:1990}.  This implies that $k_{\text{TST}}$ is an adequate scale to
normalize the steady-state escape rate $k_{A \to C}$ (see Fig.\
\ref{bellShape}),
\begin{equation}
k_{A \to C}=\kappa k_{\text{TST}}, \quad \kappa \le 1.
\label{eq:kappa}
\end{equation} 

\subsection{Extensions of the classical model}
\label{sec:nonMarkovMod}

Ever since Kramers published his work a great variety of extensions were
carried out for his classical model. Among other things Kramers's
classical one-dimensional treatment was extended to a multidimensional
system for both limiting regimes of the damping rate $\beta$
\cite{Brinkman:1956,Landauer:1961}. Furthermore corrections of the
escape rate in the spatial-diffusion regime arising from anharmonicities
of the potential \cite{TalknerR1993,Talkner1993,Talkner:1995}, the influence of a
non-Gaussian white noise \cite{Sukhor2007,Grabert2008,Baura2011}
and quantum effects like quantum tunneling
\cite{Ishioka1980,Wolynes:1981} were investigated \cite{HanggiNM:1982}.

All these extensions are of Markovian nature, meaning that there is a clear-cut separation
between the angular frequency $\omega_a$ in the initial potential well
and the correlation time $\tau_{\mathrm{corr}}$ \cite{HanggiNM:1982}, related to the thermal bath, of the
form,
\begin{equation}
\tau_{\mathrm{corr}}\ll\frac{2 \pi}{\omega_a}.
\label{eq:CondNonMarkov}
\end{equation}
If there exists such a clear separation between the relevant time scales the classical Markovian LE, Eq.\ \eqref{eq:langevinEq1}, is appropriate to describe the time-evolution of a Brownian particle being subjected to an external potential $V(x)$. However, this might not be the case for various applications \cite{HanggiNM:1982}.
Whenever $\tau_{\mathrm{corr}}$ is of the order of $2\pi\omega_a^{-1}$
or even larger, the classical escape rates $k_{A \to C}$ for weak and
strong friction $\beta$ (see Eqs.\ \eqref{eq:lowFric},
\eqref{eq:intermediate-to-strongFric}, \eqref{eq:strongFric}) derived by
Kramers are no longer applicable.  In this case a non-Markovian
treatment of Kramers's escape rate problem is required
\cite{HanggiNM:1982}. In contrast to the classical model (see previous
subsection) the Brownian motion in the asymmetric double-well potential
(see Fig.\ \ref{KramersPot}) is described by the GLE \cite{Cortes:1985},
complemented by the external potential field $V(x)$,
\begin{equation}
\begin{split}
\dot{x} &= v , \\
\dot{v} &=-\frac{1}{m}\frac{dV(x)}{dx}-\frac{1}{m}\int_0^t\Gamma(t-t')v(t')\dd t' +\frac{\xi}{m}, \quad \left\langle \xi(t) \right\rangle=0,
\end{split}
\label{eq:genlang1}
\end{equation}
whereby it should be recalled that the centered noise $\xi(t)$ and the
dissipation kernel $\Gamma(t)$ are related by the second
fluctuation-dissipation theorem
\begin{equation}
\Gamma(|t|)=\frac{1}{k_{\text{B}}T}\left<\xi(0)\xi(t)\right>.
\label{eq:dissFluc}
\end{equation}
As a note, Eq.\ \eqref{eq:langevinEq1} results from Eq.\ \eqref{eq:genlang1} if
$\Gamma(t-t')=2 \gamma \delta(t-t')$.  

Again there are two limiting regimes as a function of the damping rate
$\beta$, the weak- and the strong-friction regime. As in
the classical treatment the weak-friction regime is governed by energy
diffusion - or equivalently action diffusion - described by
\cite{Carmeli:1983}
\begin{equation}
\frac{\partial P(I,t)}{\partial t}=\frac{\partial}{\partial I}
\left\lbrace 2\pi\epsilon(I)\left[2\pi k_{\text{B}}T\frac{\partial}{\partial I}+\omega(I) \right]P(I,t) \right\rbrace,
\label{eq:IDiffusionNM}
\end{equation}
where $\omega(I)$ is specified by the potential $V(x)$ and
$\epsilon(I)$ is defined as \cite{Carmeli:1984,Grote:1982}.
\begin{equation}
%	\epsilon(I)=2M\sum_{n=1}^{\infty}\left|x_n\right|^2\ReN\left[ \tilde{\Gamma}_n(\omega(I))\right] 
\epsilon(I)=\frac{1}{\omega^2(I)}\int_{0}^{\infty}\Gamma(t)\left\langle v(0)v(t) \right\rangle\dd t.
\label{eq:epsilon}
\end{equation}
Hereby $v(t)$ is to be obtained by solving \eqref{eq:genlang1} without
dissipation kernel $\Gamma(t)$ and noise $\xi(t)$ \cite{Carmeli:1984}
for constant energy $E(I)$ and $\left\langle v(0)v(t) \right\rangle$ corresponds to the average over the initial phase $\phi_0$, where as in Subsec.\ \ref{sec:classMod}
relation \eqref{eq:E-I} applies.

From the diffusion equation for the action, Eq.\ \eqref{eq:IDiffusionNM}, the mean first passage time $\tau_{\mathrm{MFP}}(I_0,I)$
to reach a final action $I$, starting from an initial action $I_0$ can be derived
(\cite{Carmeli:1983} and references therein),
\begin{equation}
\tau_{\mathrm{MFP}}(I_0,I)=\frac{1}{k_{\text{B}}T}\int_{I_0}^{I}
\left\lbrace \frac{\exp\left[ \frac{E(x)}{k_{\text{B}}T}\right] }{\epsilon(x)}\int_{0}^{x} \exp\left[-\frac{E(y)}{k_{\text{B}}T} \right] \dd y\right\rbrace \dd x. 
\label{eq:MFP}
\end{equation} 
The steady-state escape rate $k_{A \to C}$ in the weak-friction regime
is then obtained by averaging the mean first passage time
$\tau_{\mathrm{MFP}}(I,I_b)$ with regard to the steady-state
distribution $P_{\mathrm{SS}}(I)$ inside the initial well \cite{Carmeli:1983}
\begin{equation}
k_{A \to C}=\left[ \int_{0}^{I_B} P_{\mathrm{SS}}(I)\tau_{\mathrm{MFP}}(I,I_b) \dd I \right]^{-1}.
\label{eq:lowFricNM_temp}
\end{equation}
Supposing the well is deep enough, it can be assumed that
$P_{\mathrm{SS}}(I)$ is Boltzmann distributed. Inserting the Boltzmann distribution together 
with Eq.\ \eqref{eq:MFP} in Eq.\ \eqref{eq:lowFricNM_temp} a very compact approximate 
formula for the steady-state escape rate $k_{A \to C}$ in the weak-friction regime
is obtained \cite{Carmeli:1983}:
\begin{equation}
	k_{A \to C}=\frac{\omega_a\epsilon(I_b)\omega(I_b)}{k_{\text{B}}T}\exp\left[-\frac{E_{\text{b}}}{k_{\text{B}}T}\right],
	\label{eq:lowFricNM}
\end{equation}
where $\epsilon(I_b)$ is to be computed via Eq.\ \eqref{eq:epsilon}.

In the intermediate-to-strong-friction regime the corresponding
diffusion equation is referred to as generalized Fokker-Planck equation
(GFPE) and given by Eq.\ \eqref{eq:genFokkerPlanck}, which will be
discussed in detail in Subsec.\ \ref{sec:derivation}. The associated
escape rate $k_{A \to C}$ is \cite{HanggiNM:1982}
\begin{equation}
k_{A \to C}=
%	\frac{\omega_a}{2\pi}\frac{\sqrt{\left(-\frac{\bar{\beta}}{2}+\sqrt{\frac{\bar{\beta}^2}{4}+\bar{\omega}^2}\right)}}
%	{\omega_b}\exp\left[-\frac{Eb}{k_{\text{B}}T}\right]=
\frac{\lambda_{\mathrm{NM}}}{\omega_b}\frac{\omega_a}{2\pi}\exp\left[-\frac{E_{\text{b}}}{k_{\text{B}}T}\right],
\label{eq:intermediate-to-strongFricNM}
\end{equation}
where $\lambda_{\mathrm{NM}}$ is defined as
\begin{equation}
\lambda_{\mathrm{NM}}=\sqrt{\frac{\bar{\beta}^2}{4}+\bar{\omega}^2}-\frac{\bar{\beta}}{2}
\label{eq:lamNM}
\end{equation}
and the subscript NM represents the non-Markovian case. The meaning of $\bar{\beta}$ and $\bar{\omega}$
and the relation \eqref{eq:lamNM} will be specified in Subsec.\
\ref{sec:derivation}. This result for
the steady-state escape rate $k_{A \to C}$ in case of a non-Markovian
treatment of Kramers's classical escape rate problem is formally
identical to the appropriate result for the classical model (see
Eq.\ \eqref{eq:intermediate-to-strongFric}). One only needs to
exchange $\lambda_\mathrm{M}$ with $\lambda_{\mathrm{NM}}$ or the bare
damping $\beta$ and frequency $\omega_b$ with their non-Markovian
analogues $\bar{\beta}$ and $\bar{\omega}$ \cite{HanggiNM:1982}. For
correlation function $C_1$, Eq.\ \eqref{eq:C1}, the computation of
$\lambda_{\mathrm{NM}}$ is indicated in \ref{sec:escRateC1}.

\subsection{Derivation of Kramers's escape rate in the spatial-diffusion
  regime (intermediate-to-strong friction)}
\label{sec:derivation} 

In this subsection the appropriate quasi-steady-state escape rate
$k_{A \to C}$ from the reactant well A to the product well C (see Fig.\
\ref{KramersPot}) in the intermediate-to-strong-friction regime, also
referred to as spatial-diffusion regime, will be explicitly derived,
following what was done in Ref.\ \cite{Hanggi:1990,Kramers:1940} in the
Markovian case and Ref.\ \cite{HanggiNM:1982} in the non-Markovian case.

For the following considerations it is possible to handle the
quasi-steady-state rate as a real steady-state rate without influencing
the underlying physics, provided that the condition
$E_{\text{b}}\gg k_{\text{B}}T$ holds \cite{Carmeli:1984}. To that end
the initial A-well is provided with a source, feeding it with particles
at energies much smaller than the barrier height $E_{\text{b}}$ and the
B-well with a sink, removing particles that traversed the barrier
\cite{Hanggi:1990,Carmeli:1984}.

Before starting with the actual derivation it should be emphasized, that
the steady-state escape rate in the spatial-diffusion regime is
essentially characterized by the dynamics around the top of the barrier
at $x_b$ \cite{HanggiNM:1982}. In both cases the main task will be to
determine the stationary probability density $\rho(x,v)$, obeying
various boundary conditions - which will be specified later - for the
stationary current $j$. For a given probability density $\rho(x,v)$ it
is then easy to compute the population $n_a$ of the Brownian particles
in the initial A-Well, given by
\begin{equation}
n_a=\int_{-\infty}^{x_b}\rho(x,v)\dd x \dd v
\label{eq:na}
\end{equation}
and the current $j_b$ with respect to the barrier top at $x=x_b$,
obtained by
\begin{equation}
j_b=\int_{-\infty}^{\infty}v\rho(x_b,v) \dd v.
\label{eq:j}
\end{equation}
Inserting the appropriate solutions of \eqref{eq:na} and \eqref{eq:j} into
Eq.\ \eqref{eq:fluxKramers} the steady-state escape rate
$k_{A\to C}$ from the A- to the C-well is readily calculated.

\subparagraph{Markovian case} The Markovian Brownian motion in an
external potential field $V(x)$ is described by the LE, Eq.\
\eqref{eq:langevinEq1}. This equation can be transformed into its
corresponding FPE, Eq.\ \eqref{eq:FokkerPlanckEquation}.
%\begin{equation}
%\frac{\partial P(x,v,t)}{\partial t}=\left[-v\frac{\partial}{\partial x}-\frac{\partial}{\partial v}\left(-\frac{\gamma}{m}v - 
%\frac{V'(x)}{m}\right)+\frac{D}{2m^2}\frac{\partial^2}{\partial v^2}\right]P(x,v,t).
%\label{eq:FokkerPlanckEquation}
%\end{equation}
As already mentioned above, the essential dynamics of the
spatial-diffusion regime is restricted to the vicinity of the barrier
top. Expanding the potential $V(x)$ around $x_b$, i.e.
\begin{equation}
V(x)\approx V(x_b)-\frac{1}{2}m\omega_b^2(x-x_b)^2, \quad x\approx x_b,
\label{eq:potLimb}
\end{equation}
the corresponding FPE reads 
\begin{equation}
\left[-v\frac{\partial}{\partial x}-\frac{\partial}{\partial v}\left(-\frac{\gamma}{m}v + 
\omega_b^2(x-x_b)\right)+\frac{D}{2m^2}\frac{\partial^2}{\partial v^2}\right]\rho(x,v)=0, \quad x\approx x_b,
\label{eq:FP_barrier}
\end{equation}
where the in general dynamic probability density $P(x,v,t)$ is replaced
by the stationary probability density $\rho(x,v)$ in search of a
stationary escape rate.

To determine a general solution $\rho(x,v)$ for Eq.\ \eqref{eq:FP_barrier}
Kramers then used the ansatz \cite{Hanggi:1990,Kramers:1940}
\begin{equation}
\rho(x,v)=\frac{1}{Z}\Xi(x,v)\exp\left[-\frac{\frac{1}{2}mv^2+V(x)}{k_{\text{B}}T}\right].
\label{eq:rhoGen}
\end{equation}
Following what Kramers did two limiting cases for $\rho(x,v)$, leading
to several boundary conditions for $\Xi(x,v)$, have to be considered.
Inside the well in a small area around the bottom located at $x_a$ (see
Fig.\ \ref{KramersPot}), the particles are assumed to be
thermalized. This is a reasonable requirement given that
$E_{\text{b}}\gg k_{\text{B}}T$. Hence, the probability density around
$x\approx x_a$ is well approximated by a Boltzmann distribution,
\begin{equation}
\rho(x,v)=\frac{1}{Z}\exp\left[-\frac{\frac{1}{2}mv^2+V(x)}{k_{\text{B}}T}\right], \quad x\approx x_a.
\label{eq:rhoLim1}
\end{equation}
Comparing both expressions, Eqs.\ \eqref{eq:rhoGen} and
\eqref{eq:rhoLim1}, the first boundary condition for $\Xi(x,v)$ is
identified as
\begin{equation}
\Xi(x,v) \approx 1, \quad x\approx x_a.
\label{eq:lim1}
\end{equation}
Furthermore the probability density $\rho(x,v)$ is supposed to vanish
beyond the barrier at $x=x_b$, i.e.
\begin{equation}
\rho(x,v)\approx 0, \quad x> x_b,
\end{equation}
since the particles are removed by a sink leading to
\begin{equation}
\Xi(x,v)\approx 0, \quad x> x_b.
\label{eq:lim2}
\end{equation}
For $\Xi(x,v)$ to obey these two limits (Eqs.\ \eqref{eq:lim1} and
\eqref{eq:lim2}) Kramers assumed it to be only dependent on a linear
combination of position and velocity \cite{Hanggi:1990,Kramers:1940},
i.e.
\begin{equation}
\Xi(x,v)=\Xi(z), \quad z=v-b(x-x_b),
\label{eq:xiLinComb}
\end{equation}
where $b$ denotes a yet undetermined constant.  By inserting the general
expression for the probability density $\rho(x,v)$, Eq.\
\eqref{eq:rhoGen}, into the FPE of $\rho(x,v)$ around $x_b$, Eq.\
\eqref{eq:FP_barrier} the appropriate FPE for $\Xi(x,v)$ is obtained:
\begin{equation}
\left[-v\frac{\partial}{\partial x}-\left(\frac{\gamma}{m}v + 
\omega_b^2(x-x_b)\right)\frac{\partial}{\partial v}+\frac{D}{2m^2}\frac{\partial^2}{\partial v^2}\right]\Xi(x,v)=0.
\label{eq:FP_barrier1}
\end{equation}
Using furthermore relation \eqref{eq:xiLinComb} the FPE for $\Xi(x,v)$,
Eq.\ \eqref{eq:FP_barrier1}, can be converted into the
corresponding FPE for $\Xi(z)$,
\begin{align}
%\left[vb  \frac{\partial \xi}{\partial z}-\left[\frac{\gamma}{m}v+\omega_b^2(x-x_b) \right]\frac{\partial}{\partial z}+\frac{D}{2m^2}\frac{\partial^2}{\partial z^2}\right]\Xi(z) &=0, \\
\left[\left(\left(b-\frac{\gamma}{m}\right)v-\omega_b^2(x-x_b)\right)\frac{\partial}{\partial z}+\frac{D}{2m^2}\frac{\partial^2}{\partial z^2}\right]\Xi(z) &=0,
\end{align}
where the relations
\begin{align}
\frac{\partial \Xi}{\partial x} &=\frac{\partial\Xi}{\partial z}\underbrace{\frac{\partial z}{\partial x}}_{=-b}=-b\frac{\partial\Xi}{\partial z},\\
\frac{\partial \Xi}{\partial v} &=\frac{\partial\Xi}{\partial z}\underbrace{\frac{\partial z}{\partial v}}_{=1}=\frac{\partial\Xi}{\partial z},\\
\frac{\partial^2 \Xi}{\partial v^2} &=\frac{\partial}{\partial v}\left(\frac{\partial\Xi}{\partial v}\right)=
\frac{\partial}{\partial v}\left(\frac{\partial\Xi}{\partial z}\right)= \frac{\partial}{\partial z}\left(\frac{\partial\Xi}{\partial v}\right)
= \frac{\partial}{\partial z}\left(\frac{\partial\Xi}{\partial z}\right)=\frac{\partial^2 \Xi}{\partial z^2}
\end{align}
have been applied. To proceed further by requiring that
\begin{equation}
\left(b-\frac{\gamma}{m}\right)v-\omega_b^2(x-x_b)=\lambda z,
\label{eq:propOrdDGL}
\end{equation}
Kramers transformed Eq.\ (\ref{eq:FP_barrier1}) into the ordinary differential equation
\begin{equation}
\left[\lambda z\frac{\partial}{\partial z}+\frac{D}{2m^2}\frac{\partial^2}{\partial z^2}\right]\Xi(z)=0, \quad \forall x\approx x_b, \,v.
\label{eq:ordDGL}
\end{equation}
Eqs.\ \eqref{eq:propOrdDGL} and \eqref{eq:xiLinComb}
determine the two constants $b$ and $\lambda$:
\begin{align}
&\left(b-\frac{\gamma}{m}\right)v-\omega_b^2(x-x_b)=\lambda z =\lambda(v-b(x-x_b)), \\
&\Rightarrow\left(b-\frac{\gamma}{m}\right)v-\omega_b^2(x-x_b)=\lambda v-\lambda b(x-x_b).
\end{align}
Therefore by comparison of coefficients one finds
\begin{align}
\lambda &= b-\frac{\gamma}{m}, \label{eq:lamb} \\
\lambda b &= \omega_b^2, \label{eq:b}
\end{align}
which leads to a quadratic relation for $b$ by insertion of Eq.\
\eqref{eq:lamb} into Eq.\ \eqref{eq:b}
\begin{equation}
b^2-\frac{\gamma}{m}b-\omega_b^2=0.
\end{equation}
Calculating the roots results in
\begin{equation}
b_{\pm}=\frac{\beta}{2}\pm\sqrt{\left(\frac{\beta}{2}\right)^2+\omega_b^2},
\end{equation}
where $\beta=\frac{\gamma}{m}$. Replacing then $b$ by $b_{\pm}$ in
Eq.\ \eqref{eq:lamb} $\lambda_{\pm}$ is obtained by
\begin{equation}
\lambda_{\pm}=-\frac{\beta}{2}\pm\sqrt{\left(\frac{\beta}{2}\right)^2+\omega_b^2}.
\end{equation}
Now that $\lambda_{\pm}$ are well defined, the next objective is to solve
the ordinary differential equation, Eq.\ \eqref{eq:ordDGL}, for
$\Xi(z)$.  Using the ansatz
\begin{equation}
\zeta=\frac{\partial \Xi}{\partial z},
%&\Rightarrow \frac{\partial \zeta}{\partial z}=\frac{\partial^2 \Xi}{\partial z ^2}
\end{equation}
differential equation \eqref{eq:ordDGL} can be transformed into
\begin{equation}
\frac{\partial \zeta}{\partial z}=-\frac{\lambda z}{A}\zeta, 
\label{eq:DGL_zeta}
\end{equation}
where $A=\frac{D}{2m^2}$. By integration of Eq.\ \eqref{eq:DGL_zeta} the
solution for $\zeta$ is given by
\begin{equation}
\zeta=\zeta_0\exp\left[-\frac{\lambda z^2}{2A}\right].
\end{equation}
To receive $\Xi$ another integration has to be performed
\begin{equation}
\Xi(z)=\zeta_0 \int_{-\infty}^{z}\exp\left[-\frac{\lambda s^2}{2A}\right]\dd s.
\label{eq:Xi}
\end{equation}
Due to boundary conditions \eqref{eq:lim1} and \eqref{eq:lim2} the integration of Eq.\ \eqref{eq:Xi} over all $z$ has to be equal to one which therefore determines the integration constant to be   
%Extending the upper integration border of Eq.\ \eqref{eq:Xi} to
%infinity, demanding the result to approach unity inside the A-well (see
%boundary condition \eqref{eq:lim1}), the integration constant $\zeta_0$
%is readily obtained:
\begin{equation}
\zeta_0=\sqrt{\frac{\lambda_+}{2\pi A}},
\end{equation} 
where $\lambda$ in Eq.\ \eqref{eq:Xi} is identified with the
positive root $\lambda_+$ for the integral to be convergent
\cite{Hanggi:1990}.  Finally, $\Xi$ is given in the following form:
\begin{equation}
\Xi(z)=\sqrt{\frac{\lambda_+}{2\pi A}} \int_{-\infty}^{z}\exp\left[-\frac{\lambda_+ s^2}{2A}\right]\dd s.
\label{eq:xiFinal}
\end{equation}
The next objective will be to determine the population of the A-well
$n_a$ and the current $j_b$ over the barrier top to subsequently derive
Kramers's result for the spatial-diffusion regime. Insertion of the
result for $\Xi$, Eq.\ \eqref{eq:xiFinal}, in Kramers's ansatz for
the probability density $\rho(x,v)$ \eqref{eq:rhoGen} and expanding the
potential $V(x)$ around $x_a$, i.e
\begin{equation}
V(x)\approx V(x_a)+\frac{1}{2}m\omega_a^2(x-x_a)^2, \quad x\approx x_a,
\label{eq:potLima} 
\end{equation}
$n_a$ is readily obtained calculating \eqref{eq:na} using Eqs.\
\eqref{eq:rhoLim1} and \eqref{eq:potLima}:
\begin{equation}
\begin{split}
n_a &=\int\limits_{-\infty}^{\infty}\rho(x,v)\dd x \dd v, \quad x\approx x_a\\
%& \overset{y=x-x_a}{\approx} \frac{1}{Z}\int_{-\infty}^{\infty}\exp\left[-\frac{\frac{mv^2}{2}+V(x_a)+\frac{m\omega_a^2 y^2}{2}}{k_{\text{B}}T}\right] \dd y \dd v \\
%&\approx \frac{1}{Z}\exp\left[-\frac{V(x_a)}{k_{\text{B}}T}\right]\underbrace{\left(\int_{-\infty}^{\infty}\exp\left[-\frac{mv^2}{2k_{\text{B}}T}\right]\dd v\right)}_{=\frac{\sqrt{\pi}}{\sqrt{\frac{m}{2k_{\text{B}}T}}}}
%\underbrace{\left(\int_{-\infty}^{\infty}\exp\left[-\frac{m\omega_a^2y^2}{2k_{\text{B}}T}\right]\dd y\right)}_{=\frac{\sqrt{\pi}}{\sqrt{\frac{m\omega_a^2}{2k_{\text{B}}T}}}} \\
&\approx \frac{1}{Z}\frac{k_{\text{B}}T}{m}\frac{2\pi}{\omega_a}\exp\left[-\frac{V(x_a)}{k_{\text{B}}T}\right].
\end{split}
\label{eq:nFinal}
\end{equation}
Computation of the integral \eqref{eq:j}, using the expansion of the
potential $V(x)$ around $x_b$ evaluated at $x_b$, Eq.\
\eqref{eq:potLimb}, and Eq.\ \eqref{eq:rhoGen} yields
\begin{equation}
\begin{split}
j_b &=\int_{-\infty}^{\infty}v\rho(x_b,v) \dd v \\
%&=\frac{1}{Z}\sqrt{\frac{\lambda_+}{2\pi A}}\int_{-\infty}^{\infty}\left(v\exp\left[-\frac{\frac{mv^2}{2}+V(x_b)}{k_{\text{B}}T}\right]
%\int_{-\infty}^{v}\exp\left[-\frac{\lambda_+z^2}{2A}\right]\dd z\right)\dd v \\
%&=\frac{1}{Z}\sqrt{\frac{m'}{\pi}}\exp\left[-\frac{V(x_b)}{k_{\text{B}}T}\right]
%\int_{-\infty}^{\infty}\left(v\exp\left[-kv^2\right]
%\underbrace{\int_{-\infty}^{v}\exp\left[-m'z^2\right]\dd z}_{=\frac{\sqrt{\pi}\left(1+\mathrm{erf}(v\sqrt{m'})\right)}{2\sqrt{m'}}}\right)\dd v \\
&=\frac{1}{Z}\sqrt{\frac{m'}{\pi}}\exp\left[-\frac{V(x_b)}{k_{\text{B}}T}\right]\frac{\sqrt{\pi}}{2\sqrt{m'}}
\underbrace{\int_{-\infty}^{\infty}v\exp\left[-kv^2\right]\dd v}_{=0}\\
&+\frac{1}{Z}\sqrt{\frac{m'}{\pi}}\exp\left[-\frac{V(x_b)}{k_{\text{B}}T}\right]\frac{\sqrt{\pi}}{2\sqrt{m'}}
\underbrace{\int_{-\infty}^{\infty}v\exp\left[-kv^2\right]\mathrm{erf}(v\sqrt{m'})\dd v}_
{=\frac{\sqrt{m'}}{k\sqrt{k+m'}}} \\
&=\frac{1}{Z}\exp\left[-\frac{V(x_b)}{k_{\text{B}}T}\right]\frac{\sqrt{m'}}{2k\sqrt{k+m'}},
\end{split}
\label{eq:jFinal}
\end{equation}
where to the third equality sign $k=\frac{m}{2k_{\text{B}}T}$ and
$m'=\frac{\lambda_+}{2A}$ were substituted.  Resubstitution of
$A=\frac{D}{2m^2}$ - making use of relation Eq.\ \eqref{eq:gamma} -,
%$A=\frac{\beta k_{\text{B}}T}{m}$, 
$k$ and $m'$ in \eqref{eq:jFinal}
yields
\begin{align}
j_b 
% &=\frac{1}{Z}\exp\left[-\frac{E_{\text{b}}}{k_{\text{B}}T}\right]\frac{\sqrt{\frac{\cancel{m}\lambda_+}{\cancel{2k_{\text{B}}T}\beta}}}
% {\cancel{2}\frac{m}{\cancel{2}k_{\text{B}}T}\sqrt{\frac{\cancel{m}}{\cancel{2k_{\text{B}}T}}+\frac{\cancel{m}\lambda_+}{\cancel{2k_{\text{B}}T}\beta}}} \\
% &=\frac{1}{Z}\frac{k_{\text{B}}T}{m}\exp\left[-\frac{E_{\text{b}}}{k_{\text{B}}T}\right]\sqrt{\frac{\frac{\lambda_+}{\beta}}{{1+\frac{\lambda_+}{\beta}}}} \\
% &=\frac{1}{Z}\frac{k_{\text{B}}T}{m}\exp\left[-\frac{E_{\text{b}}}{k_{\text{B}}T}\right]\sqrt{\frac{\lambda_+}{\beta+\lambda_+}} \\
% &=\frac{1}{Z}\frac{k_{\text{B}}T}{m}\exp\left[-\frac{E_{\text{b}}}{k_{\text{B}}T}\right]\sqrt{\frac{\lambda_+}{b_+}} \\
% &=\frac{1}{Z}\frac{k_{\text{B}}T}{m}\exp\left[-\frac{E_{\text{b}}}{k_{\text{B}}T}\right]\sqrt{\frac{\lambda_+^2}{\omega_b^2}} \\
=\frac{1}{Z}\frac{k_{\text{B}}T}{m}\frac{\lambda_+}{\omega_b}\exp\left[-\frac{V(x_b)}{k_{\text{B}}T}\right].
\label{eq:jFinal1}
\end{align}

Finally Kramers's result for the steady-state escape rate $k_{A\to C}$,
indicated in Subsec.\ \ref{sec:classMod}, is obtained by means of
Eq.\ \eqref{eq:fluxKramers}, using Eqs.\ \eqref{eq:nFinal} and
\eqref{eq:jFinal1} and defining $\lambda_+:=\lambda_\mathrm{M}$:
\begin{align}
k_{A\to C} 
%      &=\frac{j_b}{n_a}
%	  =\frac{\cancel{\frac{1}{Z}\frac{k_{\text{B}}T}{m}}\frac{\lambda_+}{\omega_b}\exp\left[-\frac{E_{\text{b}}}{k_{\text{B}}T}\right]}
%	  {\cancel{\frac{1}{Z}\frac{k_{\text{B}}T}{m}}\frac{2\pi}{\omega_a}\exp\left[-\frac{V_0}{k_{\text{B}}T}\right]} \\
&=\frac{\lambda_\mathrm{M}}{\omega_b}\frac{\omega_a}{2\pi}\exp\left[-\frac{E_{\text{b}}}{k_{\text{B}}T}\right],
%	  &=\frac{\lambda_+}{\omega_b}k_{\text{TST}}
\end{align}
where $E_{\text{b}}=V(x_b)-V(x_a)$.

\subparagraph{Non-Markovian case}

In case of colored noise the non-Markovian Brownian motion around the
barrier in an asymmetric double-well potential $V(x)$ can be described
by the GLE \eqref{eq:genlang1}, introducing the new notation $y=x-x_b$:
\begin{equation}
\begin{split}
\dot{y} &= \dot{x}= v, \\
\dot{v} &=\omega_b^2 y-\frac{1}{m}\int_0^t\Gamma(t-t')v(t')\dd t'+\frac{\xi(t)}{m},
\label{eq:genLangDeriv}
\end{split}
\end{equation}
where $V(x)$ is expanded around $x_b$ yielding 
\begin{equation}
V(y)=V(x_b)-\frac{1}{2}m\omega_b^2y^2,
\end{equation}
and $\xi(t)$ is a centered stationary Gaussian process
\begin{equation}
\left<\xi(t)\right>=0,
\end{equation}
obeying the second fluctuation-dissipation theorem (see Eq.\
\eqref{eq:dissFluc}). The corresponding GFPE around $x\approx x_b$ for the probability density
$P(x,v,t)$ of the system, described by \eqref{eq:genLangDeriv}, is given by \cite{HanggiNM:1982,Adelman:1976}
\begin{equation}
\begin{split}
\frac{\partial P(x,v,t)}{\partial t} =&\left[-v\frac{\partial}{\partial y}-\frac{\partial }{\partial v}\left(-\bar{\beta}(t)v+\bar{\omega}_b^2(t)y\right)
+\bar{\beta}(t)\frac{k_{\text{B}}T}{m}\frac{\partial^2}{\partial v^2}\right]P(x,v,t) \\
& + \frac{k_{\text{B}}T}{m\omega_b^2}\left(\bar{\omega}_b^2(t)-\omega_b^2\right)\frac{\partial^2 P(x,v,t)}{\partial v \partial y}
\label{eq:genFokkerPlanck}
\end{split}
\end{equation}
with
\begin{align}
\bar{\beta}(t) &=-\frac{\dot{a}(t)}{a(t)}, \\
\bar{\omega}_b^2(t) &=-\frac{b(t)}{a(t)},
\end{align}
where
\begin{align}
a(t) &=\chi_y(t)\dot{\chi}_v(t)-\dot{\chi}_y(t)\chi_v(t), \\
b(t) &=\dot{\chi}_y(t)\ddot{\chi}_v(t)-\ddot{\chi}_y(t)\dot{\chi}_v(t)
\end{align}
and 
\begin{equation}
\chi_y(t)=1+\omega_b^2\int_0^t\chi_v(\tau)\dd \tau.
\label{eq:chi_y_temp}
\end{equation}
In the latter equation $\chi_v(t)$ is given by the inverse Laplace
transform (LT)
\begin{equation}
\chi_v(t)=\mathcal{L}^{-1}\left[\frac{1}{s^2-\omega_b^2+\frac{\tilde{\Gamma}}{m}s}\right],
\end{equation}
where $\tilde{\Gamma}(s)$ is the LT of the dissipation kernel
$\Gamma(t)$. 
%In contrast to the Markovian case an explicit derivation of
%the GFPE is renounced here, since it is a very technical and expensive
%task, leading to no real insights. 
For a detailed derivation reference is made to
Ref.\ \cite{Adelman:1976}.  Nonetheless a brief motivation and explanation of
distinct terms of the above GFPE shall be given next. Comparing the
classical FPE and the GFPE (Eqs.\ \eqref{eq:FokkerPlanckEquation}
and \eqref{eq:genFokkerPlanck}) several similarities are
remarkable. Except for an additional diffusive term the GFPE corresponds
to the classical FPE, where the damping rate $\beta$ and the frequency
$\omega_b$ are replaced by a time dependent damping rate
$\bar{\beta}(t)$ and a time dependent frequency $\bar{\omega}_b(t)$.
Furthermore, both functions depend on the frequency $\omega_b$ and the
dissipation kernel $\Gamma(t)$ \cite{Adelman:1976}. In the Markovian
limit, where $\Gamma(t)=2\gamma\delta(t)$, $\bar{\beta}(t)=\beta$ and
$\bar{\omega}_b(t)=\omega_b$ the classical FPE is obtained.

The next step is to show, where relation \eqref{eq:chi_y_temp} is
derived from. Given the GLE \eqref{eq:genLangDeriv} and performing its
Laplace transform one obtains (using Eqs.\ \eqref{eq:exp},
\eqref{eq:faltung}, \eqref{eq:abl})
\begin{align}
s\tilde{Y}-y_0 &=\tilde{V}, \label{eq:LT1}\\
s\tilde{V}-v_0 &=\omega_b^2\tilde{Y}-\frac{\tilde{\Gamma}}{m}\tilde{V}+\frac{\tilde{\Xi}}{m} \label{eq:LT2},
\end{align}
where capital letters with tilde denote the Laplace transforms of the corresponding quantities.
Inserting the first relation, Eq.\ \eqref{eq:LT1}, into the second one, Eq.\ \eqref{eq:LT2}, and subsequently solving the resulting expression for
$\tilde{Y}$ yields
\begin{align}
s\left(s\tilde{Y}-y_0\right)-v_0&=\omega_b^2\tilde{Y}-\frac{\tilde{\Gamma}}{m}\left(s\tilde{Y}-y_0\right)+\frac{\tilde{\Xi}}{m},\\
% &\Rightarrow \left(s^2-\omega_b^2+\frac{1}{m}\tilde{\Gamma}s\right)\tilde{Y}=y_0\left(1+\frac{1}{m}\tilde{\Gamma}\right)
% +v_0+\frac{\tilde{\Xi}}{m}\\
&\Rightarrow \tilde{Y}=\frac{y_0\left(s+\frac{\tilde{\Gamma}}{m}\right)
	+v_0+\frac{\tilde{\Xi}}{m}}{s^2-\omega_b^2+\frac{\tilde{\Gamma}}{m}s}. \label{eq:yLT}
\end{align}
The inverse Laplace transform of Eq.\ \eqref{eq:yLT} then leads to \cite{Adelman:1976}
\begin{equation}
y(t)=y_0\chi_y(t)+v_0\chi_v(t)+\frac{1}{m}\int_0^t\chi_v(t-t')\xi(t')\dd t',
\end{equation}
where $\chi_y(t)$ and $\chi_v(t)$ are defined by 
\begin{align}
\chi_y(t)&=\mathcal{L}^{-1}
\left[\frac{\left(s+\frac{\tilde{\Gamma}}{m}\right)}{s^2-\omega_b^2+\frac{\tilde{\Gamma}}{m}s}\right]=
\frac{\left<y(t)y_0\right>}{\left<y_0^2\right>}, \label{eq:chi_y}\\
\chi_v(t)&=\mathcal{L}^{-1}\left[\frac{1}{s^2-\omega_b^2+\frac{\tilde{\Gamma}}{m}s}\right]=
\frac{\left<y(t)v_0\right>}{\left<v_0^2\right>}=\frac{m}{k_{\text{B}}T}\left<y(t)v_0\right>. \label{eq:chi_v}
\end{align}
By an analogous procedure the solution for $v(t)$ can be determined \cite{Adelman:1976}
\begin{equation}
v(t)=y_0\dot{\chi}_y(t)+v_0\dot{\chi}_v(t)+\frac{1}{m}\int_0^t\dot{\chi}_v(t-t')\xi(t')\dd t',
\end{equation}
where $\dot{\chi}_y(t)$ is given by
\begin{equation}
\dot{\chi}_y(t)=\omega_b^2\mathcal{L}^{-1}\left[\frac{1}{s^2-\omega_b^2+\frac{\tilde{\Gamma}}{m}s}\right]
\label{eq:dotChi_y}
\end{equation}
and $\dot{\chi}_v(t)$ is defined as
\begin{equation}
\dot{\chi}_v(t)=\mathcal{L}^{-1}\left[\frac{s}{s^2-\omega_b^2+\frac{\tilde{\Gamma}}{m}s}\right].
\end{equation}
Comparing Eqs.\ \eqref{eq:chi_v} and \eqref{eq:dotChi_y} the above
connection between $\chi_y(t)$ and $\chi_v(t)$ (see Eq.\
\eqref{eq:chi_y_temp}) is obtained,
\begin{equation}
\dot{\chi}_y(t) = \omega_b^2\chi_v(t) \, \Rightarrow \,\chi_y(t)=1+\omega_b^2\int_0^t\chi_v(\tau)\dd \tau,
\end{equation}
where the relation $\chi_y(0)=1$ has been employed, which follows from Eq.\ \eqref{eq:chi_y}. 

Turning now again to the actual task of this section, namely the
derivation of the steady-state escape rate, the first objective will be
to determine the stationary probability density $\rho(x,v)$. As in the
original derivation Kramers's ansatz
\begin{equation}
\rho(x,v)=\frac{1}{Z}\Xi(x,v)\exp\left[-\frac{\frac{mv^2}{2}+V(x)}{k_{\text{B}}T}\right]
\label{eq:KramersApproach}
\end{equation}
is used, where the same boundary conditions apply (see Eqs.\
\eqref{eq:lim1} and \eqref{eq:lim2}).  Inserting Eq.\
\eqref{eq:KramersApproach} into the GFPE \eqref{eq:genFokkerPlanck}, the
corresponding GFPE for $\Xi$ is obtained,
\begin{equation}
\begin{split}
v\frac{\partial \Xi}{\partial y} +  \bar{\omega}_b^2y\frac{\partial \Xi}{\partial v}=&\frac{k_{\text{B}}T}{m}\bar{\beta}\frac{\partial^2 \Xi}{\partial v^2}-\bar{\beta}v\frac{\partial \Xi}{\partial v}\\ &+\frac{k_{\text{B}}T}{m\omega_b^2}\left(\bar{\omega}_b^2-\omega_b^2\right)
\left[\frac{m\omega_b^2y}{k_{\text{B}}T}\frac{\partial \Xi}{\partial v}-\frac{mv}{k_{\text{B}}T}\frac{\partial \Xi}{\partial y}+\frac{\partial^2 \Xi}{\partial y \partial v}\right],
\end{split}
\end{equation}
whereby the time dependent functions $\bar{\beta}(t)$ and
$\bar{\omega}_b^2(t)$ from Eq.\ \eqref{eq:genFokkerPlanck} have been
substituted by the stationary quantities $\bar{\beta}$ and
$\bar{\omega}_b^2$, defined by
\begin{equation}
\bar{\beta}=\lim\limits_{t\to\infty}\bar{\beta}(t), \quad 	\bar{\omega}_b^2=\lim\limits_{t\to\infty}\bar{\omega}_b^2(t).
\end{equation}
Again $\Xi(y,v)$ is demanded to depend on a linear combination of y and
v,
\begin{equation}
\Xi(y,v)=\Xi(z), \quad z=v-by,
\end{equation}
where $b$ is again a yet undetermined constant.  With
\begin{align}
\frac{\partial \Xi(z)}{\partial v}&=\frac{\partial \Xi}{\partial z}\frac{\partial z}{\partial v}=\frac{\partial \Xi}{\partial z}, \\
\frac{\partial \Xi(z)}{\partial y}&=\frac{\partial \Xi}{\partial z}\frac{\partial z}{\partial y}=-b\frac{\partial \Xi}{\partial z}, \\
\frac{\partial^2 \Xi(z)}{\partial v^2}&=\frac{\partial}{\partial v}\left(\frac{\partial \Xi}{\partial z}\right)
=\frac{\partial}{\partial z}\left(\frac{\partial \Xi(z)}{\partial v}\right)=\frac{\partial}{\partial z}\left(\frac{\partial \Xi}{\partial z}\right)
=\frac{\partial^2 \Xi}{\partial z^2}, \\
\frac{\partial^2 \Xi(z)}{\partial y \partial v}&=\frac{\partial}{\partial y}\left(\frac{\partial \Xi}{\partial z}\frac{\partial z}{\partial v}\right)
=\frac{\partial}{\partial z}\left(\frac{\partial \Xi}{\partial y}\right)=\frac{\partial}{\partial z}\left(-b\frac{\partial \Xi}{\partial z}\right)=-b\frac{\partial^2 \Xi}{\partial z^2}
\end{align}
the GFPE for $\Xi(z)$ is given by
\begin{equation}
\begin{split}
\left(-vb+ \bar{\omega}_b^2y\right)\frac{\partial \Xi}{\partial z} = &\frac{k_{\text{B}}T}{m}\bar{\beta}\frac{\partial^2 \Xi(z)}{\partial z^2} - \bar{\beta}v\frac{\partial \Xi}{\partial z}+\frac{k_{\text{B}}T}{m\omega_b^2}\left(\bar{\omega}_b^2-\omega_b^2\right)\frac{m\omega_b^2y}{k_{\text{B}}T}\frac{\partial \Xi}{\partial z}\\
&+
\frac{k_{\text{B}}T}{m}c\left[\frac{mv}{k_{\text{B}}T}b\frac{\partial \Xi}{\partial z}-b\frac{\partial^2 \Xi}{\partial z^2}\right], 
\end{split}
% -vb\Xi'_z + \bar{\omega}_b^2y\Xi'_z &= \frac{k_{\text{B}}T}{m}\bar{\beta}\Xi''_{zz} - \bar{\beta}v\Xi'_z+\frac{k_{\text{B}}T}{m\omega_b^2}\left(\bar{\omega}_b^2-\omega_b^2\right)
% \left[\frac{m\omega_b^2y}{k_{\text{B}}T}\Xi'_z+\frac{mv}{k_{\text{B}}T}b\Xi'_z-b\Xi''_{zz}\right], \\
%  -vb\Xi'_z &= \frac{k_{\text{B}}T}{m}\bar{\beta}\Xi''_{zz} - \bar{\beta}v\Xi'_z-\omega_b^2y\Xi'_z+
%  \frac{k_{\text{B}}T}{m}c\left[\frac{mv}{k_{\text{B}}T}b\Xi'_z-b\Xi''_{zz}\right],
\end{equation}
where $c=\frac{\bar{\omega}_b^2-\omega_b^2}{\omega_b^2}$.  Rearranging
the terms provides
\begin{align}
% -vb\Xi'_z +\bar{\beta}v\Xi'_z+\omega_b^2y\Xi'_z-\frac{k_{\text{B}}T}{m}c\frac{mv}{k_{\text{B}}T}b\Xi'_z &= \frac{k_{\text{B}}T}{m}
% \left[\bar{\beta}-cb\right]\Xi''_{zz},\\
% -vb\Xi'_z +\bar{\beta}v\Xi'_z+\omega_b^2y\Xi'_z-bcv\Xi'_z &= \frac{k_{\text{B}}T}{m}
% \left[\bar{\beta}-cb\right]\Xi''_{zz},\\
% -\left[b-\bar{\beta}+bc\right]v\Xi'_z+\omega_b^2y\Xi'_z &=\frac{k_{\text{B}}T}{m}
% \left[\bar{\beta}-cb\right]\Xi''_{zz},\\
-\left[b(1+c)-\bar{\beta}\right]v\frac{\partial \Xi}{\partial z}+\omega_b^2y\frac{\partial \Xi}{\partial z} &=\frac{k_{\text{B}}T}{m}
\left[\bar{\beta}-cb\right]\frac{\partial^2 \Xi}{\partial z^2}. \label{eq:beforeOrdDGL}
\end{align}
The next task will be to transform Eq.\ \eqref{eq:beforeOrdDGL} into an ordinary differential equation by demanding
\begin{equation}
\left[b(1+c)-\bar{\beta}\right]v-\omega_b^2y = \lambda z=\lambda(v-by).
\end{equation}
By comparison of coefficients the two following relations are obtained:
\begin{align}
b(1+c)-\bar{\beta} &=\lambda, \label{eq:lam1NM}\\
\omega_b^ 2 &= b\lambda \Rightarrow \lambda=\frac{\omega_b^2}{b}. \label{eq:lam2NM}
\end{align}
Inserting Eq.\ \eqref{eq:lam1NM} into Eq.\ \eqref{eq:lam2NM} results in
a quadratic relation for $b$,
\begin{equation}
b(1+c)-\bar{\beta}=\frac{\omega_b^2}{b} \Rightarrow b=\frac{\omega_b^2}{b(1+c)-\bar{\beta}}.
\label{eq:quadEq}
\end{equation}
Computing the roots of the quadratic Eq.\ \eqref{eq:quadEq} using that
\begin{equation}
1+c=1+\frac{\bar{\omega}_b^2-\omega_b^2}{\omega_b^2}=\frac{\bar{\omega}_b^2}{\omega_b^2},
\end{equation}
results in 
\begin{equation}
b_\pm=\frac{\omega_b^2}{\bar{\omega}_b^2}\left(\frac{\bar{\beta}}{2}\pm\sqrt{\frac{\bar{\beta}^2}{4}+\bar{\omega}_b^2}\right)
\label{eq:bNM}
\end{equation}
or equivalently in 
\begin{equation}
	\lambda_\pm=-\frac{\bar{\beta}}{2}\pm\sqrt{\frac{\bar{\beta}^2}{4}+\bar{\omega}_b^2}
	\label{eq:lamNeu+}
\end{equation}
by inserting Eq.\ \eqref{eq:bNM} into Eq.\ \eqref{eq:lam2NM}.
Solving now the resulting ordinary differential equation, which is
formally identical to Eq.\ \eqref{eq:DGL_zeta} in the
Markovian case, using boundary conditions \eqref{eq:lim1} and
\eqref{eq:lim2}, $\Xi(z)$ is given by
\begin{equation}
\Xi(z)=\sqrt{\frac{\lambda_+}{2\pi A}}
\int_{-\infty}^z\exp\left[-\frac{\lambda_+ s^2}{2 A}\right]\dd s,
\end{equation}
%\begin{equation}
%\Xi(z)=\sqrt{\frac{m(b_+\left(1+c\right)-\bar{\beta})}{2\pi k_{\text{B}}T\left(\bar{\beta}-b_+c\right)}}
%\int_{-\infty}^z\exp\left[-\frac{m(b_+\left(1+c\right)-\bar{\beta})}{2 k_{\text{B}}T\left(\bar{\beta}-b_+c\right)}s^2\right]\dd s,
%\end{equation}
where 
\begin{equation}
	A=\frac{k_{\text{B}}T\left(\bar{\beta}-b_+c\right)}{m}
\end{equation} 
and $\lambda_+$ denotes the positive root of Eq.\ \eqref{eq:quadEq}, which needs
to be employed for the integral term to be convergent. 
%The exact form of$b_+$ will be specified later.
As soon as $\Xi(z)$ and therefore
$\rho(z)$ is known the population $n_a$ in the A-well and the stationary
current over the potential barrier $j_b$ can be computed. Calculating
the integral for $n_a$ in the non-Markovian case yields the same result
as in the Markovian case (see Eq.\ \eqref{eq:nFinal}), since the
stationary probability density $\rho(z)$ around $x_a$ is identical. In a
region around the top of the barrier at $x=x_b$, however, the density
$\rho(z)$ is significantly different from its Markovian
analog. Nonetheless even for the computation of $j_b$ the results from
the Markovian treatment can be used. Only the quantity $m'$ needs to be
replaced by $n$ defined by
\begin{equation}
n=\frac{m(b_+(1+c)-\bar{\beta})}{2k_{\text{B}}T(\bar{\beta}-b_+c)}.
\label{eq:n}
%	\Rightarrow \frac{2k_{\text{B}}Tn}{m}=\frac{b(1+c)-\bar{\beta}}{(\bar{\beta}-bc)}.
\end{equation}
As in the Markovian case a temporary result is obtained by
\begin{align}
j_b 
% &=\int_{-\infty}^{\infty}v\rho(x_b,v) \dd v \\
%    &=\frac{1}{Z}\sqrt{\frac{m(b\left(1+c\right)-\tilde{\beta})}{2\pi k_{\text{B}}T\left(\tilde{\beta}-bc\right)}} \\
%    &\int_{-\infty}^{\infty}\left(v\exp\left[-\frac{\frac{mv^2}{2}+E_{\text{b}}}{k_{\text{B}}T}\right]
%    \int_{-\infty}^{v}\exp\left[-\frac{m(b\left(1+c\right)-\tilde{\beta})}{2\pi k_{\text{B}}T\left(\tilde{\beta}-bc\right)}z^2\right]\dd z\right)\dd v \\
%    &=\frac{1}{Z}\sqrt{\frac{n}{\pi}}\exp\left[-\frac{E_{\text{b}}}{k_{\text{B}}T}\right]
%    \int_{-\infty}^{\infty}\left(v\exp\left[-lv^2\right]
%    \underbrace{\int_{-\infty}^{v}\exp\left[-nz^2\right]\dd z}_{=\frac{\sqrt{\pi}\left(1+erf(v\sqrt{n})\right)}{2\sqrt{n}}}\right)\dd v \\
%    &=\frac{1}{Z}\sqrt{\frac{n}{\pi}}\exp\left[-\frac{E_{\text{b}}}{k_{\text{B}}T}\right]\frac{\pi}{2\sqrt{n}}
%    \underbrace{\int_{-\infty}^{\infty}v\exp\left[-lv^2\right]\dd v}_{=0}\\
%    &+\frac{1}{Z}\sqrt{\frac{n}{\pi}}\exp\left[-\frac{E_{\text{b}}}{k_{\text{B}}T}\right]\frac{\pi}{2\sqrt{n}}
%    \underbrace{\int_{-\infty}^{\infty}v\exp\left[-lv^2\right]erf(v\sqrt{n})\dd v}_
%    {=\frac{\sqrt{n}}{l\sqrt{l+n}}} \\
&=\frac{1}{Z}\exp\left[-\frac{V(x_b)}{k_{\text{B}}T}\right]\frac{\sqrt{n}}{2k\sqrt{k+n}}.
%    =\frac{1}{Z}\exp\left[-\frac{E_{\text{b}}}{k_{\text{B}}T}\right]
%    \frac{1}{2\frac{m}{2k_{\text{B}}T}}
%    \sqrt{\frac{n}{\frac{m}{2k_{\text{B}}T}+n}} \\
%    &=\frac{1}{Z}\frac{k_{\text{B}}T}{m}\exp\left[-\frac{E_{\text{b}}}{k_{\text{B}}T}\right]\sqrt{\frac{\frac{2k_{\text{B}}Tn}{m}}{1+\frac{2k_{\text{B}}Tn}{m}}}
\label{eq:jFinalNM}
\end{align}
Reinserting relation \eqref{eq:n} into the temporary result for $j_b$, Eq.\ \eqref{eq:jFinalNM}, yields 
\begin{align}
j_b 
% &=\frac{1}{Z}\frac{k_{\text{B}}T}{m}\sqrt{\frac{\frac{b(1+c)-\bar{\beta}}{(\bar{\beta}-bc)}}{1+\frac{b(1+c)-\bar{\beta}}{(\bar{\beta}-bc)}}} 
% \exp\left[-\frac{E_{\text{b}}}{k_{\text{B}}T}\right]\\
%   &=\frac{1}{Z}\frac{k_{\text{B}}T}{m}\sqrt{\frac{b(1+c)-\bar{\beta}}{\left(\bar{\beta}-bc\right)\left(1+\frac{b(1+c)-\bar{\beta}}{(\bar{\beta}-bc)}\right)}} 
% \exp\left[-\frac{E_{\text{b}}}{k_{\text{B}}T}\right]\\
% &=\frac{1}{Z}\frac{k_{\text{B}}T}{m}\sqrt{\frac{b(1+c)-\bar{\beta}}{\cancel{\bar{\beta}}-\cancel{bc}+b(1+\cancel{c})-\cancel{\bar{\beta}}}} 
% \exp\left[-\frac{E_{\text{b}}}{k_{\text{B}}T}\right]\\
&=\frac{1}{Z}\frac{k_{\text{B}}T}{m}\frac{\lambda_+}{\omega_b} \exp\left[-\frac{V(x_b)}{k_{\text{B}}T}\right].
\label{eq:jFinalNM1}
\end{align}
%\begin{align}
%j_b 
%% &=\frac{1}{Z}\frac{k_{\text{B}}T}{m}\sqrt{\frac{\frac{b(1+c)-\bar{\beta}}{(\bar{\beta}-bc)}}{1+\frac{b(1+c)-\bar{\beta}}{(\bar{\beta}-bc)}}} 
%% \exp\left[-\frac{E_{\text{b}}}{k_{\text{B}}T}\right]\\
%%   &=\frac{1}{Z}\frac{k_{\text{B}}T}{m}\sqrt{\frac{b(1+c)-\bar{\beta}}{\left(\bar{\beta}-bc\right)\left(1+\frac{b(1+c)-\bar{\beta}}{(\bar{\beta}-bc)}\right)}} 
%% \exp\left[-\frac{E_{\text{b}}}{k_{\text{B}}T}\right]\\
%% &=\frac{1}{Z}\frac{k_{\text{B}}T}{m}\sqrt{\frac{b(1+c)-\bar{\beta}}{\cancel{\bar{\beta}}-\cancel{bc}+b(1+\cancel{c})-\cancel{\bar{\beta}}}} 
%% \exp\left[-\frac{E_{\text{b}}}{k_{\text{B}}T}\right]\\
%&=\frac{1}{Z}\frac{k_{\text{B}}T}{m}\sqrt{\frac{b_+(1+c)-\bar{\beta}}{b_+}} \exp\left[-\frac{V(x_b)}{k_{\text{B}}T}\right].
%\label{eq:jFinalNM1}
%\end{align}
Finally using Eq.\ \eqref{eq:fluxKramers} together with
\eqref{eq:nFinal} and \eqref{eq:jFinalNM1} the steady-state escape rate
$k_{A\to C}$ is given by
\begin{align}
k_{A\to C} 
% &=\frac{j_b}{n_a} \\
%  &=\frac{\cancel{\frac{1}{Z}\frac{k_{\text{B}}T}{m}}\sqrt{\frac{b(1+c)-\bar{\beta}}{b}} \exp\left[-\frac{E_{\text{b}}}{k_{\text{B}}T}\right]}{\cancel{\frac{1}{Z}\frac{k_{\text{B}}T}{m}}\frac{2\pi}{\omega_a}\exp\left[-\frac{V_0}{k_{\text{B}}T}\right]},\\
&=\frac{\lambda_{\mathrm{NM}}}{\omega_b}\frac{\omega_a}{2\pi}\exp\left[-\frac{E_{\text{b}}}{k_{\text{B}}T}\right],
\label{eq:strongFricNM}
\end{align}
%\begin{align}
%k_{A\to C} 
%% &=\frac{j_b}{n_a} \\
%%  &=\frac{\cancel{\frac{1}{Z}\frac{k_{\text{B}}T}{m}}\sqrt{\frac{b(1+c)-\bar{\beta}}{b}} \exp\left[-\frac{E_{\text{b}}}{k_{\text{B}}T}\right]}{\cancel{\frac{1}{Z}\frac{k_{\text{B}}T}{m}}\frac{2\pi}{\omega_a}\exp\left[-\frac{V_0}{k_{\text{B}}T}\right]},\\
%&=\frac{\omega_a}{2\pi}\sqrt{\frac{b_+(1+c)-\bar{\beta}}{b_+}}\exp\left[-\frac{E_{\text{b}}}{k_{\text{B}}T}\right],
%\label{eq:kNM}
%\end{align}
where again $E_{\text{b}}=V(x_b)-V(x_a)$ and $\lambda_+$ was identified with $\lambda_{\mathrm{NM}}$ (see Eq.\ \eqref{eq:lamNM}). 
That is the desired result for $k_{A\to C}$ as indicated in Subsec.\ \ref{sec:nonMarkovMod}.

Subsequent there are a number of comments to be made about the just
derived escape rate $k_{A\to C}$.  It can be shown that the prefactor $\lambda_{\mathrm{NM}}$ of
Eq.\ \eqref{eq:strongFricNM} corresponds to the largest positive
root of $s^2-\omega_b^2+\frac{\tilde{\Gamma}}{m}s$
\cite{Hanggi:1983,Carmeli:1984},
%\begin{equation}
%\lambda_{\mathrm{NM}}=\sqrt{\frac{\bar{\beta}^2}{4}+\bar{\omega}_b^2}-\frac{\bar{\beta}}{2},
%\label{eq:lamNM1}
%\end{equation}
originating from the inverse Laplace transform $\chi_v(t)$ (see
Eq.\ \eqref{eq:chi_v}). For a derivation of this statement reference
is made to Ref.\ \cite{Carmeli:1984}. 
%Replacing the prefactor in Eq.\
%\eqref{eq:strongFricNM} by relation \eqref{eq:lamNM1} finally results in
%the expression for the escape rate in the
%intermediate-to-strong-friction regime (see Eq.\
%\eqref{eq:intermediate-to-strongFricNM}) indicated in Subsec.\
%\ref{sec:nonMarkovMod}.  
The entire information about the dissipation
kernel $\Gamma(t)$ is therefore completely contained in
$\lambda_{\mathrm{NM}}$.  In \ref{sec:escRateC1} 
%(see appendix \ref{sec:escRateC1}) 
it is shown how to derive $\lambda_{\mathrm{NM}}$
for correlation function $C_1$, Eq.\ \eqref{eq:C1}.  For this
correlation function the above expression,
$s^2-\omega_b^2+\frac{\tilde{\Gamma}}{m}s$, becomes a cubic function of
s.  Thus, in order to compute $\lambda_{\mathrm{NM}}$ only the cubic
roots are needed.

\section{Numerical studies}
\label{seq:numStud}

%The purpose of this Section is to numerically investigate the
%above-described steady-state current as a function of the friction
%constant $\beta$ for both, white and colored noise.  
This section is devoted to the core of this work: Kramers's escape rate problem, which was presented in the previous two sections, will be numerically investigated for both, white and colored thermal noise.
To that end the colored noise is generated by means of the numerical implementation of the algorithm, indicated in Sec.\ \ref{chap:genCN}, which is given in the Appendix of Ref.\ \cite{Schmidt:2014zpa} and the GLE, Eq.\ \eqref{eq:genlang1},
is solved, using the explicit three-step Adams-Bashforth algorithm
\cite{Faires:2003}:
\begin{align}
	y_0&=a_0, \quad y_1=a_1, \quad y_2=a_2, \\
	y_{i+1}&=y_i+\frac{h}{12}\left[23f(t_i,y_i)-16f(t_{i-1},y_{i-1})+5f(t_{i-2},y_{i-2})\right],
\end{align}
where $i=2,3,...,N-1$ and the local and global error are $\mathcal{O}(h^4)$ and $\mathcal{O}(h^3)$, respectively. Furthermore, $y$ stands representative for position and velocity in the GLE, $f(t,y)$ corresponds to its right-hand side, respectively, and $h$ is the step-size. The three values $y_0$, $y_1$ and $y_2$, where $y_0$ corresponds to the initial conditions and $y_1$ and $y_2$ are to be evaluated using Euler's method, are required to apply the above indicated three-step Adams-Bashforth method.

In what follows a first step will be to present the details of the numerical simulations regarding the used potential $V(x)$, correlation functions, initial conditions and the algorithm, which is employed to compute the escape rate . 
Afterwards it will be exemplarily shown that  the numerical simulations are able to fit the approximate analytic formulas properly.

Subsequently Kramers's steady-state escape rate as a function of the friction rate $\beta$ will be investigated for different correlation functions and compared to the appropriate analytic formulas.

\subsection{Numerical setup}

\begin{figure}
	\begin{center}
		\begin{tikzpicture}[xscale=5.5,yscale=1,samples=400]
		%geplottete funktion
		\draw [<->,ultra thick] (-2,5) node [left] {$\frac{V(x)}{k_{\text{B}}T}$} -- (-2,0) -- (0.5,0) node [below right] {$\frac{x}{x_a}$};
		\draw [-,ultra thick] (-2,0) -- (-2,-2.5);
		\draw[red, ultra thick, domain=-1.6:-0.7] plot (\x, {13.888888887*(\x+1)*(\x+1)});
		\draw[red, ultra thick, domain=-0.7:0.22] plot (\x, {2.5-13.888888887*(\x+0.4)*(\x+0.4)});
		
		%Beschriftung x-Achse 
		\draw [ultra thick] (-1.5,-0.1) -- (-1.5,0.1);
		\draw [ultra thick] (-1,-0.15) node[below]{1} -- (-1,0.15);
		\draw [ultra thick] (-0.5,-0.1) -- (-0.5,0.1);
		\draw [ultra thick] (0,-0.15) node[below]{2} -- (0,0.15);
		
		%Beschriftung y-Achse
		%		\draw [ultra thick] (-2.027,-3) node [left]{-3}-- (-1.973,-3);
		\draw [ultra thick] (-2.027,-2) node [left]{-2}-- (-1.973,-2);
		\draw [ultra thick] (-2.027,-1) node [left]{-1}-- (-1.973,-1);
		\draw [ultra thick] (-2.027,0) node [left]{0}-- (-1.973,0);
		\draw [ultra thick] (-2.027,1) node [left]{1}-- (-1.973,1);
		\draw [ultra thick] (-2.027,2) node [left]{2}-- (-1.973,2);
		\draw [ultra thick] (-2.027,3) node [left]{3}-- (-1.973,3);
		\draw [ultra thick] (-2.027,4) node [left]{4}-- (-1.973,4);
		
		%dashed lines + arrow
		\draw [<->,ultra thick] (-0.4,0) -- (-0.4,2.5);
		\node [right] at (-0.4,1.25) {$E_{\text{b}}$};
		%				\draw [-,dashed] (0.2,5) -- (0.2,-2.9);
		
		%a-well
		\node [above] at (-1,0.1) {$\omega_a$};
		
		%barrier
		\node [above] at (-0.4,2.5) {$\omega_b$};
		
		%		%Kreissegment für fluss
		%		\draw [<-,black, ultra thick] (0.3,3.5) arc [radius=0.5, start angle=20, end angle= 160];
		%		\node [above] at (-0.1875,3.8) {$k_{a\rightarrow c}$};
		
		%help lines
		%		\draw[help lines] (-2,-5) grid (2,7);
		\end{tikzpicture}
	\end{center}
	\caption{Potential $V(x)$, Eq.\ \eqref{eq:potential}, used for the
          following numerical simulations. Here $m=1.11 \,\GeV$, $T=1 \,\GeV$, 
          $\omega=5 \,\GeV$, $E_{\text{b}}=2.5 \,\GeV$, $x_a=1 \, \GeV^{-1}$ and
          $x_b=1.6 x_a$.}
	\label{potSim}
\end{figure}
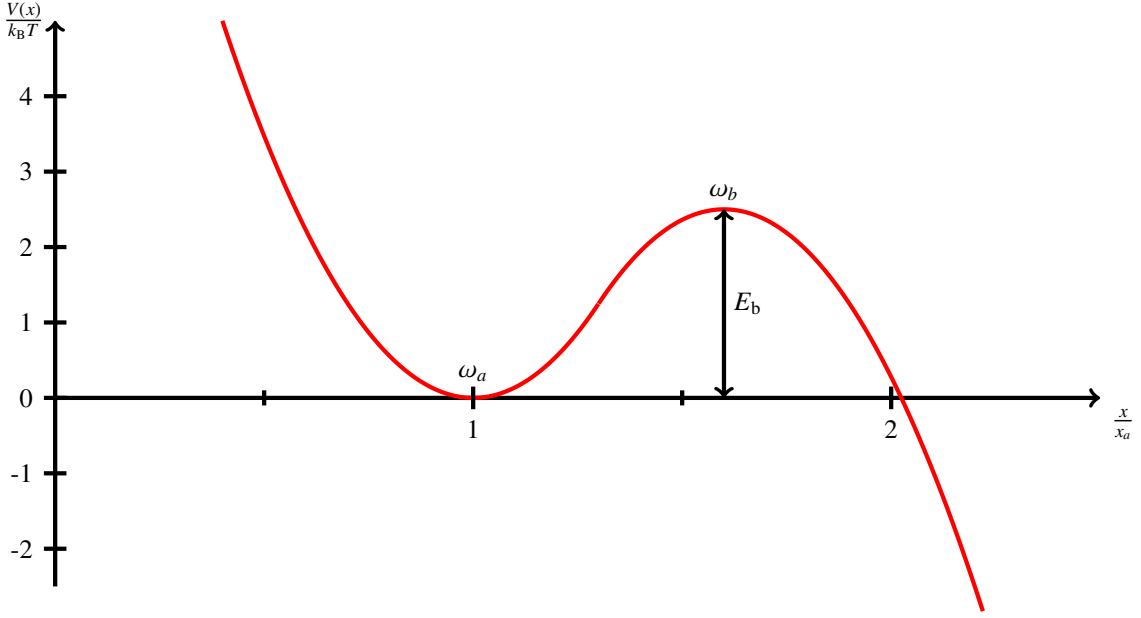

In contrast to Kramers's classical model, for the numerical simulations
a slightly idealized potential will be used. This potential (see Fig.\
\ref{potSim}) is composed of two parabolic potentials of the same
frequency $\omega_a=\omega_b=\omega$, smoothly connected at some
intermediate point $x_m$,
\begin{equation}
V(x)=
\begin{cases}
\frac{m\omega_a^2}{2}\left(x-x_a\right)^2 & \text{for} \quad x<x_m, \\
E_{\text{b}}-\frac{m\omega_b^2}{2}\left(x-x_b\right)^2 & \text{for} \quad x>x_m,
\end{cases}
\label{eq:potential}
\end{equation}
where $x_m$ is defined as
\begin{equation}
x_m=\frac{x_b+x_a}{2}.
\end{equation}

This idealized potential is to be understood as an asymmetric
double-well potential, whereby the right potential well is supposed to
be infinitely deep. In doing so anharmonic corrections
\cite{TalknerR1993,Talkner1993,Talkner:1995}, naturally arising from more realistic potentials,
can be largely neglected.

As indicated before, the simulations are performed for white and colored
noise, which are connected to an appropriate correlation function, respectively (see
Sec.\ \ref{chap:genCN}).  For the studies of this work the usual
correlation function for white noise will be used:
\begin{equation}
C_0:=\left<\xi(t)\xi(0)\right>=D\delta(t).
\label{eq:C0}
\end{equation}
In addition to that, for colored noise, the three correlation functions
$C_1$, $C_2$ and $C_3$ (see Sec.\ \ref{chap:genCN}) are covered:
\begin{align}
C_1(|t|)&=\frac{D}{2\tau}\exp\left[-\frac{|t|}{\tau}\right], \label{eq:C1Sim} \\
C_2(|t|)&=\frac{D}{a\sqrt{\pi}}\exp\left[-\left(\frac{|t|}{a}\right)^2\right], \label{eq:C2Sim} \\
C_3(|t|)&=\frac{g}{4}k_{\text{B}}T\alpha^2\left(1-\frac{\alpha}{\sqrt{m}}|t|\right)\exp\left[-\frac{\alpha}{\sqrt{m}}|t|\right].
\label{eq:C3Sim}
\end{align}
By use of the algorithm, described in Sec.\ \ref{chap:genCN}, 
a sequence of the respective colored noise can be generated from the
above given correlation functions.

\subsection{Simulations}
The starting situation of the simulations is as follows:
In each simulation, computing the evolution of a whole ensemble,
consisting of a large number of about $10^6$ realizations of the
stochastic processes $x(t)$ and $v(t)$, respectively, the particles are
initialized at the bottom of the left well at $x_0=x_a$ with velocity
$v_0=0$. The remaining relevant parameters are given as following:
\begin{equation}
E_{\text{b}}=2.5 \,\GeV, \,\, \omega_b=5 \,\GeV, \,\,
m=1.11 \,\GeV, \,\, T=1 \,\GeV, \,\,\tau_{corr}=\{0.2,\,0.4,\,1\} \,\GeV^{-1}, \label{eq:parameters}
\end{equation}
where different correlation times $\tau_{\text{corr}}$ are employed for the non-Markovian correlation functions, Eqs.\ \eqref{eq:C1Sim}, \eqref{eq:C2Sim} and \eqref{eq:C3Sim}, as also effects of growing correlation times 
shall be investigated in the following sections. The choice of magnitude of  $\tau_{corr}$
is justified due to condition \eqref{eq:CondNonMarkov}, according to which a non-Markovian description requires $\tau_{\text{corr}}\approx 1 \, \GeV^{-1}$ for the above parameters:
\begin{equation}
\tau_{\mathrm{corr}}=\frac{2\pi}{\omega}\approx 1 \, \GeV^{-1}.
\label{eq:tcorr}
\end{equation}
Vividly speaking, expression \eqref{eq:tcorr} implies that there is the fifth part
of an oscillation in about $0.2 \, \mathrm{GeV}^{-1}$ up to about one oscillation
in $1 \, \mathrm{GeV}^{-1}$.  Hence, the three cases for
$\tau_{\mathrm{corr}}$ are representative for medium ($\tau_{corr}=0.2 \,\GeV^{-1}$ and $\tau_{corr}=0.4 \,\GeV^{-1}$) and
strong ($\tau_{corr}=1 \,\GeV^{-1}$) non-Markovian
situations.

Concerning the parameters \eqref{eq:parameters} the attentive reader
will immediately notice that $m\approx T$, which implies relativistic
velocities by virtue of the equipartition theorem. Since, however the
Brownian particles used in these simulations are not ``aware'' of
relativity - as they are governed by classical Newtonian dynamics (see
Eqs.\ \eqref{eq:langevinEq1} and \eqref{eq:genlang1})- the size of the
velocity has no relevance.

Given the solutions for $x(t)$ and $v(t)$ for every realization
of the simulation the rate of particles overcoming the potential barrier
is readily obtained.

There are two possible ways to numerically determine the steady-state
rate. Both include a certain absorptive barrier $x_{\mathrm{abs}}$,
which coincides with the sink described in Subsec.\ \ref{sec:derivation}.
This absorptive barrier has to be chosen far away from the top of the
potential barrier in the right potential well to ensure that particles
that have reached the absorptive barrier will never return to the
initial well. The two ways of numerical determination of the
steady-state current now depend on what happens after reaching this
absorptive barrier.

The first method, usually referred to as population-over-method
\cite{Hanggi:1990}, is based on the re-initialization of particles,
which have overcome the absorptive barrier. This leads to a nearly
constant population in the initial well. Thereby it needs to be ensured,
that this re-initializations are not taken into account as real
backscattering, which would affect the current over the barrier. The
steady-state escape rate is then obtained, determining the current over
the barrier located at $x_b$.

The second method on the other hand gets along without any
re-initialization. Here the current is calculated concerning the
absorptive barrier. Numerically, the steady-state escape rate is
computed as follows \cite{Gontchar:2017}:
\begin{equation}
k_{A\to C}=\frac{1}{N_{\mathrm{tot}}-N_{\mathrm{abs}}}\frac{\Delta N_{\mathrm{abs}}}{\Delta t},
\end{equation}
where $N_{\mathrm{tot}}$ denotes the total number of initialized
particles, $N_{\mathrm{abs}}$ is the total number of particles, that
have already reached the absorptive border and $\Delta N_{\mathrm{abs}}$
designates the number of particles being absorbed in the course of the
time interval $\Delta t$.  It turns out that both methods yield the same
results. The second method, however, seems to be numerically more stable
as the first method requires smaller time steps $\Delta t$ for the
escape rate to be convergent. Hence, for the following numerical
discussion the second method will be used.
\begin{figure}[tb]
	\begin{center}
		%\begin{overpic}[width=0.55\textwidth]{./pictures/typCurrent/typCurr.pdf}\end{overpic} \hfill \begin{overpic}[width=0.55\textwidth]{./pictures/typCurrent/typCurr1.pdf}\end{overpic}
		\begin{overpic}[scale=0.35]{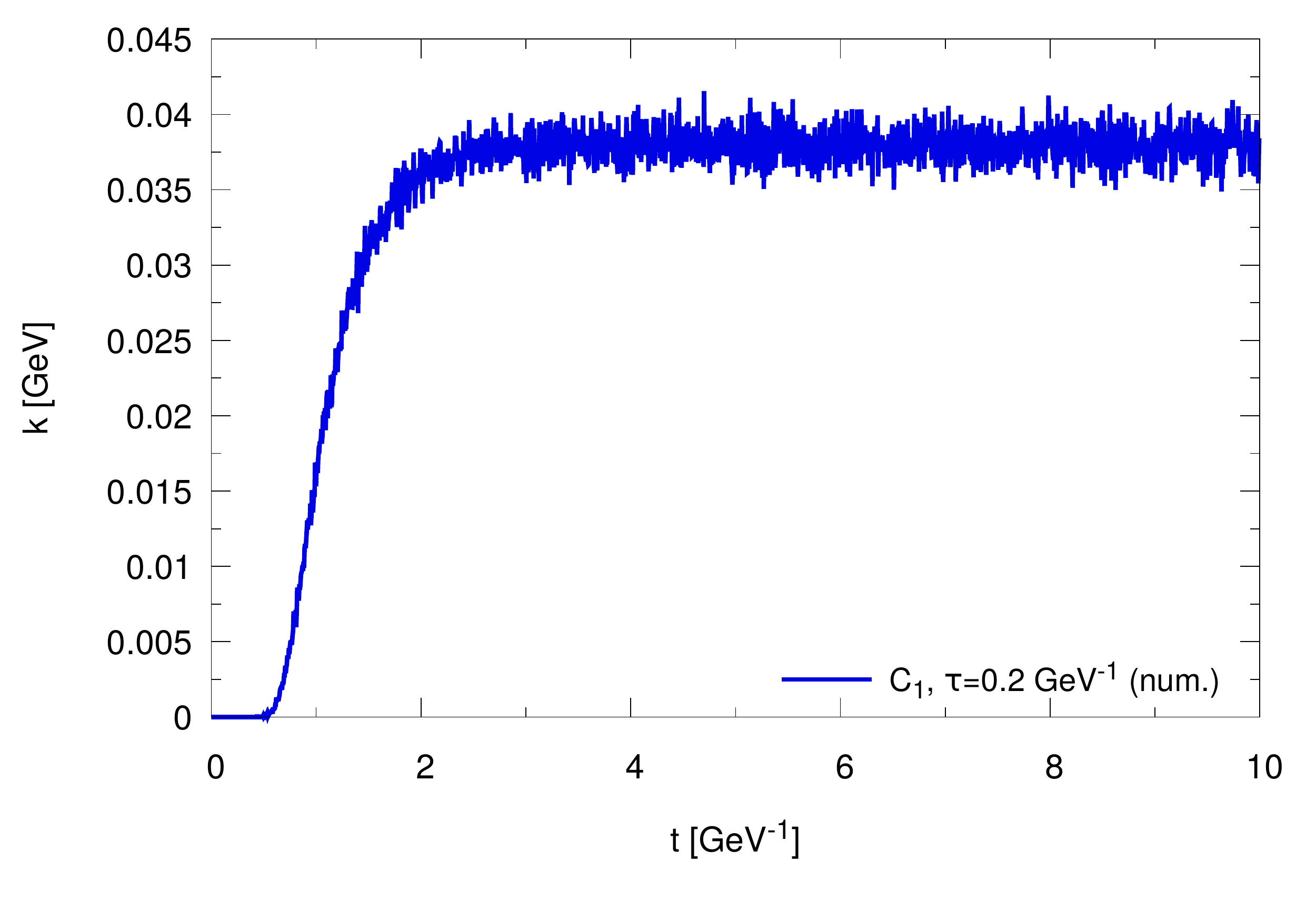}\end{overpic} \hfill \begin{overpic}[scale=0.35]{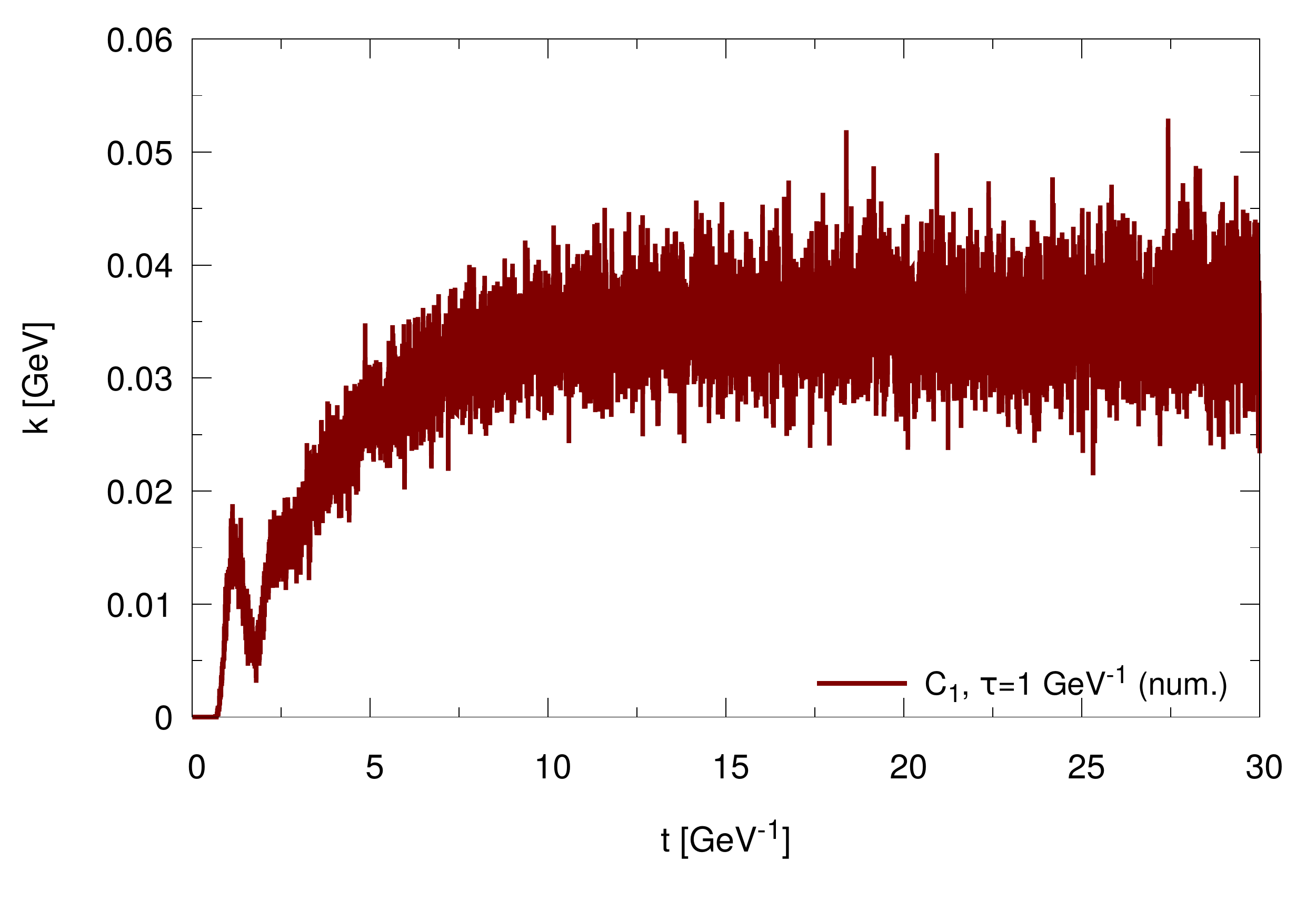}\end{overpic}
		\caption{Escape rate $k$ as a function of time for
                  correlation function $C_1$, Eq.\ \eqref{eq:C1},
                  averaged over $8\cdot 10^6$ realizations, where
                  $m=1.11 \,\GeV$, $\beta=9 \, \GeV$, $E_{\text{b}}=2.5 \, \GeV$,
                  $T=1 \, \GeV$ and $\omega=5 \,\GeV$.}
		\label{behFp} 
	\end{center}
\end{figure}

Fig.\ \ref{behFp} indicates the typical outcomes of two non-Markovian 
simulations (correlation function $C_1$, Eq.\ \eqref{eq:C1}) for different correlation times,
applying the second numerical method. Basically, the escape rate as a
function of time consists of three successive stages. After an initial
phase of a not quantifiable escape rate, during which the considered
ensemble thermalizes, a transient phase occurs. In this regime the
escape rate begins to rise moderately until in the end it takes a constant
mean value, Kramers's steady-state escape rate. In case of the
non-Markovian noise and large correlation times (i.e.
$\tau \gg \frac{2\pi}{\omega}$) a special feature occurs in the
transient phase. After an initial rise, the current significantly
decreases until it eventually starts to rise again and finally converges
to its mean value (see Fig.\ \ref{behFp}). This effective
backscattering in the transient phase is an example for the memory
effects, arising from finite correlation times \cite{Schmidt:2014zpa}. The mean value of
Kramers's escape rate is evaluated by averaging over the
quasi-stationary third stage. Dividing the evolution of rate $k$ into
$n$ bins of width $\Delta t$ and taking only into account the last $m$
steps of the third stage, Kramers's escape rate and the corresponding
standard error are evaluated, using the following equations
\cite{Gontchar:2017}:
\begin{align}
k_{A\to C}&=\frac{1}{m}\sum_{i=n-m}^{n}k_{A\to C}(t_i),  \label{eq:kmean}\\
\sigma_k&=\sqrt{\frac{1}{m(m-1)}\sum_{i=n-m}^{n}\left(k_{A\to C}(t_i)-k_{A\to C}\right)^2} \label{eq:standErr}.
\end{align}

\subsection{Parametrical dependencies}

In order to show that the used code and the algorithm to generate
colored noise, contained therein, actually work properly, it is useful
to numerically examine the occurring parametrical dependencies related
to the steady-state escape rate for the $\delta$-correlated Markovian
correlation function $C_0$ and the non-Markovian correlation function
$C_1$ and compare them to the approximate analytical results.
%, Eqs.\
%\eqref{eq:intermediate-to-strongFric}, \eqref{eq:strongFric} (Markovian escape rate)
 %and
%\eqref{eq:intermediate-to-strongFricNM} (non-Markovian escape rate), where it should 
%be recalled that $\lambda_\mathrm{NM}$ from Eq.\ \eqref{eq:intermediate-to-strongFricNM}
%is is to be identified with the largest positive root of $s^2-\omega_b^2+\frac{\tilde{\Gamma}}{m}s$,
%which is given  for correlation function $C_1$ in \ref{sec:escRateC1}. 

For this purpose the further
procedure will be the following:
While one parameter is varied, all remaining parameters will be kept
constant, to see if the isolated parameters obey the correct scaling
behavior. The parameters to be studied are the temperature $T$, the
barrier height $E_{\text{b}}$ and the frequency $\omega_b$. The
dependence on the coupling constant $\beta$ will be investigated
separately later on.

Exemplary in what follows a comparison of numerical with analytical
results, Eqs.\ \eqref{eq:intermediate-to-strongFric} and \eqref{eq:intermediate-to-strongFricNM} with $\lambda_{\mathrm{NM}}$ given in \ref{sec:escRateC1}, for
the above-named parameters will be presented to justify the validity of
the underlying numerical algorithm.
It should be recalled that $\lambda_\mathrm{NM}$ from Eq.\ \eqref{eq:intermediate-to-strongFricNM}
is to be identified with the largest positive root of $s^2-\omega_b^2+\frac{\tilde{\Gamma}}{m}s$.
\begin{figure}[tb]
	\begin{center}
		%\begin{overpic}[width=0.47\textwidth]{./pictures/barrierDep/fluxMarkov_barrierDep.pdf}\end{overpic} \hfill \begin{overpic}[width=0.47\textwidth]{./pictures/barrierDep/fluxCorr0_barrierDep.pdf}\put(65,35){}\end{overpic}
		%\vfill
		\begin{overpic}[width=0.95\textwidth]{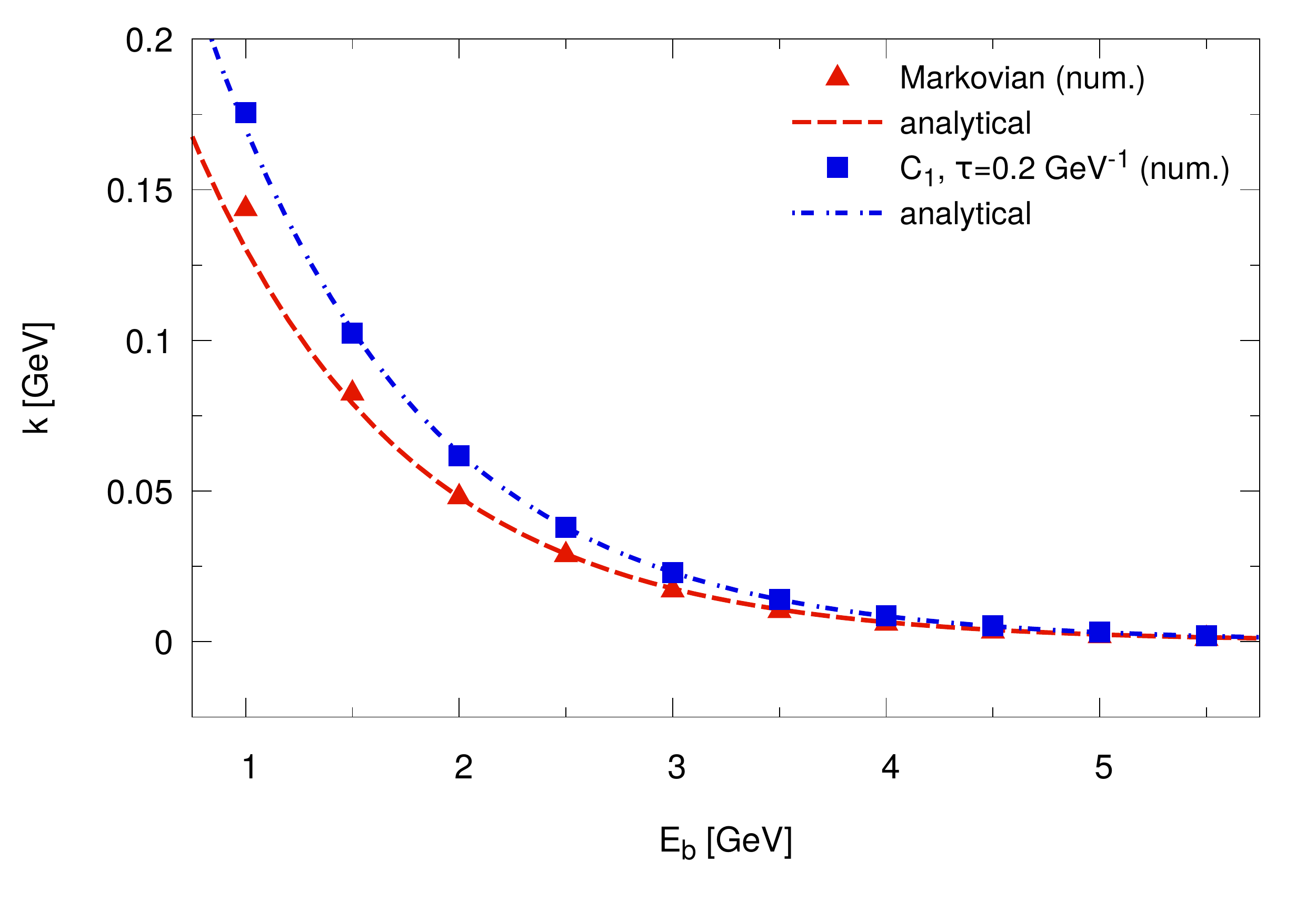}\put(65,35){}\end{overpic}
		%\begin{overpic}[scale=0.65]{./pictures/barrierDep/fluxCorr0vsMarkov_barrierDep.pdf}\put(65,35){}\end{overpic}
		%\includegraphics[scale=0.8]{./Bilder/BeffBetaVar.eps}
		\caption{Steady-state escape rate $k$ as a function of
                  the barrier height $E_{\text{b}}$, where
                  $\beta=9 \, \GeV$, $\omega_b=5 \,\GeV$ and
                  $T=1 \,\GeV$. Red: Markovian simulations, Eq.\
                  \eqref{eq:C0} (triangles), and analytical solution,
                  computed with Eq.\
                  \eqref{eq:intermediate-to-strongFric} (dashed
                  line). Blue: Non-Markovian simulations for correlation
                  function $C_1$, Eq.\ \eqref{eq:C1}, with
                  $\tau=0.2 \, \GeV^{-1}$ (squares) and analytical solution,
                  computed with Eq.\
                  \eqref{eq:intermediate-to-strongFricNM} with $\lambda_{\mathrm{NM}}$ given in \ref{sec:escRateC1} (dotted dashed
                  line).}
		\label{barrierDep}
	\end{center}
\end{figure}
\begin{figure}[tb]
	\begin{center}
		%\begin{overpic}[width=0.47\textwidth]{./pictures/tempDep/fluxMarkov_tempDep.pdf}\end{overpic} \hfill \begin{overpic}[width=0.47\textwidth]{./pictures/tempDep/fluxCorr0_tempDep.pdf}\put(65,35){}\end{overpic}
		%\vfill
		\begin{overpic}[width=0.95\textwidth]{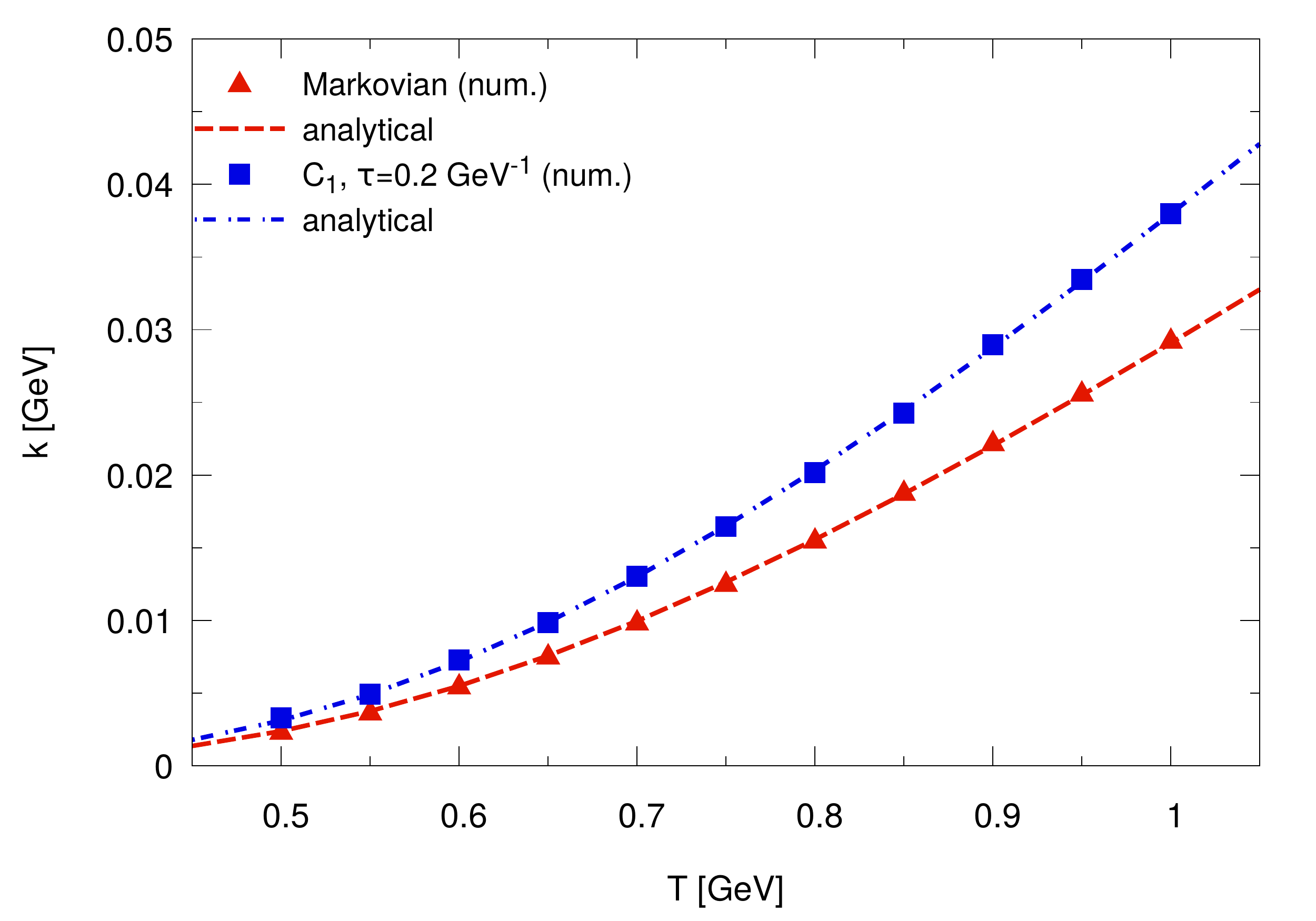}\put(65,35){}\end{overpic}
		\caption{Steady-state escape rate $k$ as a function of
                  the temperature $T$, where $\beta=9 \,\GeV$,
                  $\omega_b=5 \,\GeV$ and $E_{\text{b}}=2.5
                  \,\GeV$. Red: Markovian simulations, Eq.\
                  \eqref{eq:C0} (triangles), and analytical solution,
                  computed with Eq.\
                  \eqref{eq:intermediate-to-strongFric} (dashed
                  line). Blue: Non-Markovian simulations for correlation
                  function $C_1$, Eq.\ \eqref{eq:C1}, with
                  $\tau=0.2 \,\GeV^{-1}$ (squares) and analytical solution,
                  computed with Eq.\
                  \eqref{eq:intermediate-to-strongFricNM} with $\lambda_{\mathrm{NM}}$ given in \ref{sec:escRateC1} (dotted dashed
                  line).}
		\label{tempDep}
	\end{center}
\end{figure}
\begin{figure}[tb]
	\begin{center}
		%\begin{overpic}[width=0.47\textwidth]{./pictures/omegaDep/fluxMarkov_omegaDep.pdf}\end{overpic} \hfill \begin{overpic}[width=0.47\textwidth]{./pictures/omegaDep/fluxCorr0_omegaDep.pdf}\put(65,35){}\end{overpic}
		%\vfill
		\begin{overpic}[width=0.95\textwidth]{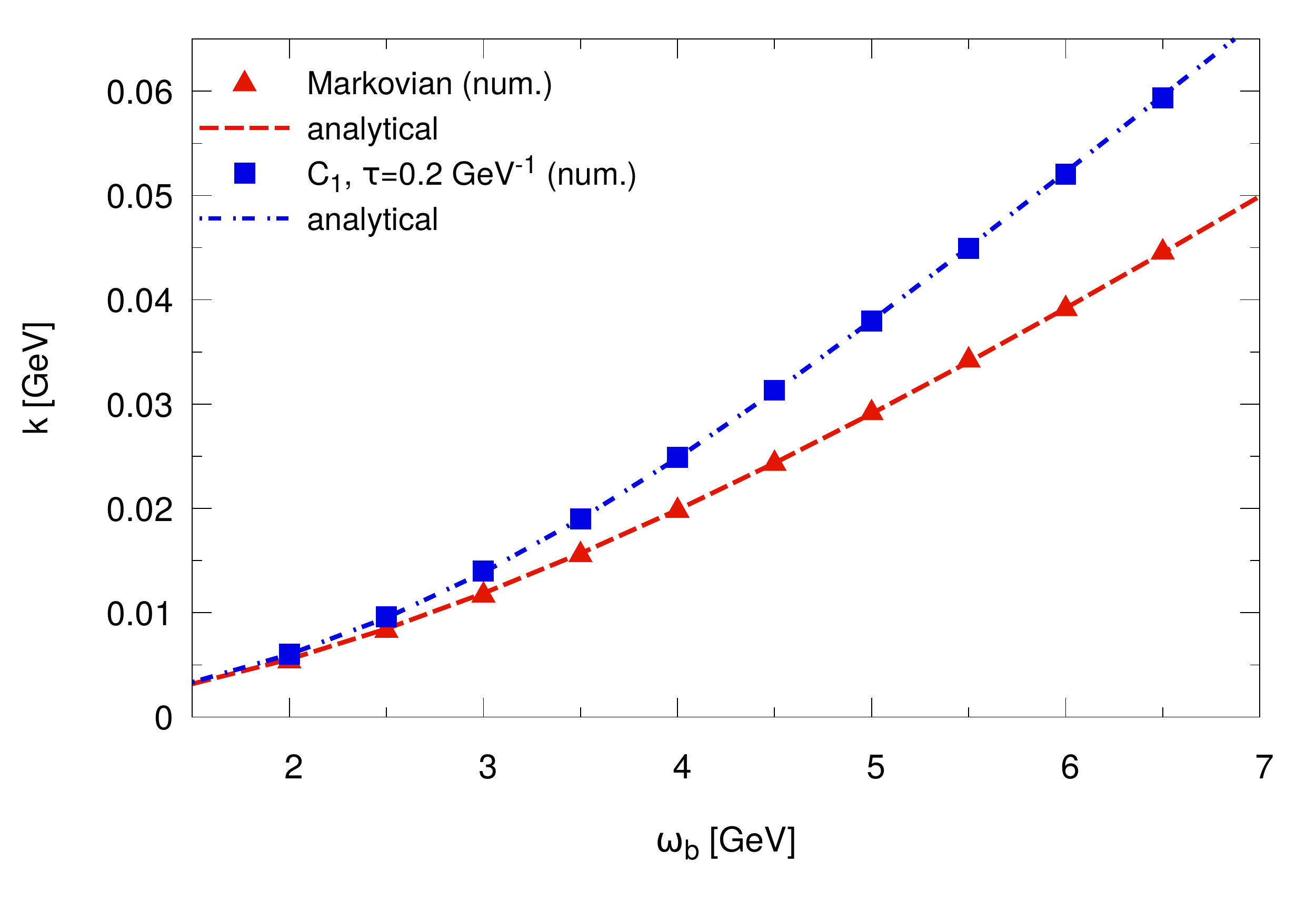}\put(65,35){}\end{overpic}
		\caption{Steady-state escape rate $k$ as a function of
                  the frequency $\omega_b$, where $\beta=9 \,\GeV$,
                  $T=1 \, \GeV$ and $E_{\text{b}}=2.5 \, \GeV$. Red:
                  Markovian simulations, Eq.\ \eqref{eq:C0} (triangles), and analytical solution, computed with
                  Eq.\ \eqref{eq:intermediate-to-strongFric} (dashed
                  line). Blue: Non-Markovian simulations for correlation
                  function $C_1$, Eq.\ \eqref{eq:C1}, with
                  $\tau=0.2 \,\GeV^{-1}$ (squares) and analytical solution,
                  computed with Eq.\
                  \eqref{eq:intermediate-to-strongFricNM} with $\lambda_{\mathrm{NM}}$ given in \ref{sec:escRateC1} (dotted dashed
                  line).}
		\label{omegaDep}
	\end{center}
\end{figure}
As can be seen in Figs.\ \ref{barrierDep}, \ref{tempDep} and
\ref{omegaDep} the expected analytical behavior (see Eqs.\
\eqref{eq:intermediate-to-strongFric} and
\eqref{eq:intermediate-to-strongFricNM}) could be recovered almost
perfectly in each case. Only for small barrier heights $E_{\text{b}}$
compared to the temperature $T$ a deviation from the analytical results is visible in
Fig.\ \ref{barrierDep}. However, this
deviation is expected as with decreasing barrier height
$E_{\text{b}}$ and simultaneous constant temperature $T$ the
approximative analytic formulas, Eqs.\
\eqref{eq:intermediate-to-strongFric} and
\eqref{eq:intermediate-to-strongFricNM} start to lose their validity due to the violation of condition \eqref{eq:Cond}. Certainly,
this deviation would also eventually appear in Fig.\ \ref{tempDep} for
higher temperatures $T$.

\subsection{Steady-state rate as a function of the damping rate $\beta$}

In the following section it will be investigated how Kramers's escape
rate behaves as a function of the coupling strength or damping rate $\beta$ 
%- or to be more precise as a function of the dimensionless parameter
%$\frac{\beta}{\omega_b}$ - 
for correlation functions $C_0$, $C_1$ and
$C_2$, Eqs.\ \eqref{eq:C0}, \eqref{eq:C1} and \eqref{eq:C2} (see also Ref.\ \cite{Carmeli:1984}), and as a
function of the dimensionless coupling strength $g$ in case of
correlation function $C_3$, Eq.\ \eqref{eq:C3}. Since in this context
the coupling strengths $\beta$ or g 
are the only varying quantities, it is sufficient to restrict the investigation 
of the Kramers's rate to the coefficient $\kappa$ of Eq.\ \eqref{eq:kappa} 
as it is solely responsible for differences in the behavior of the escape rates
regarding different correlation functions. 
To that end, all rates will be normalized to
the transition-state rate $k_{\text{TST}}$, which is always an upper
border to Kramers's escape rate $k_{A\to C}$ as already mentioned in
Subsec.\ \ref{sec:classMod} (see Fig.\ \ref{bellShape}). Doing this in case of white noise, it turns out that $\kappa$ is a function
of the dimensionless parameter $\frac{\beta}{\omega}$ in the weak-friction and of $(\frac{\beta}{\omega})^{-1}$ in the strong-friction regime  (see Fig.\ \ref{bellShape}), which, as already discussed 
in Subsec.\ \ref{sec:classMod}, also comes into play concerning the range of validity of 
the different regimes (see Fig.\ \ref{phaseDiag}). 
%different approximate escape rate formulas, Eqs.\ \eqref{eq:lowFric}, \eqref{eq:intermediate-to-strongFric},
%\eqref{eq:strongFric}, \eqref{eq:lowFricNM},
%\eqref{eq:intermediate-to-strongFricNM} (for $\lambda_{\mathrm{NM}}$ see \ref{sec:escRateC1}) and \eqref{eq:turnover}). 
This will become 
important for the comparison of numerical and analytical results.

The main
objective will be to find out about the peculiarities of a non-Markovian
compared to a Markovian correlation function in case of correlation
functions $C_1$ and $C_2$. Not only the differences between distinct
correlation functions but also the differences, relating to changes in
the correlation time will be of interest. Therefore, Kramers's escape
rate is computed for every correlation function and varying correlation
times $\tau_{\mathrm{corr}}$ ($0.2 \, \GeV^{-1}$, $0.4 \, \GeV^{-1}$ and $1 \, \GeV^{-1}$) 
within a fixed area of $\beta$-values, covering the small- and the
strong-friction regime (see also Fig.\ \ref{phaseDiag}).
%, where the
%remaining parameters are given as following:
% \begin{equation*}
% 	E_{\text{b}}=2.5 \,\GeV, \,\, \omega_b=5 \,\GeV, \,\,
% 	m=1.11 \,\GeV, \,\, T=1 \,\GeV.
% \end{equation*}
%Therefore, due to condition \eqref{eq:CondNonMarkov}, a non-Markovian description requires 
%correlation times that are  $\ge 0.2$ fm:
%\begin{equation}
%	\tau_{\mathrm{corr}}=\frac{2\pi}{\omega}\approx 0.2 \, \mathrm{fm},
%	\label{eq:tcorr}
%\end{equation} 
%which justifies the choice of correlation times $\tau_{\mathrm{corr}}$
%as given before. 
%Vividly speaking expression \eqref{eq:tcorr} implies that there is one oscillation in about $0.2 \, \mathrm{fm}$ up to about 
%five oscillations in $1 \, \mathrm{fm}$.

In what follows, one after the other the correlation functions
$C_1$ and $C_2$ are compared to the Markovian case, starting with
correlation function $C_1$. For correlation function $C_3$, however, a
comparison with the Markovian case will be omitted since no strict Markovian limit exists (see also Subsec.\ \ref{C3Discussion}).
%it is still
%under investigation whether any Markovian limiting regime exists.  
%To
%make the different curves more comparable and since the coupling strength $\beta$ or g 
%is the only variable quantity all rates are normalized to
%the transition-state rate $k_{\text{TST}}$, which is always an upper
%border to Kramers's escape rate $k_{A\to C}$, as already mentioned in
%Subsec.\ \ref{sec:classMod}.

It should be noted that when talking about weak and strong friction this
is always meant in relation to the friction value, corresponding to the
maximal escape rate. This should not be confused with the weak- and
strong-friction regimes of Kramers's escape rate problem as these
regimes do not only depend on the actual friction value but also on the
validity of certain conditions (see also Subsec.\ \ref{sec:classMod}).
\begin{figure}[h]
	\begin{center}
		\includegraphics[width=0.8\textwidth]{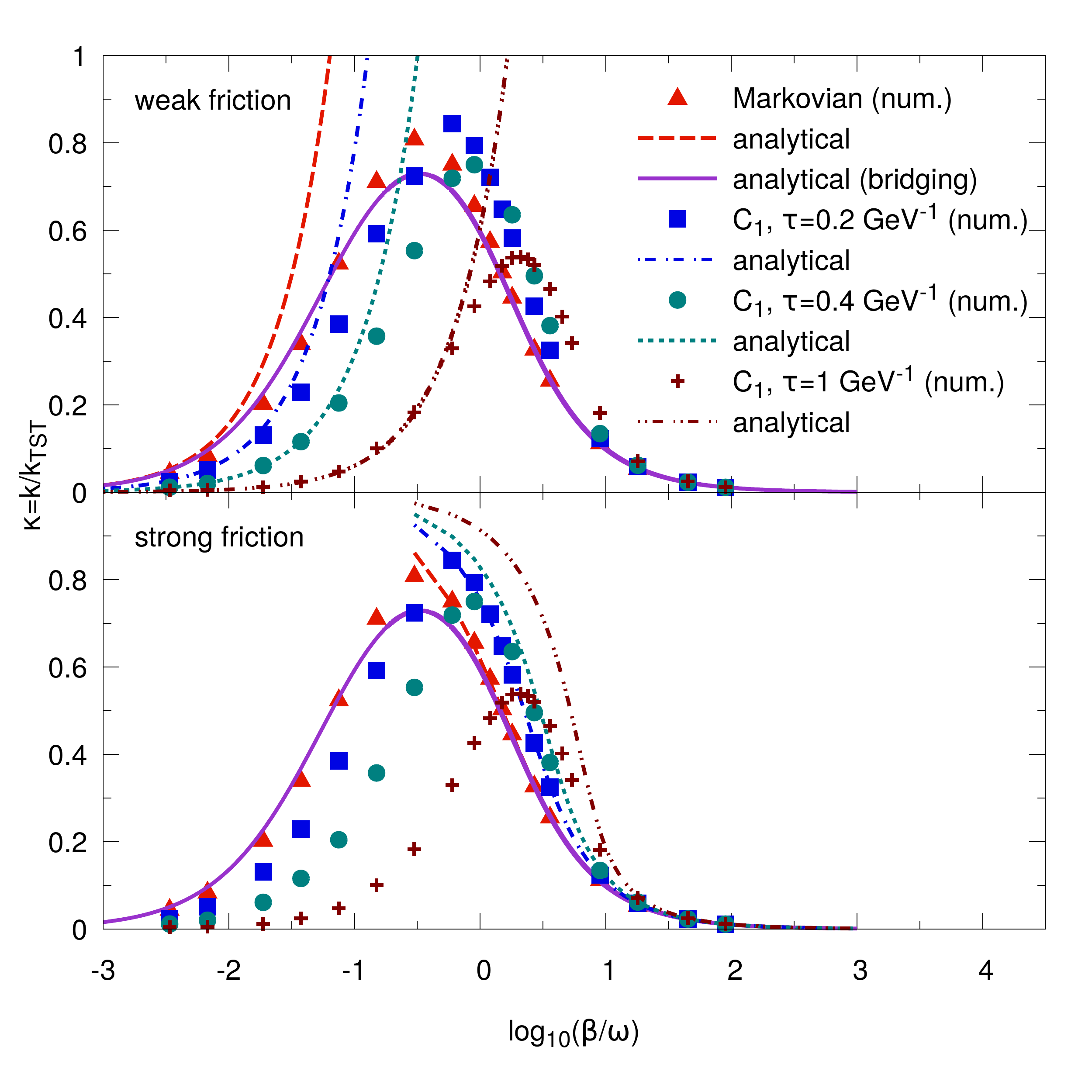}
		\caption{Comparison of the normalized steady-state
                  escape rate $\kappa$ as a function of the
                  dimensionless parameter $\frac{\beta}{\omega_b}$ for
                  correlation functions $C_0$ and $C_1$ (Eqs.\
                  \eqref{eq:C0} and \eqref{eq:C1}), where
                  $E_{\text{b}}=2.5 \,\GeV$, $\omega_b=5 \,\GeV$,
                  $m=1.11 \,\GeV$ and $T=1 \,\GeV$. Analytical results
                  are computed, using Eqs.\ \eqref{eq:lowFric} and
                  \eqref{eq:lowFricNM} in the weak-friction, Eqs.\
                  \eqref{eq:intermediate-to-strongFric} and
                  \eqref{eq:intermediate-to-strongFricNM}, with $\lambda_{\mathrm{NM}}$ given in \ref{sec:escRateC1}, in the
                  strong-friction regime and Eq.\ \eqref{eq:turnover}
                  for the bridging between strong- and weak-friction
                  regime.}
		\label{bDepC1}
	\end{center}
\end{figure}

\subsubsection{Correlation function $C_1$}

First of all, it should be recognized that the steady-state escape rate
as a function of the coupling strength $\beta$ follows the bell-shaped
course, already estimated by Kramers \cite{Hanggi:1990,Kramers:1940}, in
both the Markovian and non-Markovian case (see Fig.\ \ref{bDepC1}). In
the limit of $\beta\to 0$ or $\beta\to\infty$ the normalized escape rate
$\kappa$ tends to zero, while for some intermediate value of $\beta$
there exists a maximum. After having clarified this qualitative
similarities between the Markovian and the non-Markovian case, attention
should now be directed to the quantitative differences.

\begin{figure}[h]
	\begin{center}
		\includegraphics[width=0.95\textwidth]{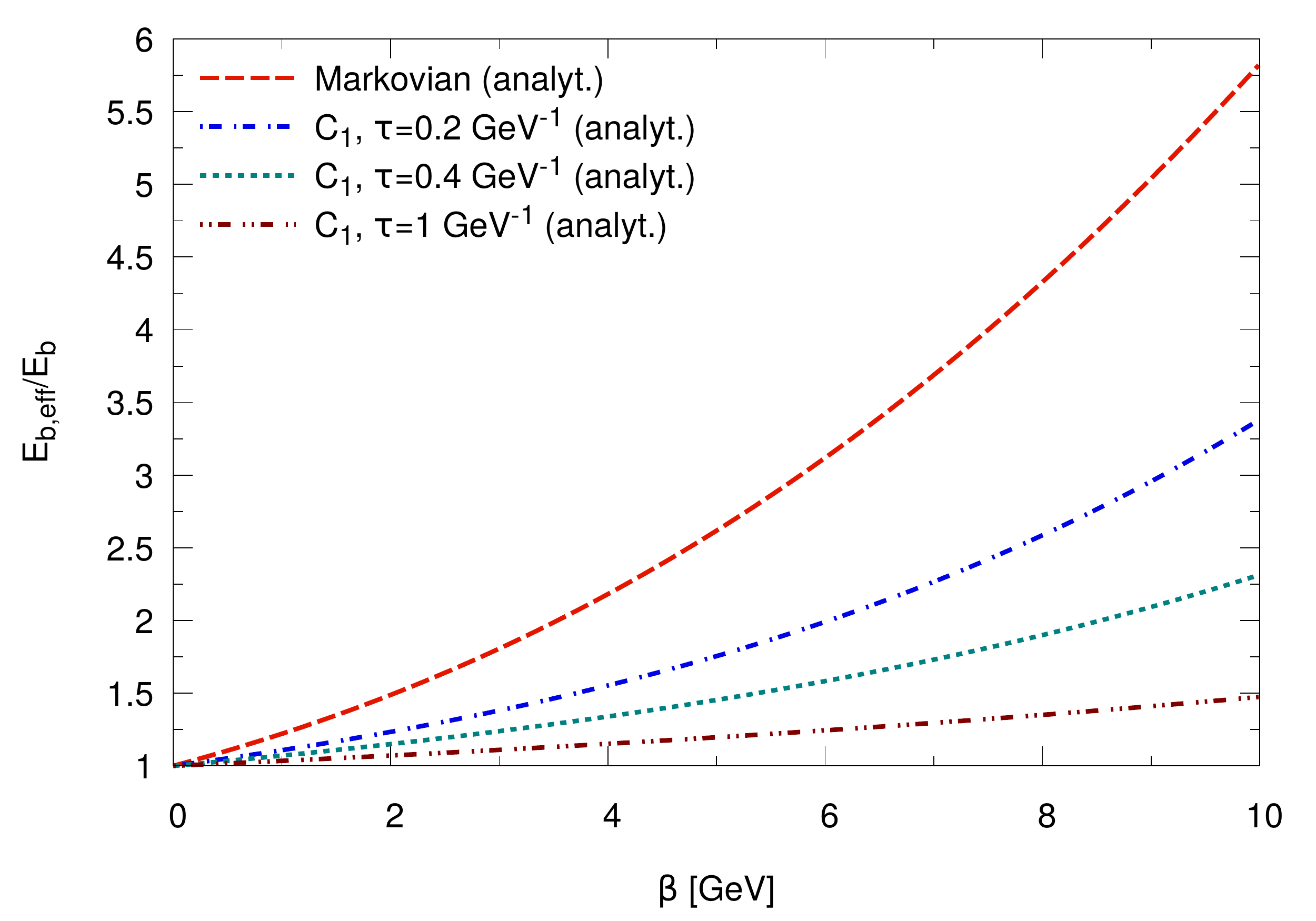}
		\caption{Effective barrier height $E_{b,\mathrm{eff}}$
                  (Eqs.\ \eqref{eq:E_effM} and \eqref{eq:E_effNM})
                  normalized to the barrier height $E_{\text{b}}$ as a
                  function of the coupling $\beta$ for different
                  correlation times $\tau$.}
		\label{fricReduc}
	\end{center}
\end{figure}

For increasing correlation times the respective curves are shifted to
the right and the values of the maxima gradually decrease. However, this
decrease of the maximal value only appears for higher correlation times.
The shift to the right, on the one hand, consequently leads to
systematically higher escape rates for strong friction in case of
increasing correlation times (see Fig.\ \ref{bDepC1}).  On the other
hand, this leads to an effective decrease of the escape rate for weak
friction.
Both, the increase and decrease of the escape rate for strong and weak
coupling $\beta$, is a consequence of an effective reduced friction for
increasing correlation times.
This effect is mentioned in Ref.\ \cite{Boilley:2006mw}, where the
influence of a non-Markovian correlation function on the diffusion over
an inverse parabolic potential is investigated. In this context an ensemble
of Brownian particles is initialized at $x_0<0$ to the left of a potential barrier,
symmetrically located around $x=0$. On that basis an
expression for the overpassing probability over the barrier for
fixed initial conditions, $x_0$ and $v_0$, in the limit of $\lambda_{\mathrm{M}}t\gg 1$ or
$\lambda_{\mathrm{NM}}t\gg 1$ is derived for correlation functions $C_0$
and $C_1$ (Eqs.\ \eqref{eq:C0} and \eqref{eq:C1}), respectively \cite{Boilley:2006mw,Schmidt:2014zpa}:
%\begin{equation}
%F_{x}(t;x_0,v_0)=\frac{1}{2}\mbox{erfc}\left(\frac{\omega}{\sqrt{\beta\lambda_\mathrm{M}}}\left[\sqrt{\frac{E_{\text{b}}}{T}}-\frac{\lambda_{\mathrm{M}}}{\omega}\sqrt{\frac{K}{T}} \right]\right),
%\end{equation}
\begin{equation}
F(t;x_0,v_0)=\frac{1}{2}\mbox{erfc}\left(\frac{\omega}{\sqrt{\beta\lambda_\mathrm{M}}}\left[\sqrt{\frac{B}{T}}-\frac{\lambda_{\mathrm{M}}}{\omega}\sqrt{\frac{K}{T}} \right]\right),
\end{equation}
%\begin{equation}
%F(t;x_0,v_0)=\frac{1}{2}\mbox{erfc}\left(\frac{\omega\sqrt{1+\lambda_{\mathrm{NM}}\tau}}{\sqrt{\beta\lambda_{\mathrm{NM}}}}\left[\sqrt{\frac{E_{\text{b}}}{T}}
%-\frac{\lambda_{\mathrm{NM}}}{\omega}\sqrt{\frac{K}{T}}\right]\right).
%\end{equation}
\begin{equation}
F(t;x_0,v_0)=\frac{1}{2}\mbox{erfc}\left(\frac{\omega\sqrt{1+\lambda_{\mathrm{NM}}\tau}}{\sqrt{\beta\lambda_{\mathrm{NM}}}}\left[\sqrt{\frac{B}{T}}
-\frac{\lambda_{\mathrm{NM}}}{\omega}\sqrt{\frac{K}{T}}\right]\right).
\end{equation}
Hereby, $K$ denotes the initial kinetic energy of a Brownian particle,
 i.e. $K=\frac{1}{2}mv_0^2$, $B$ is the height of the barrier the 
Brownian particle needs to overcome, starting from position 
$x_0$, i.e. $B=\frac{1}{2}m\omega^2x_0^2$, $\omega$ is the barrier frequency
and $\lambda_{\mathrm{M}}$ and $\lambda_{\mathrm{NM}}$ designate the
quantities, indicated in the context of the Markovian and non-Markovian
model of Kramers's escape rate problem (see Eqs.\ \eqref{eq:lamM}
and \eqref{eq:lamNM}), where $\lambda_{\mathrm{NM}}$ is derived in \ref{sec:escRateC1}. 
Given these stationary overpassing
probabilities, it is straightforward to compute an initial kinetic energy the
Brownian particle must possess to overcome the potential barrier with a
probability of 50\%, setting the expressions in parentheses to zero. For correlation function $C_0$ this is
\begin{equation}
K:=B_{\mathrm{eff}}=\left(\frac{\omega}{\lambda_{\mathrm{M}}}\right)^2B,
\label{eq:B_effM}
\end{equation}
and for correlation function $C_1$ the appropriate initial kinetic energy $K$ is given by  
\begin{equation}
K:=B_{\mathrm{eff}}=\left(\frac{\omega}{\lambda_{\mathrm{NM}}}\right)^2B.
\label{eq:B_effNM}
\end{equation}
To relate the results of Ref.\ \cite{Boilley:2006mw} to the simulations of this work, $B$ 
needs to be replaced by the barrier height $E_b$ of the composite potential, Eq.\ \eqref{eq:potential},
the Brownian particle has to overcome, starting at the bottom of the initial well (see Fig.\ \ref{potSim}), i.e.
\begin{equation}
B_{\mathrm{eff}}\approx E_{\mathrm{b,eff}} = \left(\frac{\omega}{\lambda_{\mathrm{M}}}\right)^2E_{\text{b}},
\label{eq:E_effM}
\end{equation}
\begin{equation}
B_{\mathrm{eff}}\approx E_{\mathrm{b,eff}}=\left(\frac{\omega}{\lambda_{\mathrm{NM}}}\right)^2E_{\text{b}}.
\label{eq:E_effNM}
\end{equation}
 Certainly, this 
is just an approximation but it does not change the qualitative implications:
   
Comparing the ratio $\frac{E_{b,\mathrm{eff}}}{E_{\text{b}}}$ in the
Markovian and non-Markovian limit as a function of the coupling $\beta$
it can be concluded that the effective barrier height
$E_{b,\mathrm{eff}}$ systematically reduces for increasing correlation
times and fixed $\beta$ (see Fig.\ \ref{fricReduc}).
This reduction of the effective barrier height for fixed $\beta$ and
increasing correlation times in turn is equivalent to an effectively
reduced friction. Hence, it can be assumed that the average behavior of
a considered ensemble in case of a non-Markovian noise is basically the
same as in case of a Markovian noise, but with a friction rate $\beta$
being effectively reduced (see Fig.\ \ref{fricReduc}).  

At least for the low-friction regime this effective reduction
of the friction rate $\beta$ can be directly seen from the approximate analytical formula computed with Eq.\ \eqref{eq:lowFricNM}, which will be explained in detail in Sec.\ \ref{subsec:comparison}.
Taking now the
formulas for the weak- and the strong-friction regime in case of
Kramers's classical escape rate problem (see Eqs.\
\eqref{eq:lowFric} and \eqref{eq:strongFric}) it is straightforward to understand
how increasing correlation times lead to smaller escape rates for weak
friction and higher escape rates for strong friction.  Furthermore
increasing correlation times are responsible for the shift of the
curves, since for higher correlation times higher values for $\beta$ are
required for the strong-friction regime to be valid.

\subsubsection{Correlation function $C_2$}
\begin{figure}[h]
	\begin{center}
		\includegraphics[scale=0.64]{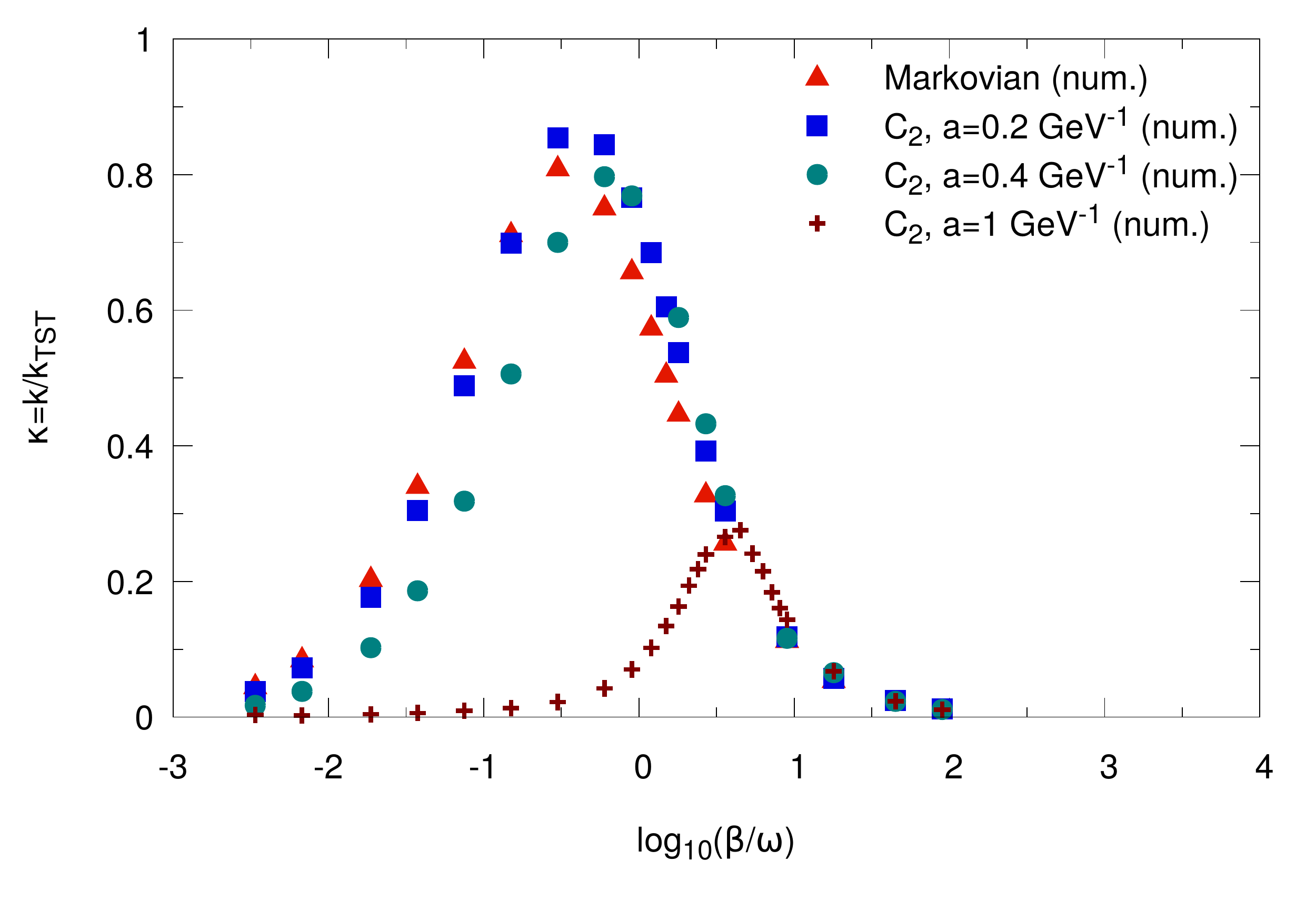}
		\caption{Comparison of the normalized steady-state
                  escape rate $\kappa$ as a function of the
                  dimensionless parameter $\frac{\beta}{\omega_b}$ for
                  correlation functions $C_0$ and $C_2$ (Eqs.\
                  \eqref{eq:C0} and \eqref{eq:C2}), where
                  $E_{\text{b}}=2.5 \,\GeV$, $\omega_b=5 \, \GeV$,
                  $m=1.11 \, \GeV$ and $T=1 \, \GeV$.}
		\label{bDepC2}
	\end{center}
\end{figure}

Again, the depicted curves for correlation function $C_2$ (Eq.\
\eqref{eq:C2}) exhibit the expected bell-shaped form (see Fig.\
\ref{bDepC2}).  As for correlation function $C_1$, the above-mentioned
effects of increasing correlation times compared to the Markovian case
are observed, i.e. the shift to the right, the decrease of the maximum,
smaller escape rates for weak friction and higher escape rates for
strong friction. In contrast to correlation function $C_1$ the shift is
comparatively tiny for smaller correlation times ($a=0.2 \, \GeV^{-1}$ and
$a=0.4 \, \GeV^{-1}$), leading to less deviation from the Markovian case (see
Fig.\ \ref{bDepC2}). For a large correlation time ($a=1 \, \GeV^{-1}$), however,
the shift is even greater than for a large correlation time ($\tau=1 \, \GeV^{-1}$) in case of correlation function $C_1$ (see Figs.\ \ref{bDepC1} and
\ref{bDepC2}).  Even though, because of a lack of analytical results for
correlation function $C_2$, no exact information exists about the
behavior of $\beta$ with regard to increasing correlation times, it is
reasonable to assume a similar behavior as for correlation function
$C_1$. However, this effective reduction of the friction for increasing
correlation times seems to be much more significant for higher
correlation times (see Fig.\ \ref{bDepC2}).

\subsubsection{Correlation function $C_3$}
\label{C3Discussion}
\begin{figure}[h]
	\begin{center}
		\includegraphics[scale=0.65]{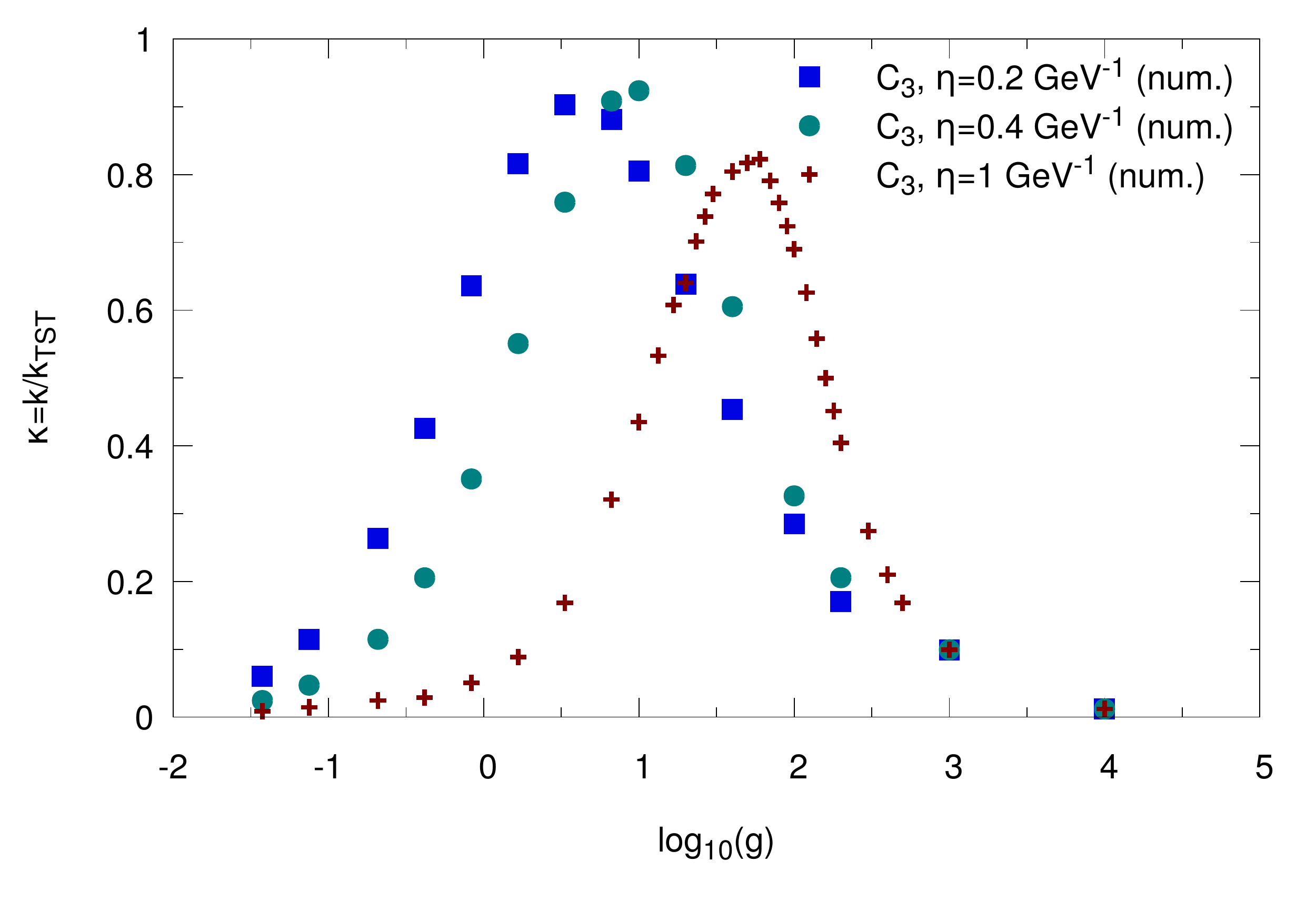}
		\caption{Normalized steady-state escape rate $\kappa$ as
                  a function of the dimensionless coupling $g$ for
                  correlation function $C_3$ (Eq.\ \eqref{eq:C3})
                  and different values of correlation time $\eta$
                  (Eq.\ \eqref{eq:eta}), where
                  $E_{\text{b}}=2.5 \, \GeV$, $\omega_b=5 \, \GeV$,
                  $m=1.11 \, \GeV$ and $T=1 \,\GeV$.}
		\label{bDepC3}
	\end{center}
\end{figure}

The numerical studies for correlation function $C_3$, Eq.\
\eqref{eq:C3}, need to be considered separately from the previous
ones. Unlike before, the steady-state escape rate is not examined as a
function of the coupling $\beta$ but of the dimensionless coupling $g$
(see Fig.\ \ref{bDepC3}).  For this particular correlation function (see
Eq.\ \eqref{eq:C3}) no strict Markovian limit exists as the Fourier
transform vanishes in the limit of $\omega \rightarrow 0$ (see Eq.\
\eqref{eq:fourierC3}).
For that reason, only correlation function $C_3$ is investigated
here for different correlation times. A number of the peculiarities
of correlation function $C_3$ is discussed in \ref{sec:C3}.

Although in many respects very different from correlation function $C_1$
and $C_2$ (see \ref{sec:C3}), even for correlation function
$C_3$ the different curves obey the above-mentioned bell-shaped
behavior. Furthermore, as for correlation functions $C_1$ and $C_2$, a
shift of the curves for increasing correlation times can be observed,
connected to the same implications as for the other correlation
functions. Different from before the value of the maximum seems to
reduce very slowly, as even for high correlation times the maximum only
lies slightly below the maxima for smaller correlation times (see Fig.\
\ref{bDepC3}). It is remarkable that for small correlation times the
steady-state escape rate comes very close to the TST-rate, much closer
than in case of correlation functions $C_1$ and $C_2$. Taking all
results together, it is again reasonable to assume that increasing
correlation times lead to an effective reduction of the actual friction
$\beta$.

\subsection{Comparison of analytical with numerical results} 
\label{subsec:comparison}

\begin{figure}[h]
	\begin{center}
		\includegraphics[width=0.95\textwidth]{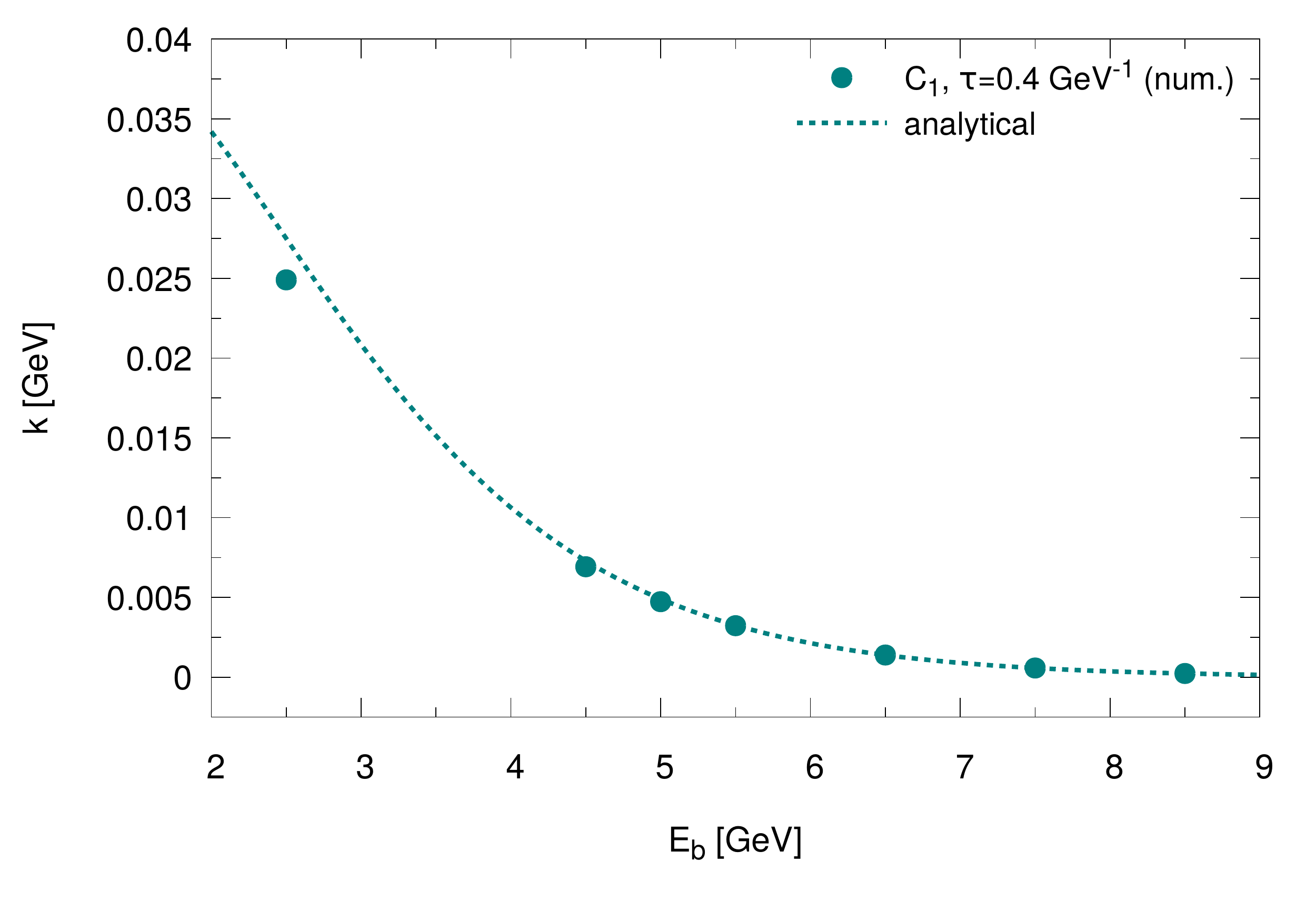}
		\caption{Steady-state escape rate $k$ as
                  a function of the barrier height $E_{\text{b}}$, where
                  $\beta=18 \,\GeV$, $m=1.11 \, \GeV$ and
                  $T=1 \,\GeV$. The analytical solution is generated
                  using Eq.\
                  \eqref{eq:intermediate-to-strongFricNM} with $\lambda_{\mathrm{NM}}$ given in \ref{sec:escRateC1}.}
		\label{tauDep}
	\end{center}
\end{figure}

The aim of this subsection is to discuss the accuracy of the numerical
results, presented above, compared to the approximate analytical solutions (see
Eqs.\ \eqref{eq:lowFric}, \eqref{eq:intermediate-to-strongFric},
\eqref{eq:strongFric}, \eqref{eq:turnover}, \eqref{eq:lowFricNM} and
\eqref{eq:intermediate-to-strongFricNM} with $\lambda_{\mathrm{NM}}$ given in \ref{sec:escRateC1}).
It should be recalled here that in the weak-friction regime the Brownian particle is subject to an almost frictionless, deterministic oscillatory movement inside the initial potential well (see Figs.\ \ref{trajectory} and \ref{potSim}) which corresponds to a harmonic oscillator. Therefore, the action $I$ at energy $E_\mathrm{b}$, a term common to the approximate analytic formulas in the weak-friction regime, Eqs.\ \eqref{eq:lowFric} and \eqref{eq:lowFricNM}, is given by:
\begin{equation}
	I(E_\mathrm{b})=\frac{2\pi E_\mathrm{b}}{\omega}.
	\label{eq:action}
\end{equation}
Starting first with the comparison in the
intermediate-to-strong-friction regime (see Fig.\ \ref{bDepC1}), for a small correlation
time, $\tau=0.2 \, \GeV^{-1}$, the
Markovian and the non-Markovian simulations (correlation functions $C_0$
and $C_1$; Eqs.\ \eqref{eq:C0} and \eqref{eq:C1}) show very good consistency with the analytical
results (Eqs.\ \eqref{eq:intermediate-to-strongFric},
\eqref{eq:strongFric} and
\eqref{eq:intermediate-to-strongFricNM} with $\lambda_{\mathrm{NM}}$ given 
in \ref{sec:escRateC1}). Deviations from the analytical
results are not greater than 2\% and within the error bars. The obtained
accuracy could be further improved by use of smaller time steps
$\Delta t$. For increasing correlation times, $\tau=0.4 \, \GeV^{-1}$ and
$\tau=1 \, \GeV^{-1}$, however, the accuracy is steadily decreasing. While the deviation of the numerical and analytical results
is about 10\% for $\tau=0.4 \, \GeV^{-1}$, the discrepancy is even greater (about
30\%) for $\tau=1 \, \GeV^{-1}$.  This growing divergence for increasing correlation times is most
likely due to fact that Eq.\ \eqref{eq:intermediate-to-strongFricNM} is
not longer applicable. In fact, it can be shown that Eq.\
\eqref{eq:intermediate-to-strongFricNM} becomes valid again for larger
barrier heights $E_{\text{b}}$.  Exemplary Fig.\ \ref{tauDep}
demonstrates how the accuracy of the numerical results for a fixed
choice of parameters is improved by increasing the barrier height
$E_{\text{b}}$.
Beginning with a deviation of about 10\% for $E_{\text{b}}=2.5$ GeV the
discrepancy reduces gradually to less than 1\% for $E_{\text{b}}=8.5$
GeV.

In the weak-friction regime the accuracy of the Markovian simulations
compared to the analytical results, evaluated with Eq.\
\eqref{eq:lowFric}, is worse than in the strong-friction regime. Here
only the first two points on the left side approximately match with the
analytical result. The deviation of the first point located to the
outermost left is about 10\%, the second point already deviates about
20\%. This higher deviation can be attributed to the fact that on the
one hand the time step $\Delta t$ is too large and on the other hand
that the condition $E_{\text{b}}\gg k_{\text{B}}T$ for Eq.\
\eqref{eq:lowFric} to be valid is not fulfilled properly. Especially
condition $E_{\text{b}}\gg k_{\text{B}}T$ seems to have a stronger effect on
the validity of Eq.\ \eqref{eq:lowFric}, which can be clarified by means
of the classical-rate phase diagram (see Fig.\ \ref{phaseDiag}).
Apparently the range of validity of Eq.\ \eqref{eq:lowFric} becomes
smaller, the smaller the ratio
$\frac{k_{\text{B}}T}{E_{\text{b}}}$. This explains the observation that
the analytical results only fit the numerical results for very small
friction values.

In contrast to that, in the non-Markovian case the accordance between
numerical and analytical results (see Eq.\ \eqref{eq:lowFricNM})
improves for increasing correlation times (see Fig.\
\ref{bDepC1}). Growing correlation times seem to enlarge the range of
validity of Eq.\ \eqref{eq:lowFricNM} step by step, leading to a
very good consistency until close to the maximum of the rate. It should
be noticed here that the approximate analytical results (see Eq.\
\eqref{eq:lowFricNM}) were evaluated under the assumption that the
initial well is an ideal harmonic oscillator. This is a reasonable
approximation for the potential field used for the simulations (see
Eq.\ \eqref{eq:potential}). In this case, computing $\epsilon(I_{\mathrm{B}})$ 
(see Eq.\ \eqref{eq:epsilon}) and inserting it into Eq.\ \eqref{eq:lowFricNM} results in 
\begin{equation}
	k_{A\to C}
	=\frac{\beta}{1+\tau^2\omega^2}\frac{ I(E_{\text{b}})}{k_{\text{B}}T}\frac{\omega_a}{2\pi}\exp\left[-\frac{E_{\text{b}}}{k_{\text{B}}T}\right],
\end{equation}
where $I(E_\mathrm{b})$ is again given by Eq.\ \eqref{eq:action}.
This corresponds to the classical steady-state escape rate in
the weak-friction regime (Eq.\ \eqref{eq:lowFric}) but with the damping 
rate $\beta$ being reduced by a factor 
%Hence, as in Kramers's classical model (see Subsec.\ \ref{sec:classMod}),
%the steady-state escape rate in the weak-friction regime is a linear
%function of the damping rate $\beta$ but reduced by a factor 
of $\frac{1}{1+\tau^2\omega^2}$, which can be essentially identified
with the Fourier transform of correlation function $C_1$ (see Eq.\
\eqref{eq:FourierC1}).  Basically, the effective damping in the weak-friction 
regime is obtained by substituting the damping $\gamma$ by
$\tilde{\Gamma}(\omega=\omega_a)/2$ in the linear harmonic approximation
as an effectively well-defined Markovian description
\cite{Greiner:1996dx,Greiner:1998vd, Xu:1999aq}.

This in fact supports the statement, at least for correlation function
$C_1$ in the low-friction regime, 
that the main difference between the Markovian and non-Markovian escape
rate is the effectively reducing friction rate for increasing correlation 
times. 
   
Summing up the results for the low- and the strong-friction regime, 
there obviously exist two opposite effects on the validity of
formulas \eqref{eq:lowFricNM} and
\eqref{eq:intermediate-to-strongFricNM} concerning increasing
correlation times. On the one hand rising correlation times lead to
improving accordance between numerical and analytical results in the
weak-friction limit. On the other hand accordance becomes worse in the
intermediate-to-strong-friction regime. To obtain a comparably good
consistency in both limiting regimes either the barrier height
$E_{\text{b}}$ has to be increased (see also Fig.\ \ref{tauDep}) or the
temperature $T$ has to be decreased.

Finally, only the comparison of the bridging formula, Eq.\
\eqref{eq:turnover} with the numerical results of the Markovian
simulations (i.e. using correlation function $C_0$, see Eq.\
\eqref{eq:C0}) remains.  First of all, it should be mentioned that the
simple ad hoc formula, Eq.\ \eqref{eq:turnover}, in fact yields the expected
bell-shaped curve. Furthermore good accordance in both limiting regimes
can be seen as expected from the construction of formula
\eqref{eq:turnover} (see Fig.\ \ref{bDepC1}). Even the points to the
left of the maximum, which were not fitted properly by the steady-state
escape rate in the weak-friction limit, Eq.\ \eqref{eq:lowFric}, are
approximately covered (see Fig.\ \ref{bDepC1}). The difference between
analytical and numerical results here is about 12\%, which is the usual
deviation between numerical and analytical results, obtained by other
researchers using different numerical approaches
\cite{Hanggi:1990}. Moreover, the second point to the outermost left is
fitted more accurately by the bridging formula, Eq.\ \eqref{eq:turnover},
compared to the analytical equation for the weak-friction escape rate,
Eq.\ \eqref{eq:lowFric}. While the discrepancy between numerical and
analytical results is about 20\% for Eq.\ \eqref{eq:lowFric}, the
difference reduces to about 10\% for Eq.\ \eqref{eq:turnover}. This
in fact seems to substantiate the above-mentioned assumption that
equation \eqref{eq:lowFric} is not longer valid for the appropriate
damping rate.

\section{Conclusions} \label{chap:concl}

In this work Kramers's steady-state escape rate has been computed
numerically as a function of the damping rate $\beta$ in the case of a
Markovian noise $C_0$, Eq.\ \eqref{eq:C0}, and three non-Markovian
noise variants, $C_1$, $C_2$ and $C_3$, cf.\ Eqs.\
\eqref{eq:C1}-\eqref{eq:C3}, solving the appropriate Markovian or 
non-Markovian GLE, Eq.\ \eqref{eq:genlang1}, with the
three-step Adams-Bashforth method, indicated in Sec.\ \ref{seq:numStud}. Hereby the numerical implementation \cite{Schmidt:2014zpa} of the algorithm, depicted in
Sec.\ \ref{chap:genCN}, is used to generate the non-Markovian noise,
given a symmetric and exponentially decaying correlation function.

A first objective then has been to verify the match between numerical
and analytical results for correlation functions $C_0$ and $C_1$, cf.\
Eqs.\ \eqref{eq:C0} and \eqref{eq:C1}.  Overall it appears that there is
good consistency between numerical and analytical results (see Subsec.\
\ref{subsec:comparison}).  Appearing deviations -- in the weak-friction
regime not larger than 10\% and in the strong-friction regime less than
2\% -- are the consequence of the invalidity of the approximative
analytic formulas, Eqs.\ \eqref{eq:lowFric},
\eqref{eq:intermediate-to-strongFric}, \eqref{eq:strongFric},
\eqref{eq:lowFricNM} and \eqref{eq:intermediate-to-strongFricNM}, where
$\lambda_{\mathrm{NM}}$ is given in \ref{sec:escRateC1}, and not
of the incorrectness of numerical results. By suitable selection of the
relevant parameters (barrier height $E_{\text{b}}$, temperature $T$,
size of time steps $\Delta t$) the accordance can be further increased
at the expense of higher computation times.

After having established that the numerical algorithm indeed works well
the main objective of this work has been to identify the differences of
Kramers's steady-state escape rate for white and colored noise for the
different correlation functions and to provide a possible explanation
for this differences.

It turns out that growing correlation times lead to a decrease of the
steady-state escape rate in the weak-friction regime and to an increase
in the intermediate-to-strong-friction regime for fixed values of the
damping rate $\beta$ for correlation functions $C_1$ and $C_2$, cf.\
Eqs.\ \eqref{eq:C1} and \eqref{eq:C2}. In the case of correlation
function $C_1$, for which analytical results exist, both effects are
identified to be the consequence of an effectively reduced friction for
increasing correlation times. Since correlation function $C_2$
qualitatively obeys the same behavior, it is reasonable to assume the
same explanation. However, this should be verified by an analytical
treatment of correlation function $C_2$.

Furthermore, special attention should be payed to correlation function
$C_3$, Eq.\ \eqref{eq:C3}. Although rather similar behavior of the
steady-state escape rate as a function of the dimensionless coupling $g$
(not $\beta$ for correlation function $C_3$) for growing correlation
times is obtained, correlation function $C_3$ obeys some special
features, compared to correlations functions $C_1$ and $C_2$, which are
discussed in \ref{sec:C3}. Next to a vanishing Fourier
transform $\tilde{\Gamma}(\omega)=0$ for $\omega=0$, solving the
GLE for a free Brownian particle, Eq.\ \eqref{eq:genlang1}, where the
potential term is neglected, with correlation function $C_3$ yields
different peculiarities:
There is a non-vanishing retarded Green's function $G_\mathrm{ret}(t)$
for $t\to\infty$, the equipartition theorem becomes invalid and the
equilibrium velocity distribution function seems to obey a Boltzmann
distribution but with a temperature being reduced by a certain factor
(see \ref{sec:C3}). However, as is shown in \ref{sec:C3},
the equipartition theorem becomes again valid for a bound Brownian particle. 

Altogether it can be stated that the numerical algorithm essentially based on 
the three-step Adams-Bashforth method and the generation of a colored, non-Markovian thermal
noise is perfectly applicable to Kramers's classical escape rate problem and can be, differently from the approximate analytical formulas \eqref{eq:lowFric},
\eqref{eq:intermediate-to-strongFric}, \eqref{eq:strongFric},
\eqref{eq:lowFricNM} and \eqref{eq:intermediate-to-strongFricNM},
employed for arbitrarily shaped potentials and correlation functions without the need of any additional corrections, resulting for example from anharmonicities
of the potential \cite{TalknerR1993,Talkner1993,Talkner:1995} (see also Sec.\ \ref{sec:nonMarkovMod}).  

%Next to this specialties it is also unknown how the Markovian limit of
%correlation function $C_3$ could look like.
%
%Further research should therefore answer the following outstanding questions:
%\begin{enumerate}
%	\item is the equipartition theorem still invalid for a bound Brownian particle?
%	\item is the velocity distribution function actually a Gaussian distribution?
%	\item what is the Markovian limit of correlation function $C_3$?
%\end{enumerate}
%
%The answer to the first question can be found by solving the GLE for a
%Brownian particle, being trapped in a harmonic potential. For an answer
%to the second question higher moments of the Brownian particle's
%velocity need to be computed. The Markovian limit corresponds to the
%limiting case of the appropriate correlation time going to zero.
%Therefore, to answer the third and last question the behavior of
%correlation function $C_3$ concerning a gradually decrease of the
%correlation time $\eta$ must be investigated.

\appendix
%\pagenumbering{Roman}
%\appendixpage
%\addappheadtotoc
%\appendix
%\chapter{Appendices}

\section{One-dimensional Laplace transform (LT)}
\label{chap:LT}

Dealing with initial value problems the application of Laplace
transforms is a very effective tool.  This section is devoted to
the fundamental principles of the Laplace transform.  Furthermore
several useful Laplace transforms are indicated.
%Ref.\ \cite{Doetsch:1974} and Ref.\ \cite{Dyke:1999} are used as a general
%orientation for the creation of this section. 
%Moreover this section is based on Ref.\ \cite{BThesis:2015}.

\paragraph{Definition}

Given a mapping in the form of
\begin{equation}
f :[\, 0, \infty) \rightarrow \mathbb{C}\, , \ t \mapsto f(t),    
%\begin{cases}
%0 & t<0  \\
%f(t) & t\geq 0 \\
%\end{cases} 
\end{equation}
being at least piecewise continuous and of exponential order, where the
latter means that regarding to two constants $M,\,\alpha \in \mathbb{R}$
the condition
\begin{equation}
|f(t)|\leq Me^{\alpha t}
\label{eq:expOrd}
\end{equation}
holds \cite{Dyke:1999}. Then the Laplace transform and its corresponding inverse are
given by \cite{Doetsch:1974,Dyke:1999}:
\begin{align}
\mathcal{L}[f](s)&=\int_0^\infty \! f(t)\cdot e^{-st } \, \mathrm{d}t:=F(s), \quad s\in\mathbb{C},
\label{eq:LT} \\
\mathcal{L}^{-1}[F](t)&=\lim\limits_{\omega \rightarrow \infty} \frac{1}{2\pi i}\int_{s-i\omega}^{s+i\omega} \! F(s)\cdot e^{st} \, \mathrm{d}s=
\begin{cases}
0 & t<0  \\
f(t) & t\geq 0 \\
\end{cases}.
\label{eq:ILT}
\end{align}
The LT of function $f$ exists for $\ReN (s)> \ReN (\alpha)$ due to condition \eqref{eq:expOrd}

\paragraph{Properties}

In accordance with their definitions in Eqs.\ \eqref{eq:LT} and
\eqref{eq:ILT} the LT and its corresponding inverse are linear
transformations. Let there be two functions $g(t)$ and $f(t)$, for which
both the Laplace transforms and their corresponding
back-transforms exist. Then for two arbitrary constants
$a,\,b \in \mathbb{C}$ the following relations hold:
\begin{equation}
\mathcal{L}[a\cdot f(t)+b \cdot g(t)]=a\cdot\mathcal{L}[f](s)+b\cdot\mathcal{L}[g](s), 
\label{eq:linLT}
\end{equation}
\begin{equation}
\mathcal{L}^{-1}[a\cdot F(s)+b \cdot G(s)]=a\cdot\mathcal{L}^{-1}[F](t)+b\cdot\mathcal{L}^{-1}[G](t). 
\label{eq:linILT}
\end{equation}

\paragraph{Useful transformations}
Let there exist two Laplace transformable functions $g(t)$ and $f(t)$,
then the following applies:
\begin{enumerate}
\item \textbf{Exponential function, $a \in \mathbb{C}$ (arbitrary)}  

\begin{equation}
\mathcal{L}[e^{at}](s)=\int_0^\infty \! e^{-(s-a)t } \, \mathrm{d}t=\frac{1}{s-a}, \quad \ReN(s)>\ReN(a)
\label{eq:exp}
\end{equation}
\item \textbf{Convolution}
\begin{equation}
\mathcal{L}\left[\int_0^t \!  f(t-\tau)g(\tau)\, \mathrm{d}\tau\right](s)=\int_0^\infty \! \left(\int_0^t \!  f(t-\tau)g(\tau)\, \mathrm{d}\tau\right) e^{-st}\, \mathrm{d}t=F(s)G(s)
\label{eq:faltung}
\end{equation}

\item \textbf{Time derivative}
\begin{equation}
\mathcal{L}\left[\frac{\dd}{\dd t}f(t)\right](s)=\int_0^\infty \! \left(\frac{\dd}{\dd t}f(t)\right)e^{-st } \, \mathrm{d}t =sF(s)-f_0
\label{eq:abl}
\end{equation}
\end{enumerate}

\section{$\lambda_{\mathrm{NM}}$ for correlation function
  $C_1$} 
\label{sec:escRateC1} 

In this section the prefactor $\lambda_{\mathrm{NM}}$  of the escape rate 
in the spatial-diffusion regime (see Eq.\ \eqref{eq:intermediate-to-strongFricNM}) will be
derived for correlation function $C_1$, Eq.\ \eqref{eq:C1}. For this purpose
the roots of the function
\begin{equation}
f(\lambda)=\lambda^2-\omega_b^2+\frac{\tilde{\Gamma}(\lambda)}{m}\lambda
\label{eq:polynomial}
\end{equation} 
have to be computed. Thereby, $\Gamma$ is related to the correlation
function $C_1$ by the second fluctuation-dissipation theorem (see
Eq.\ \eqref{eq:dissFluc}).
Taking correlation function $C_1$ (see Eq.\ \eqref{eq:C1}), $\Gamma$ is readily obtained as 
\begin{equation}
\Gamma(|t|)=\frac{D}{2k_{\text{B}}T\tau}\exp\left[-\frac{|t|}{\tau}\right]
=\frac{\gamma}{\tau}\exp\left[-\frac{|t|}{\tau}\right],
\end{equation}
where from the first to the second step the fluctuation-dissipation
relation has been employed. Performing the LT of $\Gamma$, using Eq.\
\eqref{eq:exp}, one receives
\begin{equation}
  \tilde{\Gamma}(\lambda) =
  \mathcal{L}\left[\frac{\gamma}{\tau}\exp\left[-\frac{|t|}{\tau}\right]\right]
  = \frac{\gamma}{\tau\lambda+1}.
\label{eq:gammaTilde}
\end{equation}
Subsequent insertion of Eq.\ \eqref{eq:gammaTilde} in function
\eqref{eq:polynomial} leads to
\begin{equation}
f(\lambda)=\lambda^2-\omega_b^2+\frac{\beta\lambda}{\tau\lambda+1}.
\label{eq:polynomial1}
\end{equation}
To obtain $\lambda_{\mathrm{NM}}$ the next task will be to identify the
roots of \eqref{eq:polynomial1}
\begin{equation}
\lambda^2-\omega_b^2+\frac{\beta\lambda}{\tau\lambda+1}=0,
\label{eq:polynomial2}
\end{equation}
using Cardano's formula.
To this end, the algorithm indicated in Ref.\ \cite{Greiner:2010} is
applied on the above equation. First, however, Eq.\
\eqref{eq:polynomial2} must be transformed into the form,
\begin{equation}
\lambda^3+a\lambda^2+b\lambda+c=0,
%\lambda^3+\frac{1}{\tau}\lambda^2+\left(\frac{\beta}{\tau}-\omega_b^2\right)\lambda-\frac{\omega_b^2}{\tau}=0.
\end{equation}
where 
\begin{align}
a&=\frac{1}{\tau}, \label{eq:a} \\
b&=\frac{\beta}{\tau}-\omega_b^2, \label{eq:b1} \\
c&=-\frac{\omega_b^2}{\tau} \label{eq:c}.
\end{align}
Dependent on the expression 
\begin{equation}
D=\left(\frac{q}{2}\right)^2+\left(\frac{p}{3}\right)^3,
\end{equation}
where 
\begin{equation}
p=b-\frac{a^2}{3}
\label{eq:p}
\end{equation}
and 
\begin{equation}
q=\frac{2a^3}{27}-\frac{ab}{3}+c,
\label{eq:q}
\end{equation} 
there are three different cases for the solution of Eq.\ \eqref{eq:polynomial2}, supposing $p\ne 0$
\cite{Greiner:2010}:
\begin{enumerate}
	\item $D>0$: One real root and two complex conjugate roots,
	\item $D=0$: Three real roots (one double root),
	\item $D<0$: Three distinct real roots.
\end{enumerate}
Subsequently, the solutions for the three different cases for $D$, using the
above relations for $a$, $b$, $c$, $p$ and $q$ (see Eqs.\ \eqref{eq:a},
\eqref{eq:b1}, \eqref{eq:c}, \eqref{eq:p} and \eqref{eq:q}), are
indicated: \newline \textbf{$D>0$:}
\begin{align}
\begin{split}
\lambda_1 &=A+B -\frac{a}{3}, \\ 
\lambda_{2,3} &=-\frac{A+B}{2}\pm\frac{A-B}{2} \ii\sqrt{3}-\frac{a}{3},
\end{split}
\label{eq:deltaP1}
\end{align}
where
\begin{align}
\begin{split}
A&=\sqrt[3]{-\frac{q}{2}+\sqrt{D}},\\
B&=\sqrt[3]{-\frac{q}{2}-\sqrt{D}}.
\end{split}
\end{align}
\newline \textbf{$D=0$:}
\begin{align}
\begin{split}
\lambda_1 &=\sqrt[3]{-4q}-\frac{a}{3}, \\ 
\lambda_{2,3} &=\sqrt[3]{\frac{q}{2}}-\frac{a}{3}. 
\end{split}
\label{eq:delta1}
\end{align}
\newline \textbf{$D<0$:}
\begin{align}
\begin{split}
\lambda_1 &=2\sqrt{-\frac{p}{3}}\cos\left(\frac{\theta}{3}\right)-\frac{a}{3}, \\ 
\lambda_2 &=-2\sqrt{-\frac{p}{3}}\cos\left(\frac{\theta}{3}-\frac{\pi}{3}\right)-\frac{a}{3}, \\
\lambda_3 &=-2\sqrt{-\frac{p}{3}}\cos\left(\frac{\theta}{3}+\frac{\pi}{3}\right)-\frac{a}{3},
\end{split}
\label{eq:deltaM1}
%  x_1 &=\sqrt{-\frac{4p}{3}}\cos\left(\frac{1}{3}\arccos\left(-\frac{q}{2}\sqrt{-\frac{27}{p^3}}\right)\right) -\frac{a}{3}, \\ \label{eq:delta-1}
%  x_2 &=-\sqrt{-\frac{4p}{3}}\cos\left(\frac{1}{3}\arccos\left(-\frac{q}{2}\sqrt{-\frac{27}{p^3}}\right)+\frac{\pi}{2}\right)-\frac{a}{3}, \\
%  x_3 &=-\sqrt{-\frac{4p}{3}}\cos\left(\frac{1}{3}\arccos\left(-\frac{q}{2}\sqrt{-\frac{27}{p^3}}\right)-\frac{\pi}{2}\right)-\frac{a}{3}. 
\end{align}
where 
\begin{equation}
\theta=\arccos\left(-\frac{q}{2}\sqrt{-\frac{27}{p^3}}\right).
\end{equation}
For a more detailed discussion of the roots of the particular function,
Eq.\ \eqref{eq:polynomial1}, reference is made to
Ref.\ \cite{Boilley:2006mw}. 
The quantity $\lambda_{\mathrm{NM}}$ is then given by the largest positive root $\lambda_1$ of Eqs.\ \eqref{eq:deltaP1}, \eqref{eq:delta1} or
\eqref{eq:deltaM1}, respectively.
From the above expressions it can be furthermore
concluded that $\lambda_{\mathrm{NM}}$ is a
function of $\beta$, $\omega_b$ and $\tau$,
i.e. $\lambda_{\mathrm{NM}}=\lambda_{\mathrm{NM}}(\beta,\omega_b,\tau)$.

\section{Peculiarities of correlation function $C_3$}
\label{sec:C3}

\begin{figure}[tb]
	\begin{center}
		%\begin{overpic}[width=0.6\textwidth]{./pictures/C3/GammaT.pdf}\end{overpic} \hfill \begin{overpic}[width=0.6\textwidth]{./pictures/C3/GammaOm.pdf}\put(65,35){}\end{overpic}
		\begin{overpic}[scale=0.35]{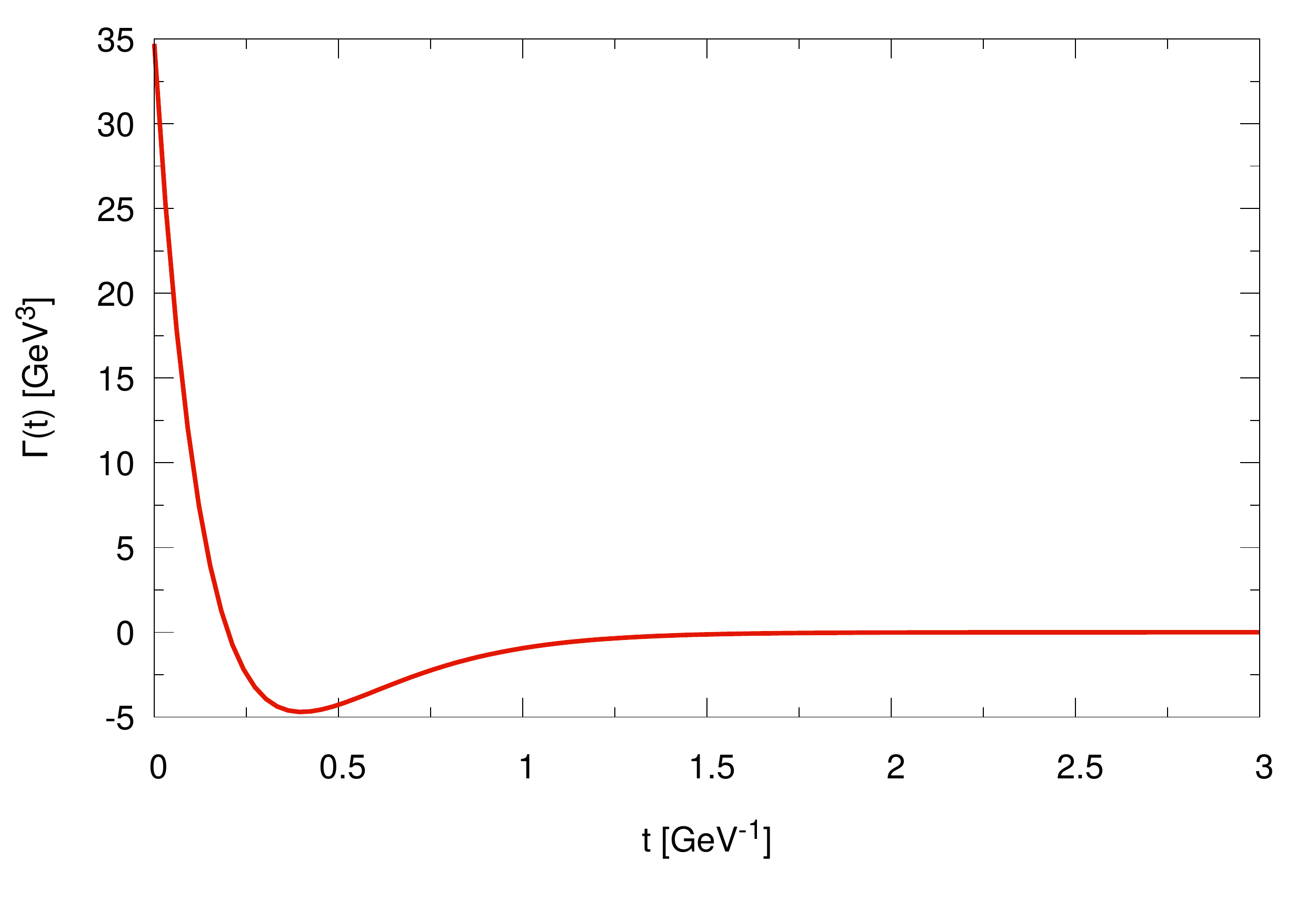}\end{overpic} \hfill \begin{overpic}[scale=0.35]{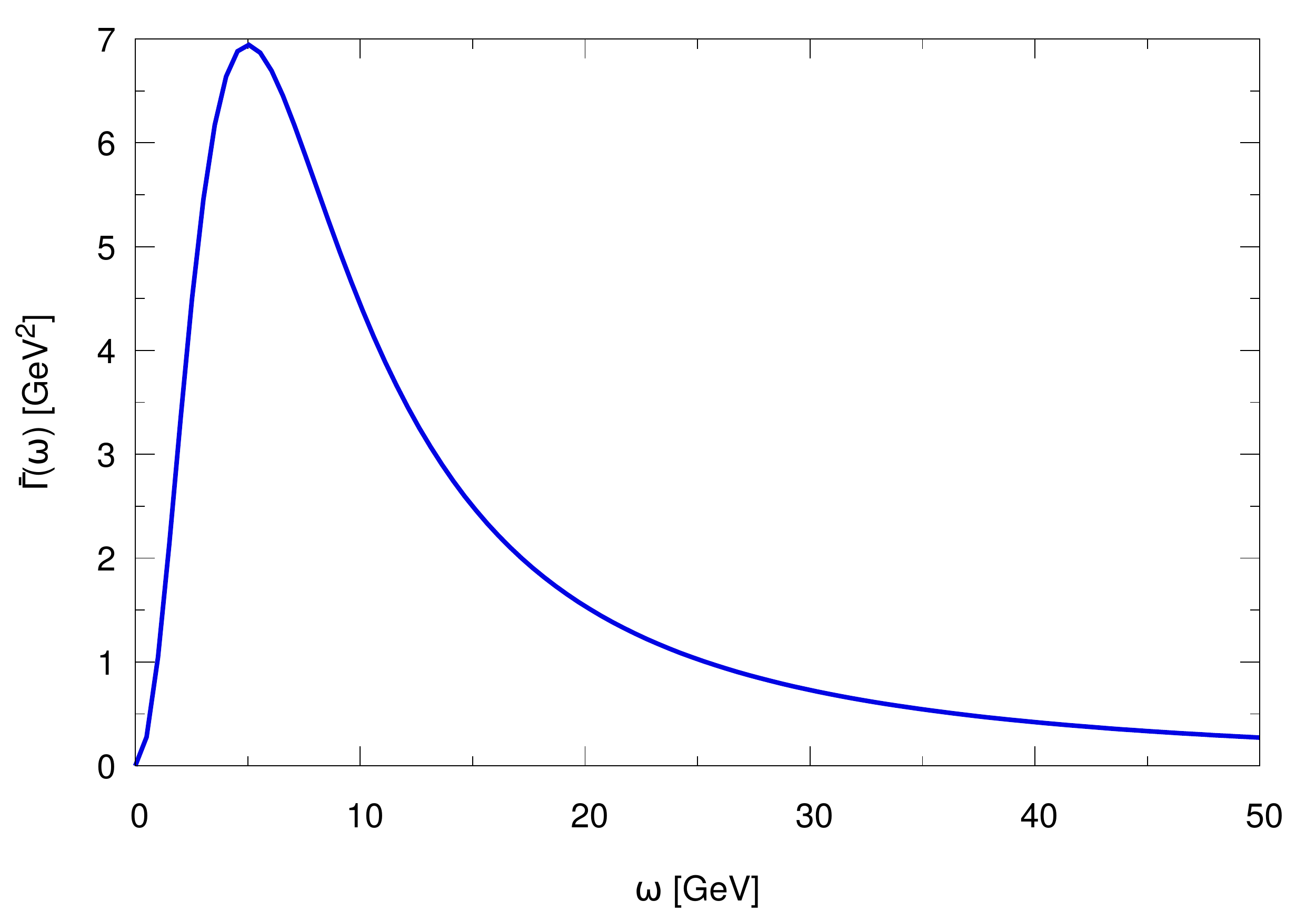}\put(65,35){}\end{overpic}
		\caption{Dissipation kernel $\Gamma(t)$ (upper figure;
                  Eq.\ \eqref{eq:Corr3}) and its Fourier
                  transform $\tilde{\Gamma}(\omega)$ (lower figure; Eq.\
                  \eqref{eq:FTCorr3}), where $g=5$, $m=1.11 \,\GeV$ and
                  $\eta=0.2 \, \GeV^{-1}$.}
		\label{C3figure}
	\end{center}
\end{figure}

An interesting dissipation kernel, bearing very special features and
being related to correlation function $C_3$, cf.\ Eq.\ \eqref{eq:C3}, via
the second fluctuation-dissipation theorem, Eq.\ \eqref{eq:dissFluc}, is
written as
\begin{equation}
\Gamma(|t|)=\frac{g}{4}\alpha^2\left(1-\frac{\alpha}{\sqrt{m}}|t|\right)\exp\left(-\frac{\alpha}{\sqrt{m}}|t|\right).
\label{eq:Corr3}
\end{equation}
Its Fourier transform is given by
\begin{equation}
\tilde{\Gamma}(\omega)=\frac{g\alpha^3\omega^2}{\sqrt{m}\left(\omega^2+\frac{\alpha^2}{m}\right)^2}
=\frac{g\alpha^3\omega^2}{\sqrt{m}\left(\omega+\frac{i\alpha}{\sqrt{m}}\right)^2\left(\omega-\frac{i\alpha}{\sqrt{m}}\right)^2},
\label{eq:FTCorr3}
\end{equation}
which was computed using Eq.\ \eqref{eq:inverseFourier}.
%Hereby the following convention for the Fourier transformation has been
%employed
%\begin{align}
%\Gamma(t) &=\frac{1}{2\pi}\int_{-\infty}^{\infty}e^{-\ii \omega t}\tilde{\Gamma}(\omega)\dd \omega, \\
%\tilde\Gamma(\omega) &=\int_{-\infty}^{\infty}e^{\ii\omega t}\Gamma(t)\dd t.
%\end{align}
From Eq.\ \eqref{eq:Corr3} the correlation time for correlation function
$C_3$ is immediately obtained,
\begin{equation}
\eta:=\frac{\sqrt{m}}{\alpha}.
\label{eq:eta}
\end{equation}
The dissipation kernel $\Gamma(t)$ and its Fourier transform
$\tilde{\Gamma}(\omega)$ are depicted in Fig.\ \ref{C3figure}.
Relating to this Fourier transform \eqref{eq:FTCorr3} the first
particular property of the underlying correlation function emerges: For
$\omega=0$ the Fourier transform of Eq.\ \eqref{eq:Corr3} equals
zero in contrast to the other two correlation functions $C_1$ and $C_2$
(see Eqs.\ \eqref{eq:C1} and \eqref{eq:C2}).  Additionally, the
dissipation kernel $\Gamma(t)$ of correlation function $C_3$ drops
significantly below zero until it reaches a minimum and increases again,
approaching zero for $t \to \infty$.  Such a dissipative kernel for the
damping is rather typical in a quantum field theoretical setting with a
self-interacting theory like a scalar $\Phi^4$-theory (see
e.g. Ref.\ \cite{Xu:1999aq}).  Further particularities arise by solving the
GLE with dissipation kernel \eqref{eq:Corr3} for a free Brownian
particle, i.e.
\begin{equation}
\dot{v}+\frac{1}{m}\int_0^t\Gamma(t-t')v(t')\dd t'=\frac{\xi(t)}{m},
\label{eq:genLang1}
\end{equation} 
using the method of Green's functions. However, before applying the
method of Green's functions to the latter equation several modifications
of it have to be made, leading to
\begin{equation}
\dot{v}+\frac{1}{m}\int_{-\infty}^{\infty}\underbrace{\Gamma(t-t')\Theta(t-t')}_{:=\ii\Pi_\mathrm{ret}(t-t')}v(t')\dd t'=\frac{\xi(t)}{m}.
\label{eq:genLang2}
\end{equation}
From Eq.\ \eqref{eq:genLang1} to \eqref{eq:genLang2} the upper
integration border has been extended to $\infty$ by including the
Heaviside function into the integral. The lower integration border can
be extended to $-\infty$, assuming that $v(t)=0$ for $t<0$.

Using now the method of Green's functions the starting point is
\begin{equation}
\dot{G}_\mathrm{ret}(t)+\frac{i}{m}\int_{-\infty}^{\infty}\Pi_\mathrm{ret}(t-t')G_\mathrm{ret}(t')\dd t'=\delta(t).
\label{eq:greens1}
\end{equation}
The Fourier transform of this equation reads
\begin{equation}
-i\omega\tilde{G}_\mathrm{ret}+\frac{i}{m}\tilde{\Pi}_\mathrm{ret}(\omega)\tilde{G}_\mathrm{ret}(\omega)=1.
\label{eq:greens2}
\end{equation}
Proceeding further, by solving Eq.\ \eqref{eq:greens2} for
$\tilde{G}_\mathrm{ret}$, the solution to the actual problem (see
Eq.\ \eqref{eq:greens1}) is obtained by performing the inverse
Fourier transform of
\begin{equation}
\tilde{G}_\mathrm{ret} 
=\frac{\ii}{\omega-\frac{1}{m}\tilde{\Pi}_\mathrm{ret}}.
\label{eq:GretOm}
\end{equation}
But before applying the inverse Fourier transform, first one has to
determine $\tilde{\Pi}_\mathrm{ret}$, defined in Eq.\ 
\eqref{eq:genLang2}, as
\begin{equation}
\ii\Pi_\mathrm{ret}(t)=\Gamma(t)\Theta(t).
\end{equation}
By use of the convolution theorem, $\ii\tilde{\Pi}_\mathrm{ret}$ is given by
\begin{align}
\ii\tilde{\Pi}_\mathrm{ret} =\frac{1}{2\pi}\int_{-\infty}^{\infty}
  \tilde{\Gamma}(\omega')\tilde{\Theta}(\omega-\omega') \dd \omega'. 
\label{eq:Piret}
\end{align}
Insertion of Eq.\ \eqref{eq:FTCorr3} into Eq.\ \eqref{eq:Piret} then leads to 
\begin{equation}
  \ii\tilde{\Pi}_\mathrm{ret}=\frac{\ii
    g\alpha^3}{2\pi\sqrt{m}}\int_{-\infty}^{\infty}
  \underbrace{\frac{\omega^{\prime 2}}{\left(\omega'+\frac{\ii
          \alpha}{\sqrt{m}}\right)^2\left(\omega'-\frac{\ii
          \alpha}{\sqrt{m}}\right)^2}\frac{1}{\omega-\omega'+\ii
      \epsilon}}_{:=f(\omega')} \dd \omega'. 
\label{eq:Piret1}
\end{equation}
The integral on the right-hand side of Eq.\ \eqref{eq:Piret1} can be
computed by means of the theorem of residues,
\begin{equation}
\ii\tilde{\Pi}_\mathrm{ret} 
=\frac{g\alpha^3}{\sqrt{m}}\lim\limits_{\omega' \to
  -\frac{\ii \alpha}{\sqrt{m}}}\frac{\dd}{\dd
  \omega'}\left(\left(\omega'+\frac{\ii \alpha}{\sqrt{m}}\right)^2f(\omega')\right). 
\label{eq:PiRet}
\end{equation}
Evaluating Eq.\ \eqref{eq:PiRet} a compact form for $\tilde{\Pi}_\mathrm{ret}$ is obtained:
\begin{equation}
  \tilde{\Pi}_\mathrm{ret}=\frac{g\alpha^2\omega}{4\left(\omega+\frac{\ii \alpha}{\sqrt{m}}\right)^2}.
\label{eq:gamRetOm}
\end{equation}
With Eq.\ \eqref{eq:gamRetOm} the Fourier transform of the retarded
Green's function \eqref{eq:GretOm} is given by
\begin{equation}
\tilde{G}_\mathrm{ret} 
=\frac{\ii \left(\omega+\frac{\ii \alpha}{\sqrt{m}}\right)^2}{\omega\left(\omega-\frac{\sqrt{g}\alpha}{2\sqrt{m}}+\frac{\ii
      \alpha}{\sqrt{m}}\right)\left(\omega+\frac{\sqrt{g}\alpha}{2\sqrt{m}}+\frac{\ii
      \alpha}{\sqrt{m}}\right)}. 
\label{eq:GretOm1}
\end{equation}
Now that all ingredients are together, the retarded Green's function
$G_\mathrm{ret}$ can be computed by inverse Fourier transform of
Eq.\ \eqref{eq:GretOm1}:
\begin{equation}
\begin{split}
G_\mathrm{ret}(t) &=\frac{1}{2\pi}\int_{-\infty}^{\infty}
\tilde{G}_\mathrm{ret}(\omega)\exp\left(-\ii \omega t\right) \dd\omega\\
&=\frac{1}{2\pi}\int_{-\infty}^{\infty} 
\frac{\ii \left(\omega+\frac{\ii
      \alpha}{\sqrt{m}}\right)^2}{\omega\left(\omega-\frac{\sqrt{g}\alpha}{2\sqrt{m}}
    + \frac{\ii \alpha}{\sqrt{m}}\right)
  \left(\omega+\frac{\sqrt{g}\alpha}{2\sqrt{m}}+\frac{\ii
      \alpha}{\sqrt{m}}\right)}\exp\left(-\ii \omega t\right) \dd\omega \\
&=\frac{1}{2\pi}(-2\pi \ii)\sum_{i=1}^3 \mathrm{res}_{\omega_i}g(\omega),
\end{split}
\end{equation}
where the third equal sign follows making again use of the residue theorem. 

This finally leads to
\begin{equation}
  G_\mathrm{ret}(t)=\frac{1}{4+g}\left[4+\left(\frac{g}{2}+\ii
      \sqrt{g}\right)e^{-\frac{\alpha}{\sqrt{m}}
      t\left(1+\frac{\ii \sqrt{g}}{2}\right)}+\left(\frac{g}{2}-\ii
      \sqrt{g}\right)e^{-\frac{\alpha}{\sqrt{m}}
      t\left(1-\frac{i\sqrt{g}}{2}\right)}\right].
\end{equation}
Once the retarded Green's function of the system is known, the solution
of the GLE \eqref{eq:genLang2} is straight forwardly computed by the convolution of
the retarded Green's function and the inhomogeneity $\frac{\xi(t)}{m}$
of Eq.\ \eqref{eq:genLang2}:
\begin{equation}
v(t)=\underbrace{\int_0^t G_\mathrm{ret}(t-t')\frac{\xi(t')}{m}\dd t'}_{:=v_{\xi}(t)} + \underbrace{G_\mathrm{ret}(t)v(0)}_{:=v_{a}(t)}.
\label{eq:vC3}
\end{equation}
Given this solution, another specialty of dissipation kernel
\eqref{eq:Corr3} can be derived. By computing $\left<v^2(t)\right>$ in
the limit $t\to\infty$ it appears that the usual form of the
equipartition theorem in one dimension, given by
\begin{equation}
\lim\limits_{t\to\infty}\frac{1}{2}m\left<v^2(t)\right>=\frac{k_{\text{B}}T}{2}, \label{eq:equipart}
\end{equation}
no longer holds.  Squaring and subsequently averaging of Eq.\
\eqref{eq:vC3} leads to
\begin{equation}
\left<v^2(t)\right>=\left<v_{\xi}^2(t)\right>+\left<v_{a}^2(t)\right>, 
\label{eq:vsqrC3}
\end{equation}
where the mixed terms vanish as the initial velocity $v_0$ and the noise
$\xi(t)$ are uncorrelated, i.e. $\left<v(0)\xi(t)\right>=0$.  In what
follows, the values of both terms on the right-hand side of Eq.\
\eqref{eq:vsqrC3} are calculated separately.
\begin{figure}[tb]
	\begin{center}
          \begin{overpic}[width=0.8\textwidth]{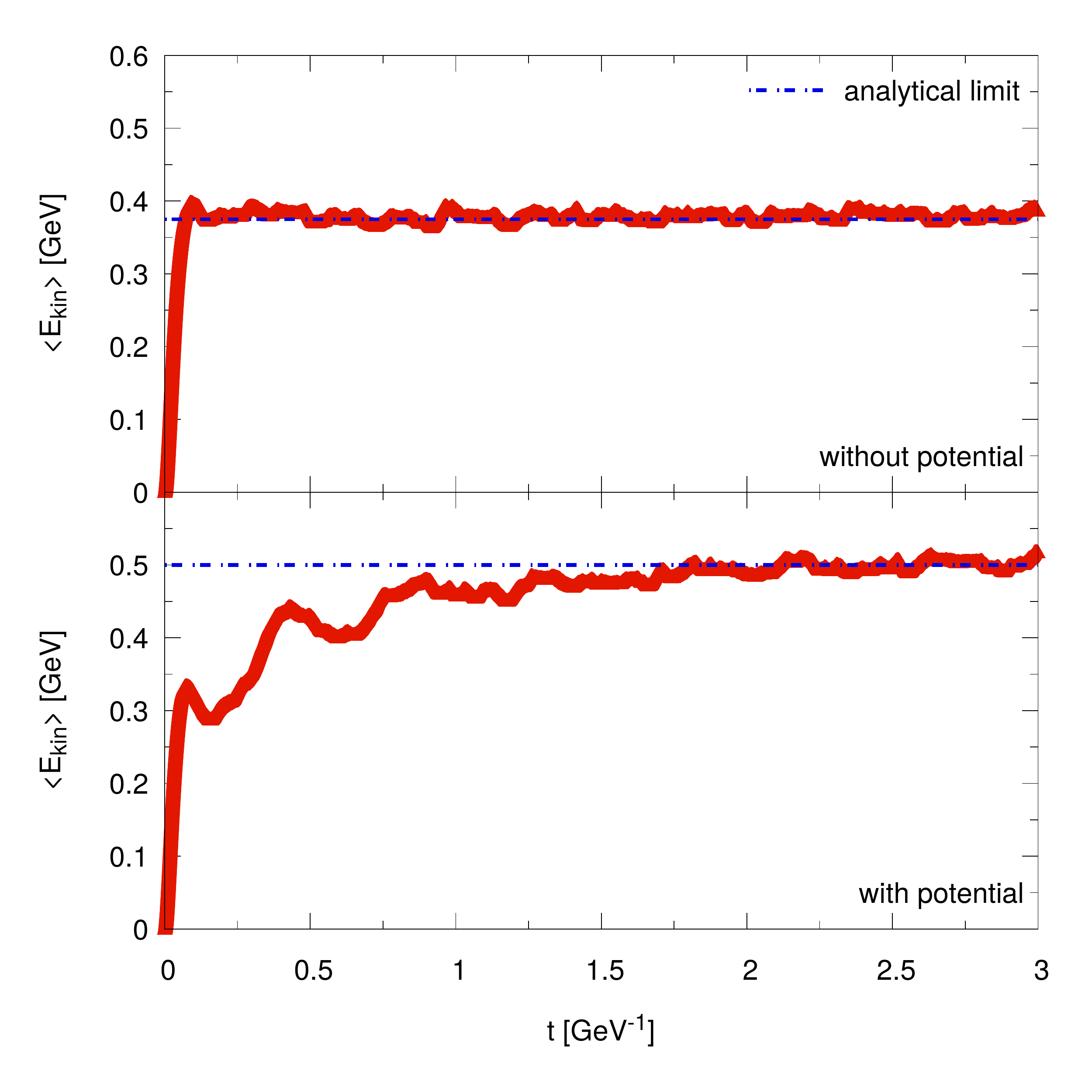}\put(65,35){}\end{overpic}
		\caption{Average kinetic energy
                  $\left<E_{\mathrm{kin}}(t)\right>$ with harmonic potential $V(x)=\frac{1}{2}m\omega_a^2x^2$ (lower figure) and without potential (upper figure) as a function of
                  the time $t$ and its limit for $t\to\infty$ (blue dotted dashed line, see
                  Eq.\ \eqref{eq:equipart} or \eqref{eq:limitEkin}), where
                  $g=4$, $m=0.1 \,\GeV$, $k_{\text{B}}=1$, $T=1 \,\GeV$,
                  $v_0=0$, $\omega_a=10 \,\GeV$ (lower figure) and step width
                  $\Delta t=3.1\cdot 10^{-5} \, \GeV^{-1}$.}
		\label{C3Ekin}
	\end{center}
\end{figure}

Starting with $\left<v_{\xi}^2(t)\right>$ the following computations have to be performed:
\begin{equation}
\begin{split}
\left<v_{\xi}^2(t)\right> &=\frac{1}{m^2}\int_0^t \dd t'\int_0^t \dd t'' G_\mathrm{ret}(t-t')G_\mathrm{ret}(t-t'')\left<\xi(t')\xi(t'')\right>\\
&=\frac{k_{\text{B}}T}{m^2}\int_0^t \dd t'\int_0^t \dd t'' G_\mathrm{ret}(t-t')G_\mathrm{ret}(t-t'')\Gamma(|t'-t''|)\\
&=\frac{k_{\text{B}}T}{m^2}\int_0^{ t} \dd \tau'\int_0^{t} \dd \tau'' G_\mathrm{ret}(\tau')G_\mathrm{ret}(\tau'')\Gamma(|\tau''-\tau'|)\\
&=\frac{k_{\text{B}}T}{m^2}\int_0^{t} \dd \tau'\int_0^{t} \dd \tau'' G_\mathrm{ret}(\tau')G_\mathrm{ret}(\tau'') \\&\times\left[\Theta(\tau'-\tau'')+\Theta(\tau''-\tau')\right]\Gamma(|\tau''-\tau'|) \\
&=2\frac{k_{\text{B}}T}{m^2}\int_0^{t} \dd \tau'\int_0^{\tau'} \dd \tau'' G_\mathrm{ret}(\tau')G_\mathrm{ret}(\tau'')\Gamma(|\tau'-\tau''|)\\
% 			   &=\frac{2T}{m^2}\frac{9m}{50}=\frac{9}{25}\frac{T}{m}, \quad t\to\infty
&=\frac{2k_{\text{B}}T}{m^2}\frac{mg(8+g)}{2(4+g)^2}=\frac{g(8+g)}{(4+g)^2}\frac{k_{\text{B}}T}{m}, \quad t\to\infty,
\end{split}
\end{equation}
where $\tau'=t-t'$ and $\tau''=t-t''$.

Furthermore, for $\left<v_{a}^2(t)\right>$ the following expression is
obtained in the limit of $t \to \infty$:
\begin{align}
\left<v_{a}^2(t)\right> &=G_\mathrm{ret}^2(t)\left<v^2(0)\right>=\frac{16}{(4+g)^2}\left<v^2(0)\right>,\quad t\to\infty.
\end{align}
Bringing together both solutions results in 
\begin{equation}
\begin{split}
\lim\limits_{t\to\infty}\left<v^2(t)\right>&=\lim\limits_{t\to\infty}\left(\left<v_{\xi}^2(t)\right>+\left<v_{a}^2(t)\right>\right) \\
&=\frac{g(8+g)}{(4+g)^2}\frac{k_{\text{B}}T}{m}+\frac{16}{(4+g)^2}\left<v^2(0)\right>,
\end{split}
\end{equation}
corresponding to the following mean kinetic energy in the limit of $t\to\infty$
\begin{equation}
\lim\limits_{t\to\infty}\frac{1}{2}m\left<v^2(t)\right>=
\frac{g(8+g)}{(4+g)^2}\frac{k_{\text{B}}T}{2}+\frac{8m}{(4+g)^2}\left<v^2(0)\right>.
\label{eq:limitEkin}
\end{equation}
Figure \ref{C3Ekin} shows that the numerical simulations in fact yield
the analytically expected behavior of the kinetic energy in the limit of
$t\to\infty$.  

\begin{figure}[tb]
	\begin{center}
          \begin{overpic}[width=0.8\textwidth]{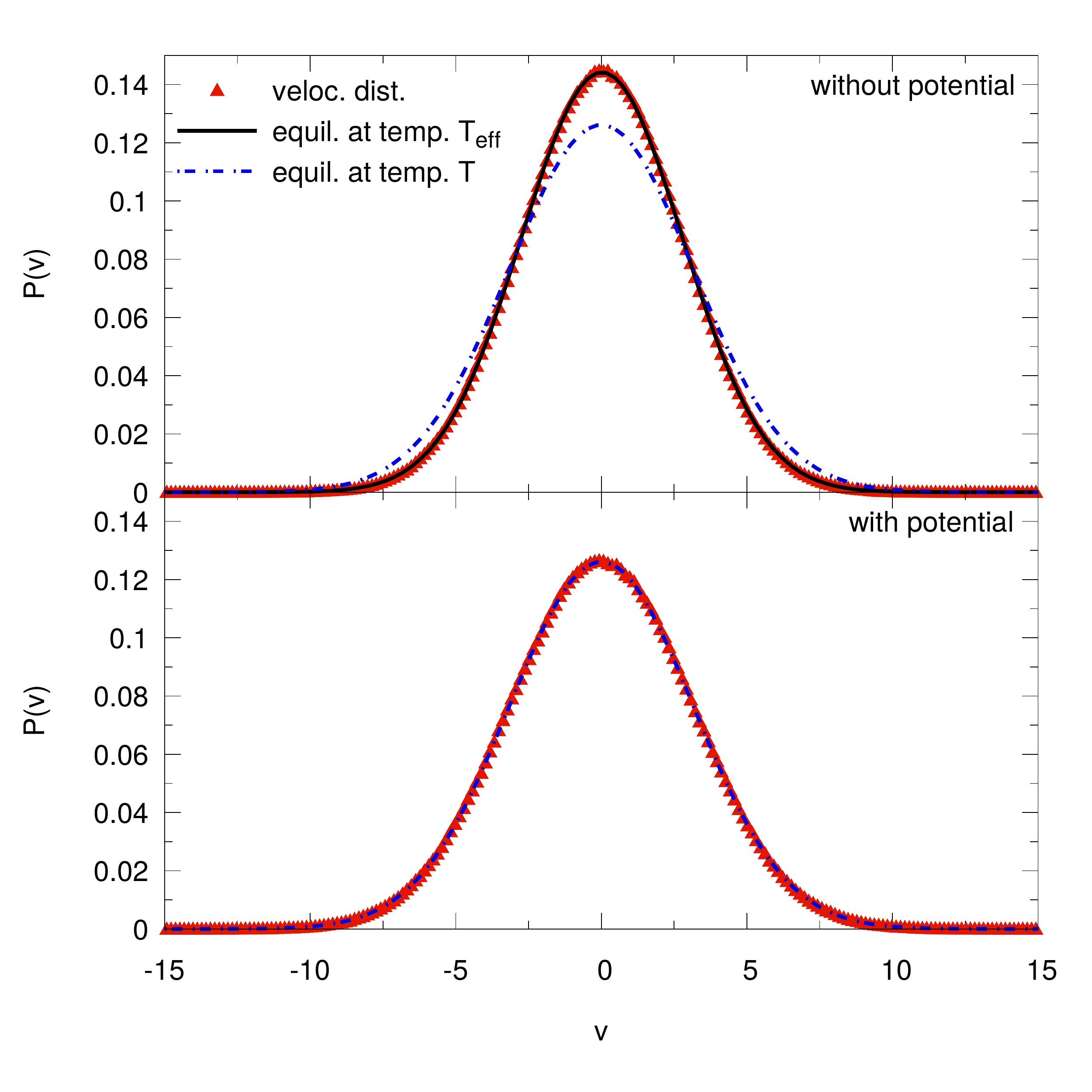}\end{overpic} 
          \caption{Velocity distribution $P(v)$ (red triangles) with  
          	harmonic potential $V(x)=\frac{1}{2}m\omega_a^2x^2$ (lower figure) and without potential (upper figure)
          	and the equilibrium distributions for temperature $T_{\mathrm{eff}}$
            (black line) (see Eq.\ \eqref{eq:Teff}) and temperature
            $T$ (blue dotted dashed line), where $g=4$, $m=0.1 \,\GeV$,
            $k_{\text{B}}=1$, $T=1 \,\GeV$, $v_0=0$, $\omega_a=10 \,\GeV$ (lower figure) and time step
            $\Delta t=3.1\cdot 10^{-5} \, \GeV^{-1}$.}
          \label{Verteilung}
	\end{center}
\end{figure}

Investigating furthermore the velocity distribution function it appears
that thermal equilibrium is established but with a temperature reduced
by approximately a factor $\frac{g(8+g)}{(4+g)^2}$, which is the
coefficient of the first term in Eq.\ \eqref{eq:limitEkin} (see
Fig.\ \ref{Verteilung}). Based on these considerations an effective
temperature $T_{\mathrm{eff}}$ can be defined as
\begin{equation}
T_{\mathrm{eff}}=\frac{g(8+g)}{(4+g)^2}T.
\label{eq:Teff}
\end{equation}
This pathological behavior of insufficient thermalization directly stems
from the fact that the Fourier transform, $\tilde{\Gamma}(\omega)$ (see
Eq.\ \eqref{eq:FTCorr3}), of correlation function $C_3$ vanishes in the
limit of $\omega \rightarrow 0 $. In contrast, for a Brownian particle
trapped in a standard oscillator potential, one can analytically prove
that the particle thermalizes for the kinetic as well as for the
potential energy. Numerical simulations of such a Brownian particle,
originally trapped at the bottom of a harmonic potential
$V(x)=\frac{1}{2}m\omega_a^2x^2$, indeed show that the usual form of the
equipartition theorem (see Eq.\ \eqref{eq:equipart}) is again valid and
thermal equilibrium with temperature $T$ instead of $T_{\mathrm{eff}}$
is recovered (see Fig.\ \ref{Verteilung}). In an analogous manner the
retarded Green's function for the position $x(t)$ will contain poles
below the real axis at $\omega = \pm \omega_a$.  For weak coupling the
effective damping is then obtained by
$\tilde{\Gamma}(\omega=\omega_a)/2$ in the linear harmonic (or
quasi-particle) approximation
\cite{Greiner:1996dx,Greiner:1998vd,Xu:1999aq}.

%\vspace*{1mm}
\pagebreak

\section*{Acknowledgment} We thank S.\ Leupold for fruitful discussions
about the dissipation kernel $C_3$ and J.\ Schmidt for providing his
$\text{C}^{++}$ implementation of colored noise. B.S.\ acknowledges
support through the Helmholtz Graduate School for Hadron and Ion
Research for FAIR (HGS-HIRe) and financial support within the framework
of the cooperation between GSI Helmholtz Centre for Heavy Ion Research
and Goethe-Universit{\"a}t Frankfurt am Main (GSI F\&E program). We are
grateful to the LOEWE Center for Scientific Computing (LOEWE-CSC) at
Frankfurt for providing computing resources.  We also acknowledge
support by the Deutsche Forschungsgemeinschaft (DFG, German Research
Foundation) through the grant CRC- TR 211 `Strong-interaction matter
under extreme conditions' - Project number 315477589 - TRR 211.

%\pagebreak

\bibliographystyle{els-hvh}

\end{document}